\documentclass[fleqn,10pt]{wlscirep}
\usepackage[utf8]{inputenc}
\usepackage[T1]{fontenc}

\usepackage[final]{changes}
\usepackage{hyperref}

\definechangesauthor[name={Boyang}, color=red]{Boyang}

\title{Physical co-presence of workers is linked with the evolution of interfirm networks}

\author[1,2]{Ketian Chen}
\author[3,4]{Hiroyasu Inoue}
\author[1,2]{Yoshihide Sekimoto}
\author[2,5]{Yoshiki Ogawa}
\author[6,7,*]{Takahiro Yabe}
\affil[1]{Graduate School of Engineering, The University of Tokyo, Tokyo, Japan}
\affil[2]{Center for Spatial Information Science, The University of Tokyo, Tokyo, Japan}
\affil[3]{Graduate School of Information Science, University of Hyogo, Kobe, Japan}
\affil[4]{Center for Computational Science, RIKEN, Kobe, Japan}
\affil[5]{Graduate School of Social Data Science, Hitotsubashi University, Tokyo, Japan}
\affil[6]{Center for Urban Science and Progress, Tandon School of Engineering, New York University, New York, USA}
\affil[7]{Department of Technology Management and Innovation, Tandon School of Engineering, New York University, New York, USA}

\affil[*]{takahiroyabe@nyu.edu}

\begin{abstract}
Inter-firm transaction networks underpin modern economies and are critical for economic growth and resilience.
Previous studies have largely explained the formation of inter-firm ties through spatial proximity.
Yet people routinely move across space, creating opportunities for encounters that extend beyond geographical proximity. 
How these opportunities relate to inter-firm transaction relationships remains poorly understood. 
Here, we combine large-scale mobile phone mobility data with a multi-year panel of transactions among 196,317 firms in Tokyo to quantify encounter opportunities between workers and examine their association with inter-firm transaction ties. 
We find that pairs of areas with greater opportunities for encounters one another are more likely to have transaction links and new link formation between previously unconnected firms.
This association is particularly pronounced in information- and service-intensive industries, reflecting the importance of encounter opportunities for exchange involving intangible knowledge and service. 
Indirect connections through shared neighbors in the encounter network are also positively associated with subsequent transaction formation across industries, but not within industries.
These findings suggest that urban policies shaping connectivity formed through human mobility could play a significant role in the development and resilience of economic networks.
\end{abstract}

\begin{document}

\flushbottom
\maketitle
%
%
\thispagestyle{empty}


\section*{Main}\label{sec1}

Firm-to-firm transaction networks are fundamental to the organization of modern economies. The production of goods or services depends on a chain of upstream suppliers and downstream buyers, and the structure of these chains shapes resource allocation and the propagation of shocks across the economy \cite{bernard2019production,acemoglu2012network,inoue2019firm}. For individual firms, trade relationships shape firm performance and exposure to shocks transmitted through production networks
\cite{bernard2022origins,barrot2016input,carvalho2021supply,yang2024post}. Understanding how these inter-firm ties emerge, persist and dissolve is therefore central to explaining the organization and dynamics of the real economy. 

From a theoretical perspective, endogenous production-network models characterize partner selection as an optimization problem in which firms choose trading partners by weighing production costs, technology and risks \cite{dhyne2023endogenous,elliott2022supply,oberfield2018theory,taschereau2026cascades}. In practice, however, firms cannot costlessly identify and establish relationships with suitable partners because of the existence of frictions. Studies of production networks and international trade have highlighted the frictions involved in partner selection, including search costs \cite{rauch2001business,atkin2017exporting}, trust building \cite{macchiavello2015value}, acquiring information about market opportunities \cite{allen2014information,steinwender2018real}, and geographic and travel costs \cite{bernard2019production}. However, direct empirical evidence remains limited on how firms overcome these frictions and how such processes translate potential matches into realized economic links. Addressing this question requires shifting attention from the firm-level characteristics of existing relationships to the processes through how firms become connected by overcoming frictions.

An important part of this process may occur through interactions between the people who work for firms. Economic relationships are embedded in the interpersonal networks through which information, trust, and reputation develop \cite{granovetter1985economic}. Establishing and sustaining trading relationships therefore depend on communication between people, who must exchange information, assess one another’s reliability, and coordinate their activities to mitigate frictions. Although digital channels support communication across distance, face-to-face interaction remains particularly valuable when communication involves complex and tacit knowledge \cite{leamer2007flat,storper2004buzz} or requires close collaboration \cite{battiston2021face,emanuel2023power,inoue2025human}. Consistent with this argument, studies of business travel suggest that in-person contact facilitates information exchange and relationship building in trade \cite{cristea2011buyer, soderlund2023importance}. Recent studies show that the observed co-location of individuals from different organizations can benefit cross-organizational collaboration \cite{chai2019temporary} and facilitate knowledge flows between firms \cite{atkin2022}. This evidence motivates examining whether encounter opportunities between workers are also associated with the formation of transaction ties between their firms.

Recent advances in large-scale human mobility data enable the observation of human movement at unprecedented scale and resolution \cite{xu2025using}. These data provide new insights into the spatial organization of economic activity by revealing where individuals work, consume, and travel \cite{miyauchi2025mobility,kreindler2023measuring}. Previous studies have shown that mobility patterns can closely track economic indicators, including sales and production \cite{matsumura2024nowcasting,yanotti2026walking}, business resilience \cite{yabe2025behaviour}, regional income \cite{kreindler2023measuring,blumenstock2015predicting} and property values \cite{wang2024infrequent}. On the other hand, mobility data can measure opportunities for encounters across social groups and the segregation people experience in daily life \cite{nilforoshan2023human,moro2021mobility,yabe2023behavioral}. These approaches connect mobility to economic activity and interpersonal exposure, respectively. Bringing them together makes it possible to ask whether workers’ movements reveal opportunities for contact that are associated with economic connections between organizations.

In this study, we examine whether mobility-based encounters between workers are associated with the formation of inter-firm trading relationships. Using large-scale anonymized mobile phone data containing more than 300,000 users with daily location records, we infer workers’ workplace locations and use their spatial and temporal co-presence to construct an encounter network between geographic cells containing firms, which we link to transaction relationships aggregated from records covering more than 190,000 firms in Tokyo, Japan. A doubling of meeting opportunities are associated with 1.4\% increase in odds of transaction link existence, and 1.9\% increase in odds of subsequent subsequent link formation among those with no prior trading relationship. We further examine how these associations vary over time, across geographic distance, and across industry structures. Beyond pairwise encounter opportunities, we examine indirect connections (i.e., 2-hop) through shared neighbors in the encounter network. These indirect connections are positively associated with subsequent transaction formation across industries, but not within industries. These findings suggest that worker encounters constitute an observable micro-level process linking urban mobility to the formation of economic connections between firms.

\section*{Results}\label{sec2}

\subsection*{Measuring encounter opportunities between workers at scale}\label{subsec_mp}

Our central hypothesis is that interactions between people representing different firms facilitate inter-firm transactions. Such interactions may help firms overcome frictions in identifying, evaluating, and coordinating with potential partners. If so, opportunities for encounters that may foster such interactions should be observable before a transaction relationship is formed. We use physical encounters in the city between workers of different firms as a measurable proxy for these opportunities. We construct a large-scale measure of the cross-firm encounter network in Tokyo using two data sources. The first dataset is an anonymized smartphone GPS dataset provided by Blogwatcher, Inc. The dataset consists of GPS location records in Japan collected through more than 140 smartphone applications from users who opted in to location sharing. We restricted the GPS sample to users observed in Tokyo between July and September 2019 with location records spanning more than 90 distinct days, yielding 329,149 active users, which corresponds to approximately 3.5\% of Tokyo's working-age population (aged 20--69). Raw trajectories are converted into stay points and only stays that exceed 10 minutes are retained \cite{hariharan2004project}. We further examined GPS sampling rates across municipalities and found consistent coverage across areas, with a sampling rate of approximately 2.2\% on average (Supplementary Table S1).

Our second data source is the inter-firm transaction dataset provided by Teikoku Databank, a private corporate credit research firm maintaining one of the most comprehensive repositories of company information in Japan. Its firm database covers approximately 70\% of Japanese companies and records each firm's headquarters location, industry classification, and a unique firm code. The transaction dataset records the firm codes of buyers and suppliers for each reported transaction annually, allowing us to observe the formation and persistence of inter-firm economic ties over time. We therefore represent each reported buyer–supplier relationship as a binary link indicating whether two firms have a transaction relationship in a given year. Using these records, we construct a panel dataset covering 196,317 Tokyo-headquartered firms and 3,293,848 annual buyer–supplier link observations from 2016 to 2022.

To translate individual mobility traces into a cross-firm encounter network, we first identify commuting workers from the mobile phone location records. Because Tokyo's extreme building density and verticality make building-level workplace inference highly sensitive to GPS positioning errors, we aggregate firm locations into hexagonal cells using the H3 cells at resolution 10 (average edge length of approximately $76~\mathrm{m}$). We designate cells containing at least one registered firm headquarters as firm cells. We defined a working day as a weekday on which a user spent at least 5 hours in a firm cell. Users are classified as workers if they have at least 40 working days at their primary workplace, spanning at least 12 distinct weeks within the three-month observation period (Methods). The identified worker population was robust to changes in the minimum daily workplace stay duration, number of working days, and number of distinct weeks required for worker identification (Supplementary Figure S4). To assess the spatial representativeness of the identified worker sample, we compare the estimated number of workers with the 2020 Population Census of Japan \cite{statisticsbureau2020census}, finding a high correlation ($r = 0.80$; Supplementary Figure S5). The inferred workers also exhibit stable weekday morning arrival times at their workplaces, consistent with regular full-time commuting patterns in Japan (Supplementary Figure S6).

\begin{figure}[!tb]
\centering
\includegraphics[width=\textwidth]{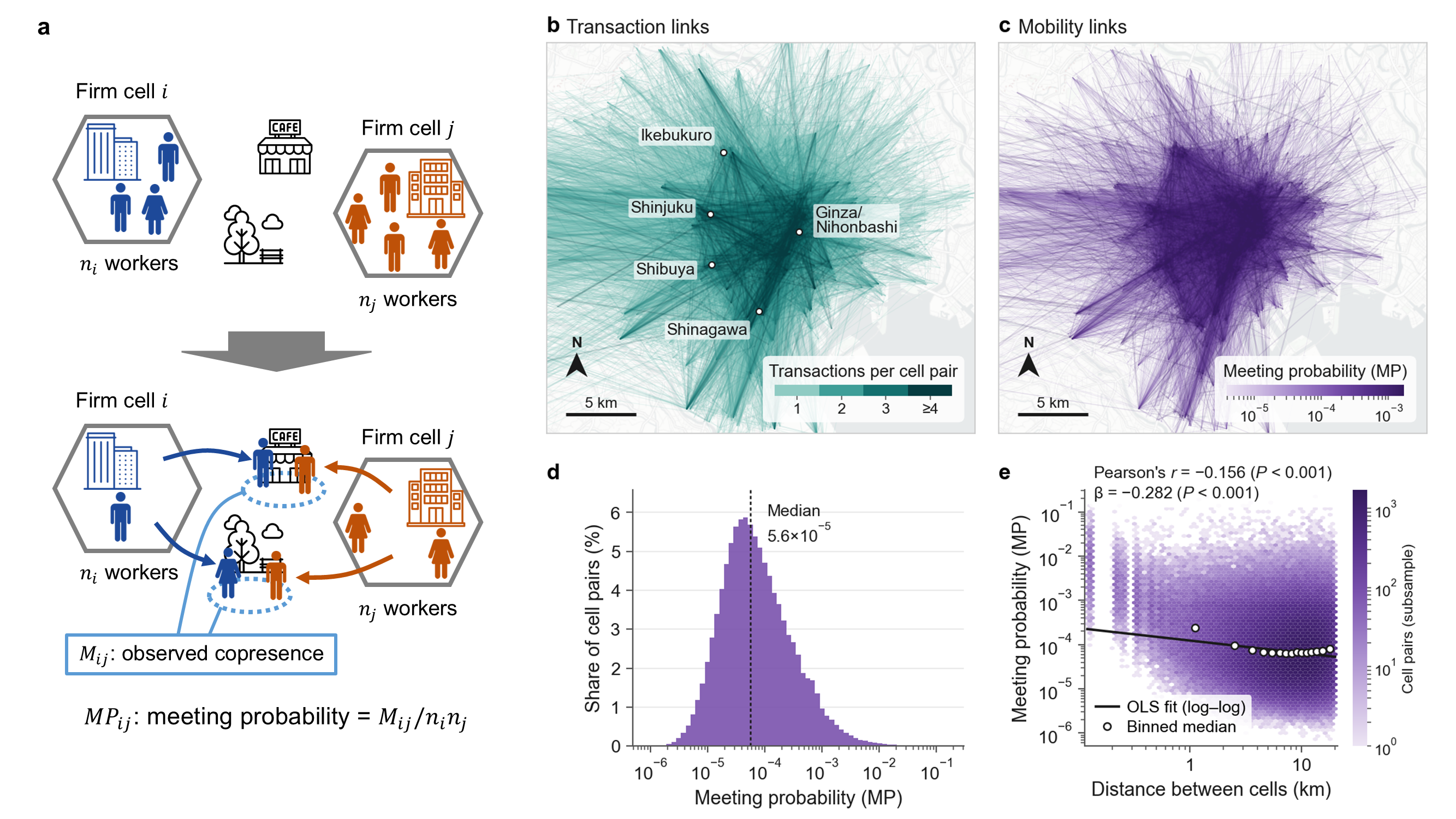}
\caption{\textbf{Measuring encounters between workers at scale.}
\textbf{a.} Conceptual illustration of measuring meeting probability. \textbf{b.} Transaction links among firm cells in Tokyo. Lines connect cells with observed inter-firm transaction relationship in 2020. \textbf{c.} Mobility links among firm locations in Tokyo. Lines connect pairs of cell with positive meeting probability ($\mathit{MP}>0$). \textbf{d.} Distribution of meeting probability ($\mathit{MP}$). The $x$-axis is displayed on a logarithmic scale. \textbf{e.} Scatter plot of meeting probability against the geographic distance between cells. Meeting probability decreases with distance, with a negative correlation (Pearson's $r=-0.156$, $p<0.001$).}
\label{fig:fig1}
\end{figure}

To quantify encounter opportunities between workers from cell pairs $(i, j)$, we use the observed spatial-temporal co-presence as a proxy \cite{atkin2022}. Because a physical encounter requires two individuals to be present at the same place and time, co-presence provides a necessary condition for encounter. We first define co-presence as $\mathrm{M}_{ij} = \sum_{c,t} n_{ict} n_{jct}$, where $c$ indexes H3 cells, $t$ indexes predefined 30-minute intervals, and $n_{ict}$ is the number of workers from firm cell $i$ whose stay overlaps with cell $c$ during interval $t$ for at least 10 minutes. Thus, $\mathrm{M}_{ij}$ counts worker-pair co-presence instances between firm cells $i$ and $j$. Because firms differ substantially in the number of workers simultaneously active throughout the day, we normalize observed co-presence by the maximum number of worker pairs that could potentially meet given their temporal activity. We then define the potential maximum meetings as $\mathrm{PMM}_{ij} = \sum_{t} \left(\sum_{c} n_{ict}\right)\left(\sum_{c} n_{jct}\right)$, which counts the number of worker pairs active in the dataset at the same time $t$, regardless of their locations. Finally, we define the meeting probability $\mathrm{MP}_{ij} = \mathrm{M}_{ij}/\mathrm{PMM}_{ij}$, where co-presence within the same location and time window constitutes a necessary condition for direct encounters. The resulting meeting probability $\mathrm{MP}_{ij}$ captures the probability that a randomly drawn worker from cell $i$ and one from cell $j$ are observed in the same location at the same time (see Figure \ref{fig:fig1}a for conceptual illustration). To assess whether the observed meeting probability (MP) is sensitive to potential measurement error, we construct three alternative MP measurements. We down-weight encounters in crowded or widely visited locations, and exclude encounters observed only once. The results in Supplementary Figure S8 shows these alternative meeting probability measurement preserve substantial features of the observed one, suggesting that the overall structure of observed meeting probability is robust to alternative specifications. Detailed definitions and results are provided in Supplementary Note 2.3. 

To assess whether observed meeting probability reflects temporal alignment beyond shared spatial mobility patterns, we construct a time-shift null model that preserves individual mobility patterns while disrupting the temporal alignment of visits. Specifically, we shift each worker’s entire daily mobility trajectory to another calendar date and recompute MP across 100 randomized realizations. This procedure preserves the locations visited and within-day mobility patterns while disrupting the original timing of encounters between workers. For example, two workers may arrange to meet at a café at noon on a particular day. Shifting their daily trajectories to different dates would eliminate this planned encounter, even though both workers still visit the same café at noon. Supplementary Figure S9 shows that the observed MP distribution is systematically higher than the null distribution. Across cell pairs, the median ratio of observed MP to null MP is 1.61, indicating that observed meeting probability is typically 61\% higher than its null expectation. These results suggest that temporal alignment contributes to observed encounter opportunities beyond what would be expected from shared mobility patterns alone. Detailed results are reported in Supplementary Note 2.4.

We next examine the spatial organization of mobility-based encounter opportunities and compare it with that of inter-firm transaction links. Figure \ref{fig:fig1}b and Figure \ref{fig:fig1}c visualize the transaction links and mobility links, respectively. Each node represents a firm cell. Transaction links connect cells containing firms with observed transaction relationships in 2020, whereas mobility links connect cell pairs with a positive meeting probability ($\mathit{MP}>0$). The two networks exhibit broadly similar spatial patterns, with dense links concentrated around multiple major business and commercial centers, consistent with Tokyo's polycentric organization. Having established where mobility links exist, we next examine how strong these links are. Figure \ref{fig:fig1}d shows the distribution of meeting probability across firm-cell. The distribution is centered around $10^{-4.5}$ but has a pronounced right tail, indicating substantial variation in encounter intensity across connected cell pairs. This variation is also visible at the node level. Firm cells differ substantially in degree, defined as the number of other cells connected through positive meeting probabilities, and node strength, defined as the sum of meeting probabilities across these connections (Supplementary Figure S7). Taken together, these distributions reveal that encounter opportunities are far from uniform, varying by several orders of magnitude in both intensity and connectivity. We finally examine how mobility-link strength varies with the geographic distance between the two firm cells. For each pair of cells $i$ and $j$, we compute the Haversine distance $d_{ij}$ and estimate how the meeting probability $MP_{ij}$ varies with distance in Figure \ref{fig:fig1}e. Meeting probability decreases with inter-cell distance (Pearson's $r = -0.156$, $p < 0.001$). The estimated log--log regression coefficient is $-0.282$, indicating that when distance doubles, meeting probability decreases by approximately 18\%. 

\subsection*{Physical encounters between workers are associated with economic ties}\label{subsec_base}

\begin{figure}[!tb]
    \centering
    \includegraphics[width=\textwidth]{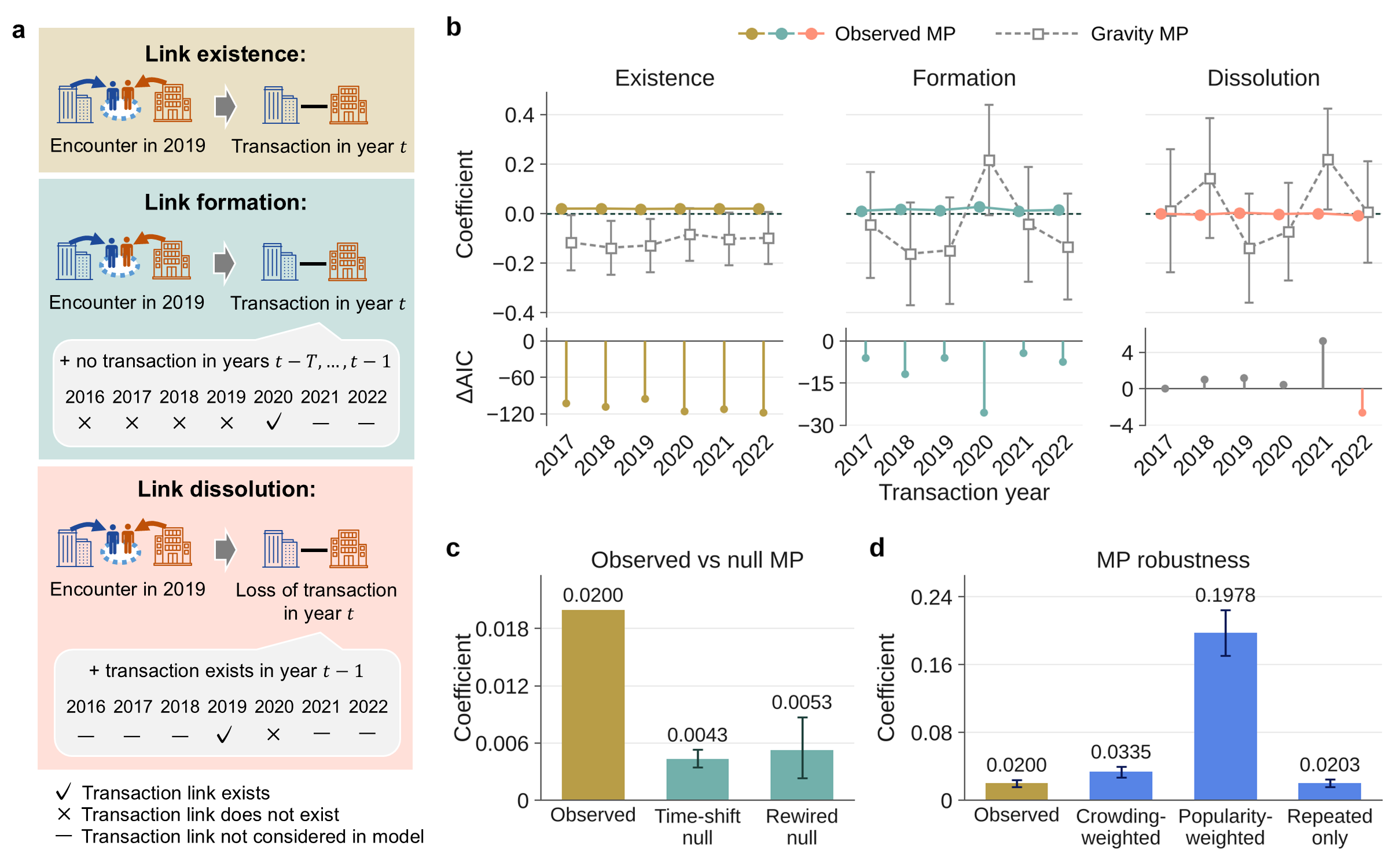}
    \caption{\textbf{Encounter effects on inter-firm transaction ties.}
\textbf{a.} Conceptual definitions of transaction outcomes.
\textbf{b.} Associations between meeting probability and inter-firm transaction ties using observed and gravity MP.
Upper panels show coefficients from logistic regressions of link existence, link formation, and link dissolution across years.
Points represent coefficient estimates, and error bars indicate 95\% confidence intervals.
Lower panels show the difference in Akaike information criterion ($\Delta\mathrm{AIC} = \mathrm{AIC}_{\mathrm{observed}} - \mathrm{AIC}_{\mathrm{gravity}}$), where negative values favor the observed-MP model.
\textbf{c.} Comparison of regression coefficients for observed MP and two null models (time-shift and rewiring) for link existence in 2020.
Bars represent the observed coefficient and mean coefficients across 100 randomization for each null model.
Error bars indicate the 2.5th--97.5th percentile ranges for the null models.
\textbf{d.} Robustness of the association between MP and link existence in 2020, comparing regression coefficients across four MP measures: observed, crowding-weighted, popularity-weighted, and repeated-only MP.
All regressions include the same control variables. Error bars indicate 95\% confidence intervals based on two-way clustered standard errors.}
    \label{fig:figure2}
\end{figure}


We examine whether worker encounters are associated with different inter-firm transaction outcomes. Specifically, we relate meeting probability measured from 2019 mobility data to transaction outcomes observed at different temporal lags and leads. The dependent variable is a binary indicator of whether a transaction relationship exists between two firm cells in a given year. Our main independent variable is the logarithm of meeting probability (MP). We estimate logistic regression models following Equation~\ref{eq:baseline} in Methods, controlling for relevant covariates and firm-cell fixed effects.

We use three binary transaction outcomes to distinguish different stages of inter-firm transaction relationships (Figure \ref{fig:figure2}a). The first outcome variable captures link existence, indicating whether a cell pair had at least one buyer–supplier relationship in a given year. link formation identifies the first observed transaction link between a pair with no link in any preceding year covered by the data. For example, a pair classified as a link formation at $t=1$ has no observed transaction during 2016--2019 but develops a transaction link in 2020. The third outcome captures link dissolution, defined as the first disappearance of a transaction link. For example, dissolution at $t=1$ means that a transaction link observed in 2019 disappears in 2020. 

To assess whether observed meeting probability contains information beyond that captured by spatial distance and cell size, we construct a gravity-based benchmark using the generalized gravity law, $MP_{ij}^{\mathrm{gravity}} = A n_i n_j / (d_0 + d_{ij})^{\gamma}$, where $n_i$ and $n_j$ are the numbers of workers in cells $i$ and $j$, respectively; $d_{ij}$ is the distance between the two cells; $d_0$ is a fixed distance offset; and $\gamma$ is the exponent parameter. We fit the gravity model using Poisson pseudo-maximum likelihood (PPML), obtaining $\gamma = 1.354$ with $d_0 = 0.1$ km (see Supplementary Note 2.5 for details). The gravity-based benchmark preserves the worker counts and normalization used in observed MP, but replaces observed co-presence counts with gravity-model predictions. We then estimate the transaction models using both observed MP and gravity MP. This comparison assesses whether a benchmark based on spatial proximity and cell size reproduces the associations observed for actual worker encounters (Figure \ref{fig:figure2}b).

For link existence, observed MP is positively and significantly associated with the presence of transaction links in all observation years, with stable coefficients of approximately 0.020 ($p<0.001$). This means that a doubling of meeting probability is associated with approximately a 1.4\% increase in the odds of an inter-firm transaction. Replacing observed MP with gravity MP yields negative coefficients across 2017–2022, ranging from $-0.137$ to $-0.083$. Thus, the positive association between encounters and transaction connectivity is not reproduced by the fitted gravity benchmark. Models using observed MP consistently achieve lower AIC values than those using gravity MP, with differences ranging from 95 to 118. This improvement in model fit suggests that observed meeting probability captures information about inter-firm connectivity that is not fully represented by the gravity-based benchmark.

For link formation, observed MP is positively and significantly associated with the formation of new transaction ties across all six years ($p<0.05$), with the largest effect observed in 2020, immediately after the mobility measurement (${\beta} = 0.027$, $p < 0.001$). This means a doubling of meeting probability is associated with approximately a 1.9\% increase in the odds of link formation in the following year. This positive association persists across alternative definitions requiring one to four years without a transaction link before formation (Supplementary Note 3.6). In contrast, gravity MP coefficients vary in sign and are statistically insignificant in every year. Models using observed MP consistently achieve lower AIC values. 

For link dissolution, coefficients of observed MP are small, vary in sign, and are statistically insignificant throughout the observation period ($p>0.05$). Gravity MP likewise shows no consistently significant association either. Differences in AIC are comparatively small and show no consistent advantage for either model. Together, these results suggest that meeting probability is more systematically associated with the formation of new transaction links than with the dissolution of existing ones. The results remain consistent across alternative transaction-history windows, ranging from one to four years for both link formation and dissolution (Supplementary Figure S12).

The preceding analyses establish a consistent positive association between observed MP and both the existence and formation of transaction links. We next examine whether these associations are affected by underlying spatial and network structures and check the robustness. First, we assess whether the association is sensitive to the measurement of MP. The estimated MP coefficient remains positive and significant under three alternative measures that down-weight encounters in crowded or frequently visited locations, and exclude one-time encounters (Figure \ref{fig:figure2}c). Detailed definitions and construction procedures for these alternative measures are provided in Supplementary Note 2.3. and Supplementary Table S6. 

Second, to assess whether the observed association reflects temporally synchronized encounters rather than workers simply visiting similar places, we replace observed MP with a null MP generated from a time-shift model (Supplementary Note 2.4 and 3.4). The null model preserves each worker's observed trajectory but randomly shifts the dates of those visits, thereby retaining where workers visit while disrupting their actual temporal alignment. The left and middle bars in Figure\ref{fig:figure2}d show that replacing observed MP with null MP reduces the estimated coefficient by approximately 79\%. This reduction suggests that shared visitation patterns alone do not reproduce the association observed for temporally aligned encounters. In addition, a further concern is that the observed association may simply reflect differences in the overall connectivity of firm cells. Highly connected cells may exhibit both more transaction links and higher MP with other cells. We therefore randomize transaction partners while preserving the number of links associated with each cell (Supplementary Note 3.5). Under these degree-preserving rewired networks, the MP estimates fall by 74\% compared with the observed MP in the same model (the right bar in Figure \ref{fig:figure2}c). This indicates that meeting probability is associated not only with how connected cells are, but also with who firms transact with.

\subsection*{Heterogeneity across time, distance, and industries}\label{subsec_hetero}

\begin{figure}[!tb]
    \centering
    \includegraphics[width=\textwidth]{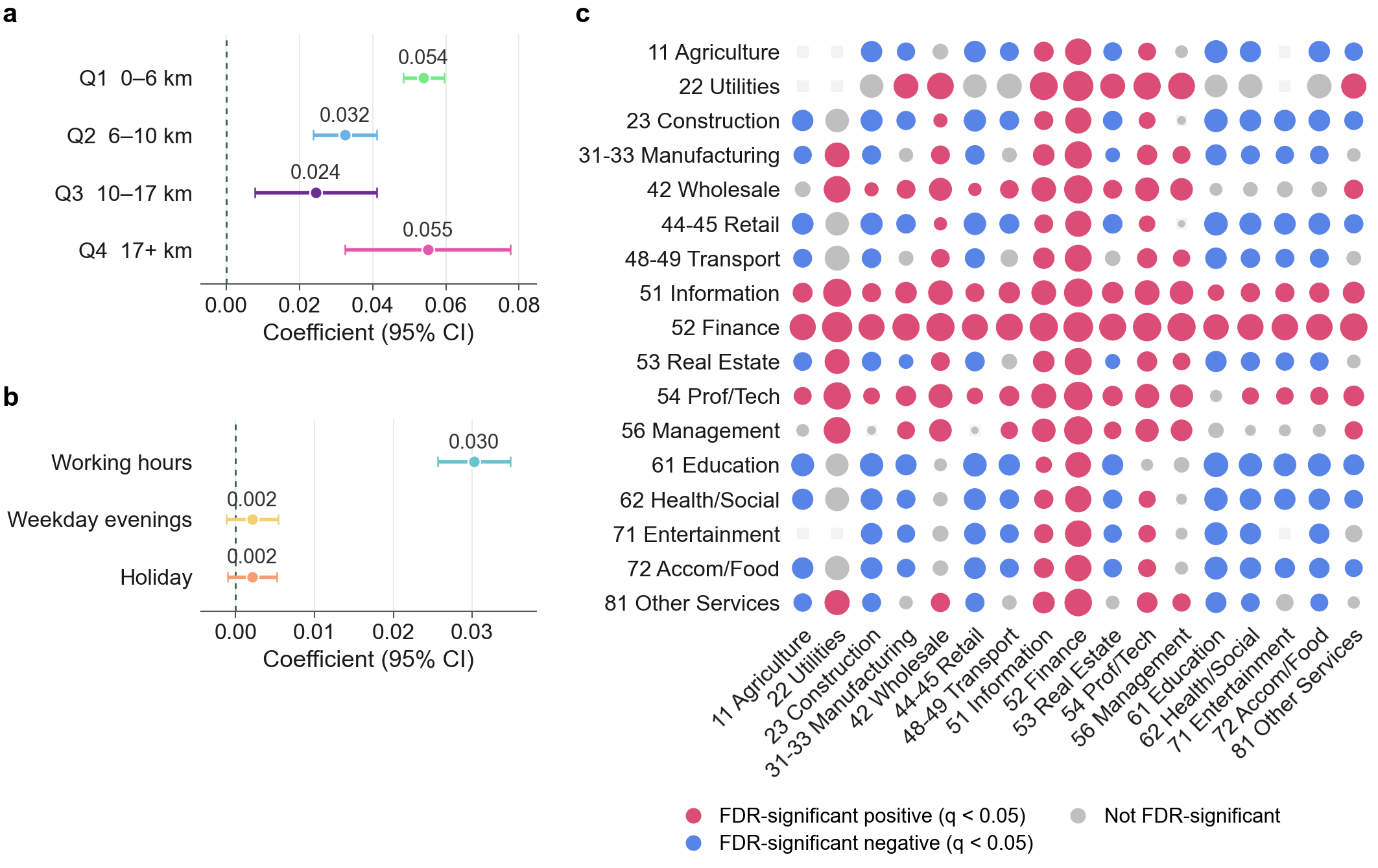}
    \caption{\textbf{Heterogeneity analysis.}
\textbf{a.} Regression coefficients for meeting probability across quartiles of inter-cell distance in the baseline model ($t = 1$). The association remains positive across all distance groups, with larger coefficients for the nearest and farthest quartiles. \textbf{b.} Regression coefficients for meeting probability during working hours, weekday evenings, and holidays in baseline model ($t = 1$). Working-hours meeting probability shows the strongest positive association with link existence. Error bars show 95\% confidence intervals. \textbf{c.} Dot matrix showing estimated coefficients for meeting probability across industry pairs. Dot size is proportional to the absolute magnitude. Red and blue dots indicate positive and negative coefficients that remain significant after FDR correction ($q < 0.05$); grey dots indicate non-significant estimates.}
    \label{fig:figure3}
\end{figure}

Geographic distance may shape how encounter opportunities are associated with inter-firm transaction ties. If meeting probability simply captures spatial proximity, its association with transaction ties should weaken as the distance between firms increases. We therefore divide cell pairs into distance quartiles and include interaction terms between meeting probability and distance-quartile indicators. A Wald test rejects the null hypothesis that the coefficients are equal across the four distance quartiles ($\chi^{2}(3)=26.99$, $p<0.001$), indicating significant heterogeneity in the association between meeting probability and transaction ties. Figure \ref{fig:figure3}a shows that the effect of meeting probability remains positive across all distance quartiles. The estimated coefficient is relatively large for the nearest cell pairs ($\beta = 0.054$, $p<0.001$), becomes smaller at intermediate distances ($\beta = 0.032$ and $0.024$, $p<0.001$), and increases again for the most distant pairs ($\beta = 0.055$, $p<0.001$). It suggests that physical encounters may be particularly relevant for geographically distant firms, potentially helping them overcome spatial frictions in establishing transaction relationships.. The full regression results are reported in Supplementary Table S9. Supplementary Table S10 and S11 show that the results remain consistent across alternative distance groupings (quintiles and deciles) and after excluding the most distant 1\% or 5\% of cell pairs (see Supplementary Note 3.7 for more details).

To examine whether the association differs by the timing of encounters, we divide meeting probability into three time-specific measures: working hours (Monday–Friday, 9:00 a.m.–6:00 p.m.), weekday evenings (Monday–Friday, 6:00 p.m.–10:00 p.m.), and Holiday (all day). This distinction allows us to assess whether the encounter effect is concentrated in periods that are more closely tied to regular economic activity. Encounters during working hours are more likely to reflect work-related activities and may therefore provide more relevant opportunities for economic interaction, whereas encounters during weekday evenings and weekends may capture a broader range of non-work activities. We included all three time-specific meeting probability measures simultaneously in the baseline logistic regression model, with the results shown in Figure~\ref{fig:figure3}b. Meeting probability during working hours was positively associated with link existence ($\beta = 0.030$, $p < 0.001$). By contrast, the coefficients were substantially smaller for weekday nights ($\beta = 0.002$, $p = 0.206$) and weekends ($\beta = 0.002$, $p = 0.179$). These results suggest that the association is concentrated in encounters occurring during regular working hours, rather than in co-presence occurring outside conventional working periods.

We next examine heterogeneity across industries. Industries differ in their knowledge bases, technologies, and patterns of inter-organizational interaction \cite{pavitt1984sectoral,malerba2002sectoral}. Face-to-face communication is particularly valuable when information is difficult to codify or requires rapid feedback \cite{storper2004buzz}. These differences suggest that the association between encounter opportunities and inter-firm transaction relationship may vary across industries. We first assign each cell a dominant industry based on the two-digit NAICS sector with the largest number of firms located in that cell. We then interact $\ln(\mathrm{MP}_{ij})$ with industry-pair indicators defined by the dominant industries of the two cells. The resulting coefficients capture industry-pair-specific associations between meeting probability and link existence in 2020. Figure \ref{fig:figure3}c reveals heterogeneity in both the sign and magnitude of the association between encounter opportunities and transaction links. Cell pairs involving information or finance and insurance exhibit FDR-significant positive coefficients across all 17 industry categories. Professional, scientific, and technical services show a similarly broad pattern of positive associations. In contrast, negative coefficients are common among pairings involving construction, retail, transportation, education, health care, and accommodation and food services. The full regression table is shown in Supplementary Table S12. The pattern therefore does not follow a simple distinction between service and non-service industries, and positive associations extend across industry boundaries rather than being confined to within-industry pairs. As an alternative, we define each cell's dominant industry based on the number of employees rather than the number of firms in each industry. We find that 83.8\% of the industry-pair coefficients retain the same sign across the two definitions (Supplementary Figure S13).

\subsection*{Indirect encounter effects on economic link formation}\label{subsec_traid}

\begin{figure}[!tb]
    \centering
    \includegraphics[width=\textwidth]{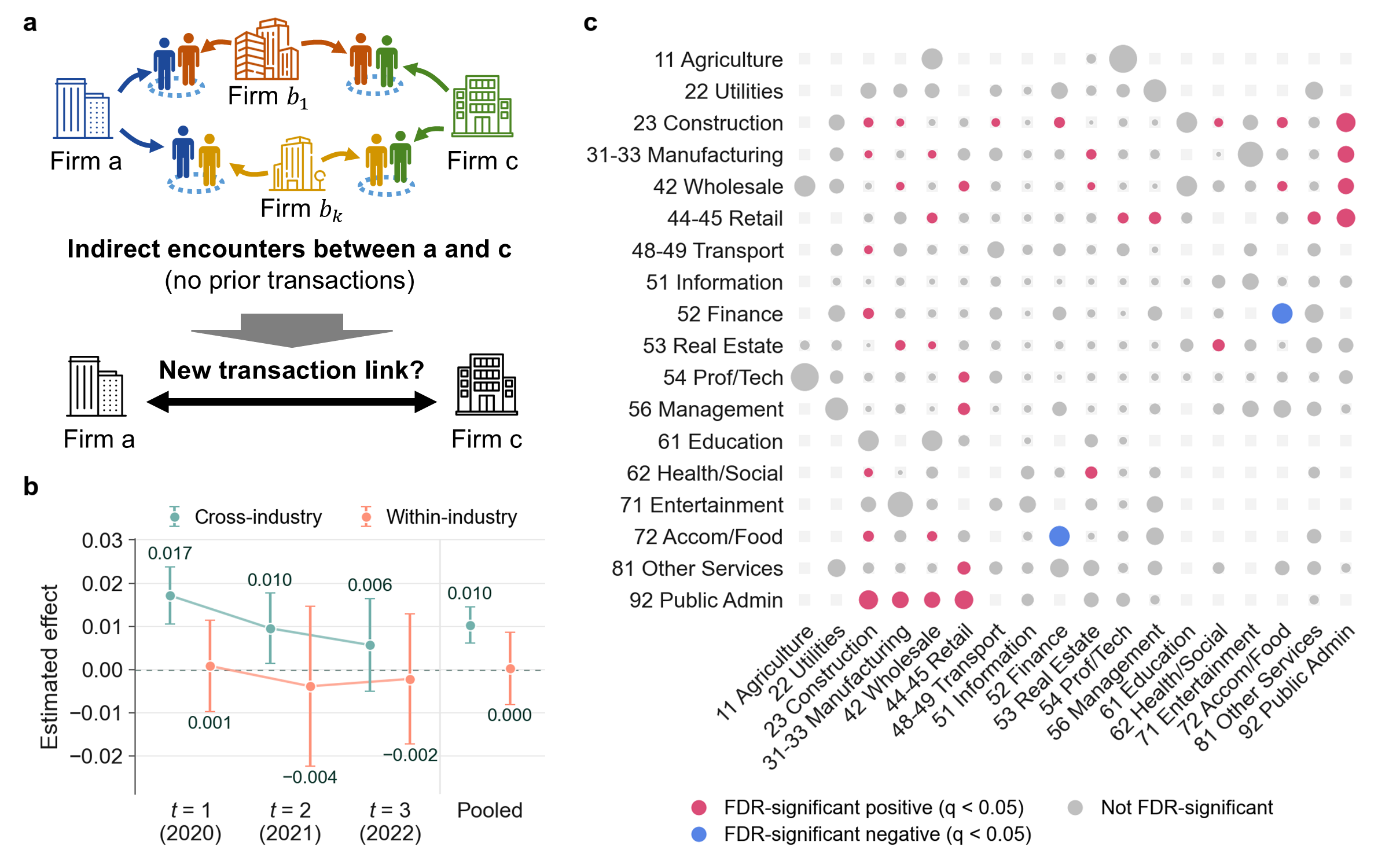}
  \caption{\textbf{Indirect encounter effects on economic link formation. } \textbf{a.} Illustration of indirect encounter exposure. Firm cells A and C are not directly connected but share a common mobility neighbour B. Indirect encounter exposure is calculated by aggregating the products of meeting probabilities along all 2-hop paths connecting the two firm cells. \textbf{b.} Estimated coefficients of indirect exposure from logistic regression models for within-industry and cross-industry cell pairs. Indirect encounter exposure is positively associated with subsequent link formation only among cross-industry cell pairs, whereas no significant association is observed for within-industry pairs. 
\textbf{c.} Dot matrix showing estimated coefficients of the indirect encounter effect for industry pairs. Each point represents the estimated coefficient for an industry pair. Dot size is proportional to coefficient magnitude, and color indicates the sign and statistical significance of the estimate.
}
\label{fig:figure4}
\end{figure}


The preceding analysis examines the association between direct encounters and transaction formation. However, firms may also establish new relationships through indirect connections involving common third parties \cite{gulati1999interorganizational,uzzi1997social,gulati1995social}. Building on this perspective, we examine whether indirect exposure through shared neighboring cells in the mobility-based encounter network is associated with transaction formation, beyond direct exposure.

We use the worker encounter network to examine whether two firm cells are more likely to form transaction relationships when their workers share encounter connections with a third firm cell. Consider two firm cells, $a$ and $c$, they are indirectly connected when both have encounter opportunities with the same intermediate cell, $b_k$. Formally, a two-hop path $a \rightarrow b_k \rightarrow c$ exists when $MP_{ab_k}>0$ and $MP_{b_kc}>0$. The intermediate cell therefore links $a$ and $c$ through two successive encounter relationships (See Figure \ref{fig:figure4}a for conceptual illustration). We restrict the sample to cell pairs with no transaction relationship observed before the mobility measurement period. This restriction allows us to examine whether indirect exposure in the encounter network is associated with subsequent link formation. We construct a measure of two-hop exposure, denoted by $\text{TwoHop}_{ac}$, defined as $\sum_{k} \text{MP}_{a b_k} \times \text{MP}_{b_k c}$. This measure aggregates all two-hop paths connecting firm cells $a$ and $c$ through shared encounter neighbors $b_k$. A shared neighbor contributes more to the measure when its meeting probability with both $a$ and $c$ are stronger. We estimate separate logistic regression models for transaction link formation one, two, and three years after the mobility measurement period. The dependent variable, $Y_{ac,t}$, equals one if a new transaction relationship between firm cells $a$ and $c$ is observed in year $t$, and zero otherwise, where $t\in{1,2,3}$ denotes the number of years after the mobility measurement period. The coefficient $\theta$ captures the association between two-hop encounter exposure and subsequent transaction link formation, conditional on the direct meeting probability between $a$ and $c$, the control variables $\mathbf{X}_{ac}$, and cell fixed effects $\delta_a$ and $\delta_c$. 

Figure \ref{fig:figure4}b shows the estimated effects of indirect encounter on subsequent transaction link formation. The association is concentrated among cell pairs linking firms from different industries. For these cross-industry cell pairs, greater indirect encounter exposure is associated with a higher likelihood of subsequent link formation in the first post-observation year ($\beta = 0.017$, $p < 0.001$) and the second year ($\beta = 0.010$, $p < 0.05$). The estimated association weakens over time and is no longer statistically significant by the third year. By contrast, indirect encounter exposure is not significantly associated with link formation among pairs linking firms from the same industry. The corresponding coefficients are small and statistically insignificant across all specifications. The pooled estimates show a similar pattern (full results are reported in Supplementary Table S13). These results are robust to controlling for the network degree and to the use of dyadic cluster-robust standard errors that allow dependence across dyads sharing either cell. Full robustness results are reported in Supplementary Note 3.10. Our findings suggest that the association between indirect exposure and transaction link formation is confined to cross-industry pairs.

Figure \ref{fig:figure4}c examines how the association between indirect encounter exposure and subsequent transaction link formation varies across industry pairs. Each dot represents the estimated coefficient for a given industry pair. Significant positive associations are concentrated among pairs involving public administration, construction, manufacturing, wholesale trade, and retail trade. These industries are closely involved in production, distribution, and institutional coordination. This pattern suggests that indirect encounter connections may be particularly relevant in sectors characterized by coordination-intensive or institutionally mediated economic relationships. 

\section*{Discussion}\label{sec3}


By linking worker mobility data with annual inter-firm transaction records in Tokyo, we show that opportunities for encounter between workplace cells are positively associated with both the presence of transaction links and their subsequent first appearance in the records. Observed co-presence adds information beyond a benchmark based on geographic distance and worker counts, suggesting that where and when workers spend time helps characterize the proximity relevant to trade. The association with link existence remains positive across distance groups, is most evident during working hours, and varies substantially across industry pairs. Indirect exposure through shared neighboring cells is also positively associated with subsequent link formation among cross-industry pairs. By contrast, we find no consistent association with link dissolution. These results connect the organization of everyday worker mobility to the emergence of inter-firm economic relationships.

These findings complement research that treats production networks as the outcome of costly partner search and matching. \cite{bernard2019production} shows that reductions in passenger travel costs following high-speed rail expansion facilitate new buyer–supplier connections, while \cite{miyauchi2024matching} finds that firms rematch with suppliers more quickly where alternative suppliers are more geographically concentrated. Experimental evidence also shows that regular meetings between business managers facilitate referrals and direct partnerships \cite{cai2018interfirm}. Across these settings, access to potential partners depends on opportunities to discover and evaluate them. Our study adds evidence from everyday urban activity. Even after accounting for distance and local characteristics, co-presence between workplace populations is associated with which locations subsequently form new transaction links. This provides a behavioral complement to accounts based on transport infrastructure, market thickness, and deliberately organized business meetings.

This perspective also helps connect the formation of production networks with theories of urban agglomeration. Sharing, matching, and learning provide distinct reasons for economic activity to concentrate spatially \cite{duranton2004micro}, and evidence on industry co-agglomeration points to roles for input–output linkages, labor pooling, and knowledge flows \cite{ellison2010causes}. Even where many firms are concentrated in one area, opportunities for contact can vary across firms. Our distance, gravity, and time-shift comparisons indicate that the locations of workplaces, the places workers visit and the timing of those visits contain different information  about economic connectivity. The positive association across distance groups suggests that the link between co-presence and transaction ties extends beyond their neighborhoods. For research on agglomeration, this suggests that the concentration of firms and the patterns of activity connecting them should be examined jointly when studying how potential partners become accessible.

The industry-level results suggest that the relevance of these connections depends on the activities involved. In Figure \ref{fig:figure3}c, meeting probability is positively associated with link existence for all industry pairings involving either information or finance and insurance, as well as for many pairings involving professional, scientific, and technical services. Several other industry pairings exhibit negative associations. This heterogeneity is consistent with accounts that emphasize the value of face-to-face communication when information is complex, changes rapidly, or is difficult to codify \cite{storper2004buzz}. Mobility-based research on knowledge spillovers likewise links opportunities for worker interaction to patent citations \cite{atkin2022}. Yet communication-intensive exchange also occurs in manufacturing, with business travel particularly strongly associated with trade in differentiated and R\&D-intensive goods \cite{cristea2011buyer}. The broader implication is that informational and coordination requirements may help explain variation across industry pairs.

The distinction between link formation and dissolution also matters. Opportunities to encounter unfamiliar partners may be especially relevant before a relationship exists, while established partners can draw on accumulated information and existing organizational arrangements. Evidence from scientific collaboration offers a parallel. Co-location facilitates the initiation of joint research, but separating previously colocated laboratories does not reduce their propensity to collaborate \cite{catalini2018microgeography}. A similar distinction appears in inter-firm knowledge networks, where geographic proximity is associated with tie formation but has no detectable association with tie persistence \cite{juhasz2018creation}. The positive association between co-presence and link formation, alongside the absence of a consistent association with dissolution, is consistent with this distinction between initiating and maintaining relationships.

The indirect-exposure results place partner formation in a wider network context. Research on strategic alliances shows that common third parties are associated with the formation of new inter-organizational relationships \cite{gulati1999interorganizational}, while models of trade allow firms to discover new partners through their existing contacts \cite{chaney2014network}. Our results extend this perspective to mobility-based exposure. Connections through common neighboring workplace cells are associated with subsequent cross-industry link formation, conditional on direct exposure. One possible interpretation is that these connections provide opportunities to learn about firms outside familiar industry circles. However, the two-hop measure does not trace information transmission or referrals, and the two legs of a path may involve different individuals in the intermediate cell.

Opportunities for contact nevertheless constitute only one part of the process through which economic relationships emerge. In a recent field experiment, \cite{cai2026firm} find that referrals generate lasting supplier–client partnerships when paired with an incentive to undertake an initial transaction, whereas information-only referrals have little effect. This distinction highlights the importance of learning about match quality through experience. Research on inter-firm relationships also documents the coordination advantages of embeddedness and the vulnerabilities associated with excessive embeddedness \cite{uzzi1997social}. These studies suggest that the consequences of greater connectivity depend on the information exchanged, the partners involved, and the wider network structure. Understanding the economic significance of mobility therefore requires examining how opportunities for contact translate into particular relationships and outcomes.

Several limitations qualify these interpretations. First, the analysis is observational. Although the association with link formation is stronger for outcomes recorded after mobility was measured, meeting probability in 2019 is also positively associated with link formations recorded before 2019. Existing relationships may generate travel, and persistent characteristics of cell pairs may influence both mobility and transactions. Cell fixed effects, observed controls, and randomized benchmarks do not eliminate these possibilities. Temporal ordering alone is therefore insufficient to establish a causal effect of encounters on transaction formation. Repeated mobility measurements and plausibly exogenous changes in opportunities for contact would help distinguish these explanations.

Second, co-presence measures an opportunity for encounter rather than an observed conversation or business meeting. We cannot determine the purpose or content of any interaction or the roles of the individuals involved in business decisions. Aggregating mobility data to workplace cells also prevents us from determining whether the individuals represented in the mobility measure are employed by the firms involved in the recorded transactions. The results therefore describe associations between co-presence among workplace populations and transaction connectivity at the cell level. In addition, we observe mobility only from July through September 2019 and only for users who regularly attend their workplaces. The data may thus incompletely capture seasonal variation, infrequent encounters, and less regular forms of work. Longer observation windows and repeated mobility measurements would help assess the stability of these patterns across seasons and years. Subsequent transaction outcomes also span the COVID-19 pandemic, during which mobility and contact patterns in Tokyo changed substantially \cite{yabe2020non}.

Finally, the transaction data primarily record major partners and identify relationships at the headquarters level. A link’s first appearance or disappearance in the records may reflect changes in reporting or in a partner’s importance, as well as the formation or termination of a relationship. The analysis also covers a single dense metropolitan area, which limits the generalizability of the findings to cities with different urban structures or business environments. We do not observe transaction values, match quality, productivity gains, or welfare effects. Consequently, an increase in the number of recorded relationships is insufficient on its own to establish an economic improvement. Linking encounter opportunities to these outcomes over longer periods and across multiple cities would help assess the broader economic consequences of the patterns documented here.

\section*{Methods}\label{sec4}

\subsection*{Mobility Data}\label{subsec41}
We utilize an anonymized mobile phone location dataset provided by Blogwatcher Inc., a commercial provider of mobile location data services. The location records were collected through more than 140 mobile phone applications with informed consent obtained from all users in Japan. Each record includes a unique anonymized user ID, latitude, longitude, and timestamps. To protect user privacy, all records were anonymized before being provided to the authors, ensuring that individual users cannot be identified from the location traces. Our analysis uses GPS users observed in Tokyo from July to September 2019. We restrict the sample to active users with location records on more than 90 distinct days, ensuring sufficient coverage of individual mobility to support the subsequent worker identification algorithm. This selection yields 329,149 users as the total sample. According to the Basic Resident Register of the Tokyo Metropolitan Government, Tokyo’s total population in January 2020 was 13,740,732. After excluding people aged 0–19 and 70 and above, who are less likely to participate in full-time employment, the working-age population is approximately 9 million. Therefore, our mobility data sample represents about 3.5 \% of the working-age population in Tokyo. Supplementary Table S1 and Supplementary Figure S1 show that the mobility sample is spatially representative, with municipality-level user counts closely tracking the underlying population distribution ($r = 0.97$). We converted the raw trajectory records into stay points using the algorithm proposed by \cite{hariharan2004project}, clustering consecutive GPS records within a 100 m radius. We then retained only stay points with a duration longer than 10 minutes (see Supplementary Note 1.1 for more details).

\subsection*{Transaction Data}\label{subsec42}
We use firm profile information and inter-firm transaction data provided by Teikoku Databank Ltd., a private corporate credit research firm that maintains one of the most comprehensive repositories of company information in Japan. These relationships are collected through TDB's corporate credit surveys, in which investigators interview firms and identify their trading partners. The data can be divided into two parts: a firm profile dataset and an inter-firm transaction dataset (Supplementary Figure S2). It is possible to link the two datasets using unique firm codes. The firm profile dataset covers approximately 1.46 million firm headquarters nationwide. The data include unique firm codes, headquarters addresses and industry classification. The inter-firm transaction data provide annual records of buyer–supplier relationships and include firm codes for both buyers and suppliers. It accounts for roughly 90\% of total sales in the Japanese economy. However, because the transaction data are recorded at the headquarters level, transactions between individual establishments cannot be observed. In this study, we focus on 196,317 firms with headquarters in Tokyo, excluding those located on the Izu--Ogasawara Islands, as these islands are geographically isolated from the contiguous Tokyo urban area considered in this study. We use inter-firm transaction data covering the period from 2016 to 2022. Linking these transaction records to firm profile data through shared firm code enables the construction of a longitudinal inter-firm transaction network.

\subsection*{Estimating encounter opportunities among workers}\label{subsec43}

We leverage large-scale mobility data with firm headquarters location data to quantify opportunities for workers from different firms to encounter one another in the city. Because Tokyo's extreme spatial density and vertical complexity make building-level workplace inference highly sensitive to identification errors, we aggregate firm locations into hexagonal cells using the Uber H3 grid system at resolution 10 (approximate cell area: 0.015 km²). Cells containing at least one registered firm location are designated as firm cells (see Supplementary Note 1.3 for more details). 

For each user, we consider stay points on weekdays, excluding Japanese national holidays. A day is classified as a working day if a user spends at least 5 hours in a firm cell during the weekday daytime window (9:00 a.m.–6:00 p.m.), excluding Japanese national holidays. We identify each user’s primary workplace as the firm cell with the largest number of working days. Users are classified as workers if they have at least 40 working days at their primary workplace, spanning at least 12 distinct weeks during the observation period. We further examine the robustness of the identified worker sample to alternative thresholds for daily dwell time, minimum working days, and minimum active weeks in Supplementary Figure S4. To validate the spatial representativeness of our worker sample, we compare the estimated worker population with the 2020 population census of Tokyo, finding a high correlation at both the basic unit block level and chō–aza  level ($r = 0.80$ and $r = 0.90$; Supplementary Figure S5). Supplementary Figure S6 shows that the identified workers also display stable commuting patterns, with regular arrival at their workplace, further supporting the reliability of the sample.

Following the approach of \cite{atkin2022}, we estimate the potential encounter opportunities based on the spatial and temporal co-presence of workers. Let $n_{ict}$ and $n_{jct}$ denote the number of workers from firm cell $i$ and firm cell $j$ present in spatial cell $c$ during the 30-minute time window $t$. We define the total number of meetings between firm cells $i$ and $j$ as $\mathrm{M}_{ij} = \sum_{c,t} n_{ict} n_{jct}$, which counts the worker-pair co-presence instances in the same location and time window. We further define the potential maximum number of meetings as $\mathrm{PMM}_{ij} = \sum_{t} \left( \sum_{c} n_{ict} \right)\left( \sum_{c} n_{jct} \right)$, which captures the total number of possible co-presence opportunities, that is, the number of times workers from both firm cells are simultaneously active in the dataset during the same time window, regardless of location. The meeting probability is then defined as $\mathrm{MP}_{ij} = \frac{\mathrm{M}_{ij}}{\mathrm{PMM}_{ij}}$. By construction, $\mathrm{MP}_{ij}$ measures the probability that a randomly selected worker from firm cell $i$ and one from cell $j$ are observed in the same location at the same time. This method captures simultaneous spatial and temporal overlap as a necessary condition for in-person encounters, without directly observing the interactions themselves.

\subsection*{Modeling the effect of worker encounters on inter-firm transactions }\label{subsec44}

To examine whether workers’ encounter opportunities are associated with inter-firm economic ties, we estimate logistic regression models (Supplementary Note 3.1). Each observation corresponds to a pair of firm cells, cell $i$ and cell $j$. The dependent variable $Y_{ij,t}$ indicates the transaction outcome between firms located in the two cells $t$ years after or before the mobility observation period. Here, $t$ denotes the temporal gap from mobility measurement year (for example, $t = 1$ corresponds to outcomes in the subsequent year). The main explanatory variable is the logarithm of meeting probability, $\ln(\mathrm{MP}_{ij})$, which captures encounter opportunities between workers from two different firm cells during the mobility observation period.

We estimate the following specification:
\begin{equation}
\mathrm{logit}\!\left[\Pr\!\left(Y_{ij,t}=1\right)\right]
=
\beta \ln(\mathrm{MP}_{ij})
+
\mathbf{X}_{ij}'\boldsymbol{\gamma}
+
\delta_i
+
\delta_j .
\label{eq:baseline}
\end{equation}

where we include cell fixed effects $\delta_i$ and $\delta_j $ to absorb unobserved heterogeneity for each cell, and $\mathbf{X}_{ij}$ is a set of dyad-level control variables: we control for geographic distance between cell centroids and industry overlap, defined as the inner product of the industry-share distributions of firms within the two cells. We also control for differences in local economic and workforce characteristics, including employment scale, establishment density, and workforce gender composition. These variables account for spatial, industrial, and socioeconomic similarities that may affect both worker encounter opportunities and inter-firm transaction probabilities. Standard errors are two-way clustered by cell $i$ and $j$.

\section*{Data availability}
The mobility data analysed in this study were obtained from Blogwatcher Inc. under license. Access to the raw data is restricted under the terms of this agreement, and the data therefore cannot be made publicly available. Information about how to request access to the data and its conditions and limitations can be found at \url{https://www.blogwatcher.co.jp/}. The firm transaction data were obtained from Teikoku Databank, Ltd. under license and are not publicly available. The data are commercially available for purchase by corporate entities from Teikoku Databank, Ltd. via \url{https://www.tdb.co.jp/}.

\section*{Code availability}
The analysis was conducted using Python. Code to reproduce the main results in the figures from the aggregated data is publicly available on GitHub \url{https://github.com/chenket/mobility-transaction-network}.

\bibliography{_references}

\section*{Acknowledgements}
We would like to thank Teikoku Databank, Ltd. and Blogwatcher Inc. for kindly providing the data. This study was carried out as part of joint research with Teikoku Databank, Ltd. K.C. acknowledges support from JST SPRING (Grant No. JPMJSP2108). H.I. acknowledges support from JSPS KAKENHI (Grant Nos. JP23K25520, JP24K00247, JP24K01108, JP25K01454, JP26K00349, and JP26K22092). Y.O. acknowledges support from JSPS KAKENHI (Grant No. JP24K01000). T.Y. acknowledges support from JSPS KAKENHI (Grant No. 24K01108). This work also partially includes the results of the Cross-ministerial Strategic Innovation Promotion Program (SIP) 3rd Phase, ``Development of Smart Mobility Platform'' promoted by the Council for Science, Technology and Innovation, Cabinet Office (Project Management Agency: New Energy and Industrial Technology Development Organization (NEDO); Project Code: JPNP23023).

\section*{Author contributions statement}
K.C. designed the algorithms, performed the analysis, and developed models and simulations. H.I. and T.Y. partially developed models and simulations. H.I., Y.S., Y.O., and T.Y. supervised the research. All authors wrote the paper. The company data were processed by K.C.and Y.O. All authors reviewed the manuscript.

\section*{Competing interests statement}

The authors declare no competing interests.

\end{document}


\maketitle

\clearpage
\tableofcontents
\clearpage

\section{Data sources and preprocessing}

\subsection{Mobility data}

In this study, we used an anonymized mobile phone location dataset provided by Blogwatcher, Inc., a Japanese company specializing in location data services. The location records were collected through more than 140 mobile phone applications with informed consent obtained from all users. Each record includes a unique anonymized user ID, latitude, longitude, and timestamps. To protect user privacy, all records were anonymized before being provided to the authors, ensuring that individual users cannot be identified from the location traces. Additional information about the dataset, including its data collection process, privacy protection measures, and limitations, is available on the Blogwatcher website (\url{https://www.blogwatcher.co.jp/terms/}).

Our analysis focuses on users observed in Tokyo from July to September 2019. We restrict the sample to active users with location records on more than 90 distinct days, ensuring sufficient coverage of individual mobility to support the subsequent worker identification algorithm. This selection yields 329,149 users as the total sample. According to the Basic Resident Register of the Tokyo Metropolitan Government, Tokyo’s total population in January 2020 was 13,740,732. After excluding people aged 0–19 and 70 and above, who are less likely to participate in full-time employment, the working-age population is approximately 9 million. Therefore, our mobility data sample represents about 3.5 \% of the working-age population in Tokyo. 

To identify meaningful visits, we convert all GPS trajectory records into stay points following the algorithm of \cite{hariharan2004project}. A stay point is defined as a spatial cluster of consecutive location pings whose maximum distance from the centroid is below the distance threshold and whose duration meets the duration threshold. In this study, we set the distance threshold to 100 meters and the minimum duration to 10 minutes. Next, each user’s home location is inferred from the stay points. For each user and each day, we identify the stay point with the longest cumulative dwell time during nighttime hours (10 p.m. to 6 a.m.) as the candidate home location. The location that is most frequently selected across the sample time range is assigned as the user’s home location. 

\begin{figure}
    \centering
    \includegraphics[width=1\linewidth]{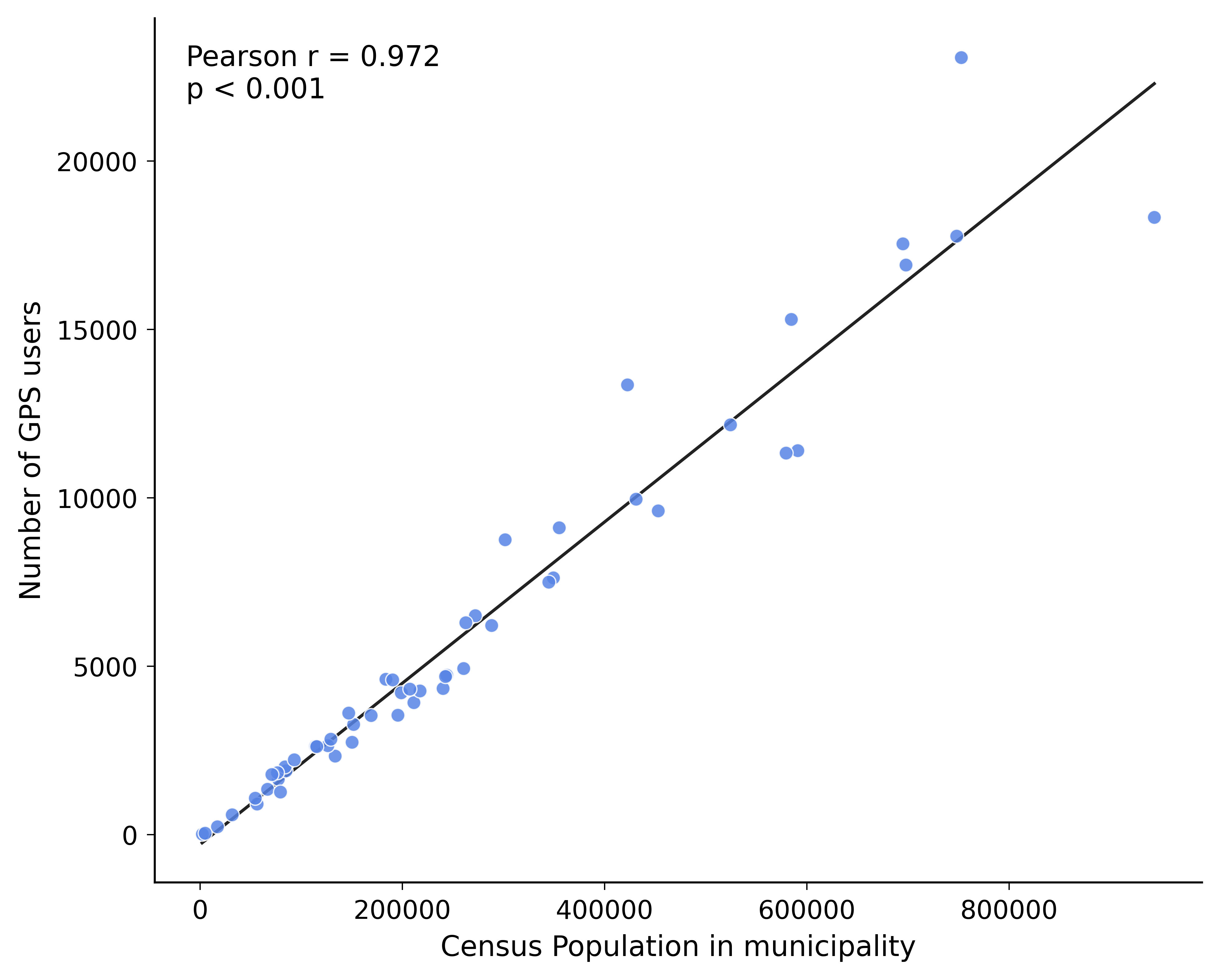}
    \caption{\textbf{Spatial representativeness of the human mobility data.} Census population (x-axis) and the number of GPS users (y-axis) are compared across municipalities. The two measures show a strong positive association (Pearson's $r=0.972$, $p<0.001$).
}
    \label{fig:gpssamling}
\end{figure}

We assessed the spatial representativeness of the mobility sample using users’ inferred home locations. Japan is administratively divided into municipalities, including cities, towns, villages, and special wards. For each municipality, we calculated the sampling rate as the number of observed mobile phone users divided by the total population reported in the 2020 census. The resulting municipality-level sampling rates were approximately 2.2\% on average, with the distribution across wards reported in Supplementary Table \ref{tab:gps-sampling-rate-ward}. The number of observed users was strongly correlated with the official population on the log--log scale (Pearson's $r=0.972$, $p<0.001$; Supplementary Figure~\ref{fig:gpssamling}). Those results suggest that the spatial distribution of the GPS sample broadly reflects the real population distribution across Tokyo.

\begin{table}[!htbp]
\centering
\caption{GPS sampling rates across municipalities in Tokyo}
\label{tab:gps-sampling-rate-ward}
\footnotesize
\begin{tabular}{llrrr}
\toprule
Code & Municipality & Census population & GPS users & Sampling rate (\%) \\
\midrule
101 & Chiyoda Ward & 66,680 & 1,348 & 2.02 \\
102 & Chuo Ward & 169,179 & 3,540 & 2.09 \\
103 & Minato Ward & 260,486 & 4,938 & 1.90 \\
104 & Shinjuku Ward & 349,385 & 7,622 & 2.18 \\
105 & Bunkyo Ward & 240,069 & 4,340 & 1.81 \\
106 & Taito Ward & 211,444 & 3,924 & 1.86 \\
107 & Sumida Ward & 272,085 & 6,503 & 2.39 \\
108 & Koto Ward & 524,310 & 12,171 & 2.32 \\
109 & Shinagawa Ward & 422,488 & 13,357 & 3.16 \\
110 & Meguro Ward & 288,088 & 6,213 & 2.16 \\
111 & Ota Ward & 748,081 & 17,780 & 2.38 \\
112 & Setagaya Ward & 943,664 & 18,336 & 1.94 \\
113 & Shibuya Ward & 243,883 & 4,732 & 1.94 \\
114 & Nakano Ward & 344,880 & 7,497 & 2.17 \\
115 & Suginami Ward & 591,108 & 11,403 & 1.93 \\
116 & Toshima Ward & 301,599 & 8,757 & 2.90 \\
117 & Kita Ward & 355,213 & 9,116 & 2.57 \\
118 & Arakawa Ward & 217,475 & 4,268 & 1.96 \\
119 & Itabashi Ward & 584,483 & 15,299 & 2.62 \\
120 & Nerima Ward & 752,608 & 23,080 & 3.07 \\
121 & Adachi Ward & 695,043 & 17,551 & 2.53 \\
122 & Katsushika Ward & 453,093 & 9,620 & 2.12 \\
123 & Edogawa Ward & 697,932 & 16,918 & 2.42 \\
201 & Hachioji City & 579,355 & 11,329 & 1.96 \\
202 & Tachikawa City & 183,581 & 4,616 & 2.51 \\
203 & Musashino City & 150,149 & 2,747 & 1.83 \\
204 & Mitaka City & 195,391 & 3,548 & 1.82 \\
205 & Ome City & 133,535 & 2,331 & 1.75 \\
206 & Fuchu City & 262,790 & 6,293 & 2.39 \\
207 & Akishima City & 113,949 & 2,623 & 2.30 \\
208 & Chofu City & 242,614 & 4,699 & 1.94 \\
209 & Machida City & 431,079 & 9,964 & 2.31 \\
210 & Koganei City & 126,074 & 2,640 & 2.09 \\
211 & Kodaira City & 198,739 & 4,215 & 2.12 \\
212 & Hino City & 190,435 & 4,593 & 2.41 \\
213 & Higashimurayama City & 151,815 & 3,276 & 2.16 \\
214 & Kokubunji City & 129,242 & 2,832 & 2.19 \\
215 & Kunitachi City & 77,130 & 1,648 & 2.14 \\
218 & Fussa City & 56,414 & 908 & 1.61 \\
219 & Komae City & 84,772 & 1,894 & 2.23 \\
220 & Higashiyamato City & 83,901 & 2,016 & 2.40 \\
221 & Kiyose City & 76,208 & 1,842 & 2.42 \\
222 & Higashikurume City & 115,271 & 2,620 & 2.27 \\
223 & Musashimurayama City & 70,829 & 1,781 & 2.51 \\
224 & Tama City & 146,951 & 3,610 & 2.46 \\
225 & Inagi City & 93,151 & 2,225 & 2.39 \\
227 & Hamura City & 54,326 & 1,082 & 1.99 \\
228 & Akiruno City & 79,292 & 1,266 & 1.60 \\
229 & Nishitokyo City & 207,388 & 4,324 & 2.08 \\
303 & Mizuho Town & 31,765 & 590 & 1.86 \\
305 & Hinode Town & 16,958 & 236 & 1.39 \\
\bottomrule
\end{tabular}
\end{table}

\subsection{Transaction data}

We use firm profile information and inter-firm transaction data provided by Teikoku Databank (TDB) Ltd. (\url{https://www.tdb.co.jp/}), a private corporate credit research firm that maintains one of the most comprehensive repositories of company information in Japan. TDB collects information on firms and their trading partners through corporate credit surveys, including on-site interviews and verification using external sources. Although there is no formal limit on the number of trading partners recorded, each report may not capture all of a firm's trading relationships. Coverage is supplemented by identifying additional trading relationships reported by other firms. The resulting database records reported supplier–customer relationships between firms. The dataset covers approximately 70\% of Japanese companies, based on a comparison with the 2021 Economic Census and accounts for more than 90 percent of all sales in the Japanese economy. It is based on establishment surveys conducted across all 47 prefectures, resulting in comprehensive nationwide coverage. 

\begin{figure}
    \centering
    \includegraphics[width=0.75\linewidth]{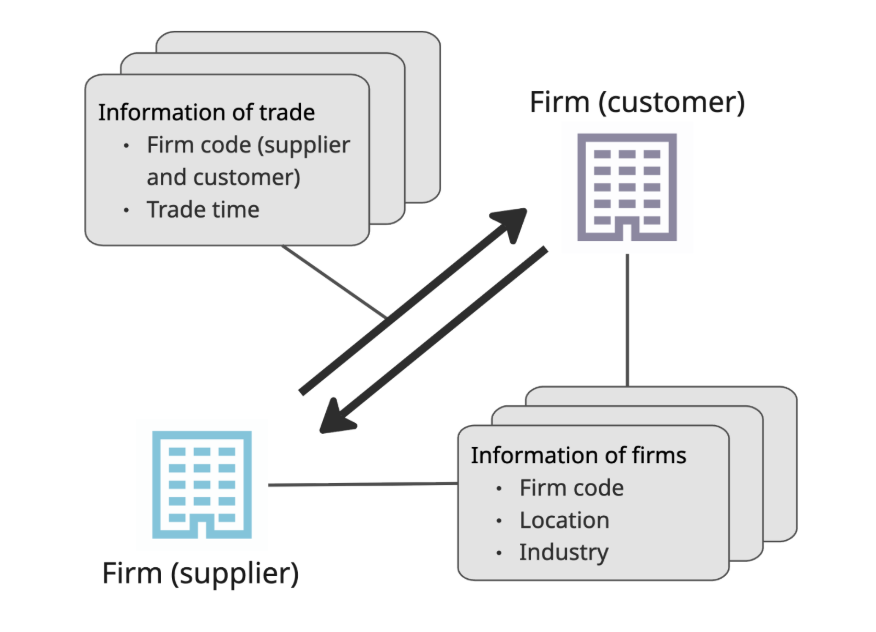}
    \caption{\textbf{Structure of the transaction data.}}
    \label{fig:tdb}
\end{figure}

The data can be divided into two parts: a firm profile dataset and an inter-firm transaction dataset (Supplementary Figure~\ref{fig:tdb}). It is possible to link the two datasets using unique firm IDs. The firm profile dataset includes unique firm IDs, headquarters addresses, and industry classifications. The inter-firm transaction data provide annual records of buyer–supplier relationships and include firm IDs for both buyers and suppliers. The transaction data are recorded at the headquarters level; transactions between individual establishments cannot be observed. In this study, we focus on 196,317 firms with headquarters in Tokyo, excluding those located on the Izu--Ogasawara Islands, as these remote islands are geographically isolated from the contiguous Tokyo urban area considered. We use inter-firm transaction data covering the period from accounting year 2016 to 2022. Linking these transaction records to firm profile data through shared firm IDs enables the construction of a longitudinal inter-firm transaction panel.

The original industry information follows the 1993 revision of the Japanese Standard Industrial Classification (JSIC), which comprises 13 broad industry divisions. These divisions aggregate economically distinct activities, particularly within service-related industries, limiting our ability to examine industry heterogeneity. To obtain a more detailed and internationally comparable classification, we harmonized the original JSIC categories with the North American Industry Classification System (NAICS). Specifically, we first mapped the JSIC categories to the International Standard Industrial Classification of All Economic Activities (ISIC) Rev. 4 and then to the corresponding NAICS sectors. Using this classification, we distinguish 20 industry sectors, enabling a more detailed analysis of differences in inter-firm encounter opportunities across industries.

\subsection{Construction of firm cells}

The urban environment in Tokyo features high spatial density and vertical complexity. Buildings are located close to one another, and many firm offices are housed in the same building. Under such conditions, inferring workplace locations at the building level based on GPS trajectories will cause errors, as even small positioning inaccuracies may cause points to fall outside the correct building footprint. Therefore, we represent firm locations using H3 hexagonal cells. Each headquarters location in the firm profile data is mapped to an H3 hexagonal cell at resolution 10, corresponding to an average edge length of approximately 75 m. Cells containing at least one firm location were defined as firm cells. The distribution of firms across firm cells is shown in Supplementary Figure \ref{fig:pie}. Single-firm cells account for 39.0\% of all firm cells, while cells containing 2 firms account for an additional 20.6\%. Together, 59.6\% of firm cells contain no more than 2 firms, and 78.8\% contain no more than 4 firms. Cells containing more than 10 firms represent only 7.9\% of the total. 

\begin{figure}
    \centering
    \includegraphics[width=0.75\linewidth]{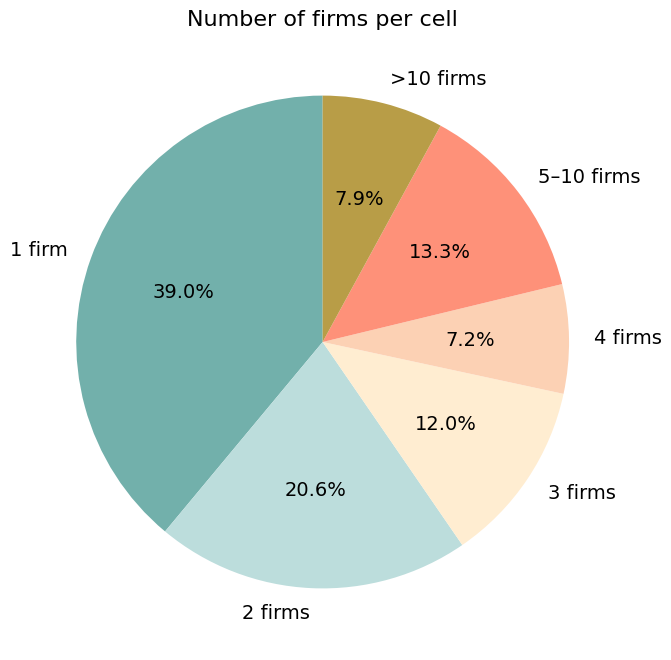}
    \caption{\textbf{Distribution of firms across firm cells.} Single-firm cells account for 39.0\% of all firm cells, while 59.6\% contain no more than two firms and 78.8\% contain four firms or fewer.}
    \label{fig:pie}
\end{figure}

\subsection{Worker identification}
\paragraph{Worker-identification procedure}

We identify workers from recurrent daytime presence at potential workplace locations. We first retain stay points observed on weekdays, excluding Japanese national holidays, between 9:00 a.m. and 6:00 p.m. This time window broadly covers typical daytime work schedules in Japan, as documented in the 2019 General Survey on Working Conditions. Stay points located within a user's inferred home location are removed. Each remaining stay point is then assigned to the corresponding firm cell. A user is considered to have attended a firm cell on a given day if their total dwell time in the cell is at least 5 hours; such a day is classified as a working day. We then define the user's primary workplace as the firm cell with the largest number of such working days during the observation period. To distinguish regular workers from occasional visitors, we retain users whose primary workplace is observed on at least 40 working days and across at least 12 distinct weeks during the three-month study period. Users who satisfy these conditions are classified as workers associated with their identified primary workplace cell. This procedure identifies 52,785 commuting workers, each assigned to a primary workplace cell.

\paragraph{Sensitivity to parameter choices}
To assess the sensitivity of worker identification to alternative parameter choices, we compare the resulting alternative worker set, $W_s$, with the baseline worker set, $W_0$. Worker set stability is measured using the Jaccard similarity:

\begin{equation}
J(W_0,W_s)
=
\frac{|W_0 \cap W_s|}{|W_0 \cup W_s|}.
\end{equation}

\begin{figure}
    \centering
    \includegraphics[width=1\linewidth]{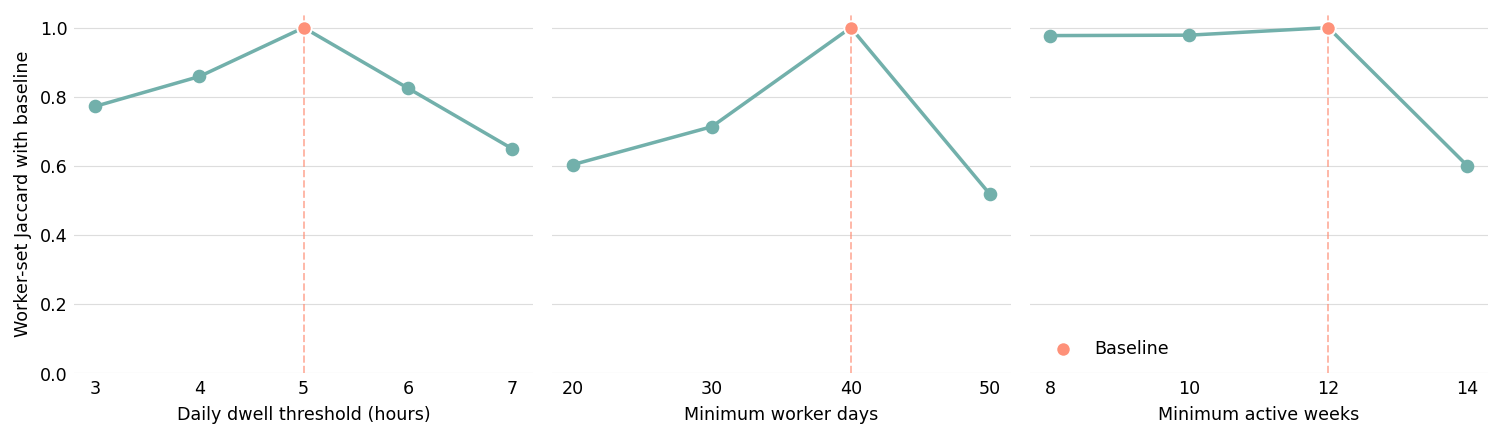}
    \caption{\textbf{Sensitivity of worker identification to alternative parameter thresholds.} Each panel reports the Jaccard similarity between the worker set identified under alternative parameters and that identified under the baseline. One parameter is varied at a time, while the other two parameters are held at their baseline values. The baseline specification requires at least 5 hours of daily workplace dwell time, 40 working days, and activity across at least 12 weeks.}
    \label{fig:para_sensitivity}
\end{figure}

A value of 1 indicates that the baseline and alternative specifications identify exactly the same workers, whereas lower values indicate greater differences in sample composition. We vary one parameter at a time while holding the remaining two at their baseline values. The alternative values are daily dwell thresholds of [3, 4, 5, 6, 7] hours, minimum working-day requirements of [20, 30, 40, 50] days, and minimum active-week requirements of [8, 10, 12, 14] weeks. Supplementary Figure~\ref{fig:para_sensitivity} shows that the identified worker sets remain substantially similar to the baseline under a range of alternative thresholds. Overall, the sensitivity analysis supports the feasibility of the baseline parameter choices. The identified worker set remains relatively stable when the daily dwell threshold is adjusted moderately around the 5-hour baseline, with Jaccard similarities of approximately 0.86 and 0.82 under the 4-hour and 6-hour specifications, respectively. The worker set is also highly stable when the minimum active-week requirement varies between 8 and 12 weeks, indicating that most workers identified under the baseline specification exhibit persistent working patterns throughout the observation period. For minimum working days, alternative thresholds around the 40-day baseline continue to preserve substantial overlap with the baseline set. Taken together, these results suggest that the baseline specification identifies a consistent core of regular commuters, with limited sensitivity to moderate changes in the identification thresholds.

\paragraph{Spatial representativeness of identified workers}

We validate the worker identification procedure along two dimensions: the spatial representativeness of the detected worker sample relative to the Population Census of Japan, and the commuting regularity of identified workers as assessed from their observed mobility patterns.

\begin{figure}
    \centering
    \includegraphics[width=1\linewidth]{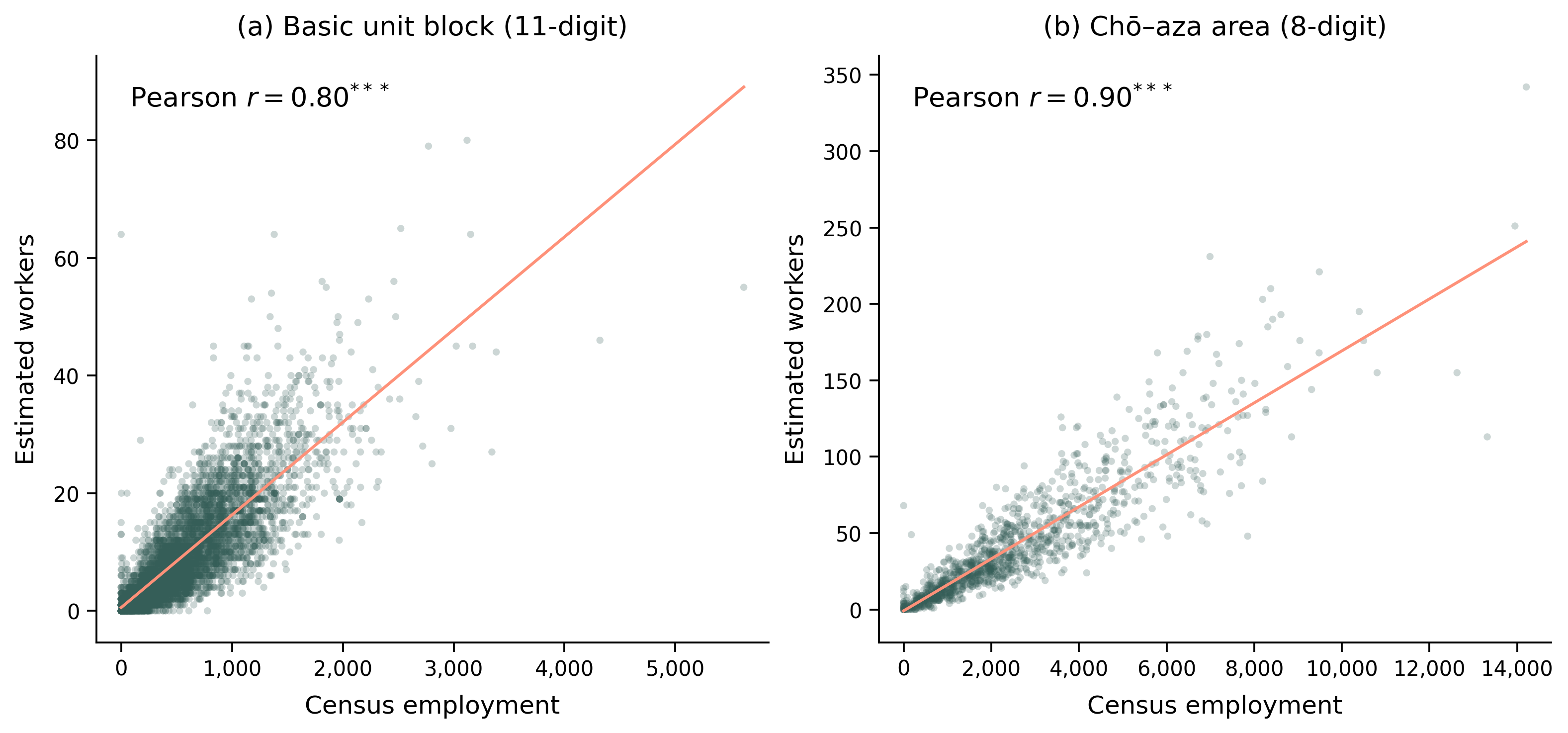}
    \caption{\textbf{Spatial representativeness of the identified worker sample.} (a) Census employment in each basic unit block, measured using the full 11-digit census area code (x-axis), versus the number of workers identified from the mobility data in the corresponding area (y-axis); Pearson's $r=0.80$. (b) The same comparison after aggregating basic unit blocks to the broader chō–aza level using the corresponding 8-digit area code; Pearson's $r=0.90$. The positive correlations at both spatial resolutions indicate that the identified worker sample broadly tracks the spatial distribution of office-based employment reported in the census.}
    \label{fig:census}
\end{figure}

We first assess the representativeness of the identified worker sample by comparing its spatial distribution with official employment statistics from the 2020 Population Census of Japan at two spatial resolutions. The first is the basic unit block, the finest census unit represented by the full 11-digit area code. The second is the broader chō–aza unit represented by an 8-digit area code. This study uses firms’ headquarters locations and does not account for their branches, stores, or factories. The relevant census benchmark should therefore reflect occupations that are more likely to be concentrated at headquarters and other office locations. We classified managers, professional and technical workers, and clerical workers as office-based occupations (2020 Population Census of Japan, Basic Tabulation on Labour Force Status). Supplementary Figure~\ref{fig:census} shows the spatial correlation between census employment counts and the number of identified workers at two different spatial resolutions. At the finer basic-unit-block level, the two measures are strongly correlated, with a Pearson correlation coefficient of \textit{r}=0.80. After aggregating the data to the broader chō–aza level, the correlation increases to \textit{r}=0.90. This result suggests that the identified worker sample provides a reasonably good representation of the spatial distribution of office-based employment in Tokyo.

\begin{figure}
    \centering
    \includegraphics[width=1\linewidth]{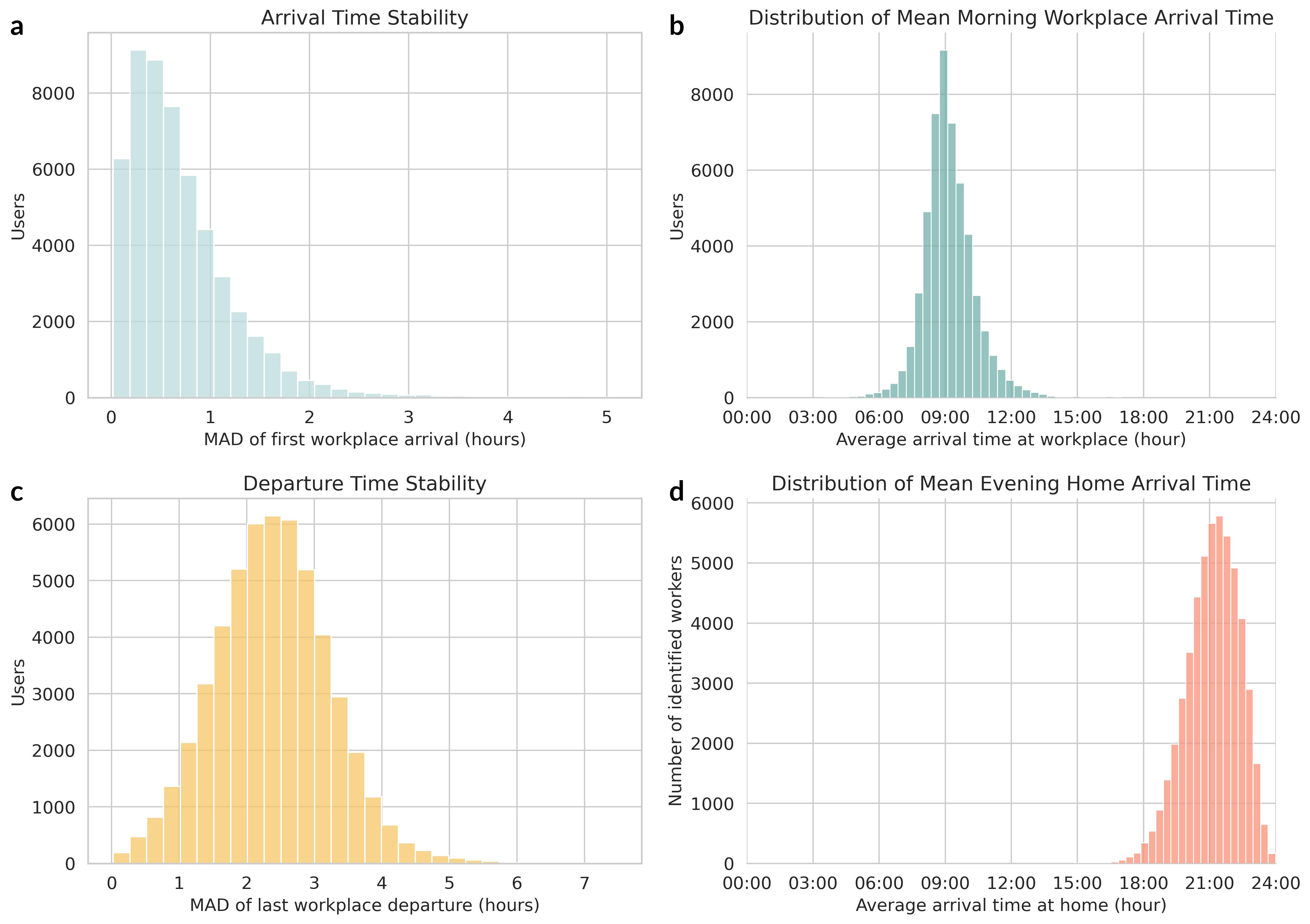}
    \caption{\textbf{Commuting patterns of identified workers.} (a) Distribution of the median absolute deviation (MAD) of each worker’s first daily arrival time at the inferred workplace. Most workers show stable morning arrival times, with a MAD of less than one hour. (b) Distribution of workers’ mean first daily arrival times at the inferred workplace. Most workers arrive around 9:00 a.m. (c) Distribution of the MAD of each worker’s last daily departure time from the inferred workplace. (d) Distribution of workers’ mean daily arrival times at the inferred home location. Most workers return home after 6:00 p.m.}
    \label{fig:commute}
\end{figure}

\paragraph{Validation of commuting regularity}

We further validate the identified worker sample by examining whether these users exhibit recurring daily commuting patterns. Supplementary Figure~\ref{fig:commute}a shows that most workers have relatively small variation in their first daily workplace arrival times, indicating that they repeatedly reach the inferred workplace at similar times across working days. Supplementary Figure~\ref{fig:commute}b further shows that mean workplace arrival times are concentrated in the morning, with the distribution peaking at approximately 9:00 a.m. Workplace departure times are more variable, as shown in Supplementary Figure~\ref{fig:commute}c, which is consistent with greater day-to-day variation in the end of the working day. Supplementary Figure~\ref{fig:commute}d shows that mean arrival times at the inferred home location are concentrated in the evening, primarily between approximately 8:00 p.m. and 11:00 p.m. The sequence of regular morning arrival at the inferred workplace and evening return to the inferred home location is consistent with daily commuting in Tokyo. These patterns are consistent with regular commuting and support the classification of these users as workers associated with the inferred workplace cells.

\section{Construction and validation of meeting probability}
\subsection{Measuring meeting probability (MP)}

Following the approach of \cite{atkin2022}, we construct a meeting probability based on the spatial and temporal co-presence of workers. Let $n_{ict}$ and $n_{jct}$ denote the number of workers from firm cell $i$ and firm cell $j$ present in cell $c$ during a 30-minute time window $t$.

We define total meetings between firm cells $i$ and $j$ as
\[
M_{ij} = \sum_{c,t} n_{ict} \, n_{jct},
\]
which counts the number of co-presence events in which workers from $i$ and $j$ are located in the same cell at the same time.

We further define potential maximum meetings as
\[
PMM_{ij} = \sum_{t} \left( \sum_{c} n_{ict} \right)\left( \sum_{c} n_{jct} \right),
\]
which captures the maximum number of possible co-presence opportunities, that is, the number of times workers from both firm cells are simultaneously active in the same time window, even if not in the same place. This normalization assumes that workers active within the same time window could potentially interact regardless of their exact locations.

The meeting probability is then defined as
\[
MP_{ij} = \frac{M_{ij}}{PMM_{ij}}.
\]
By construction, $MP_{ij}$ measures the probability that a randomly selected worker from firm cell $i$ is co-present in both space and time with a randomly selected worker from firm cell $j$. This method measures simultaneous spatial and temporal overlap as a necessary condition for in-person encounters, without directly observing the interactions themselves.

\subsection{Network structure of MP}
\begin{figure}
    \centering
    \includegraphics[width=1\linewidth]{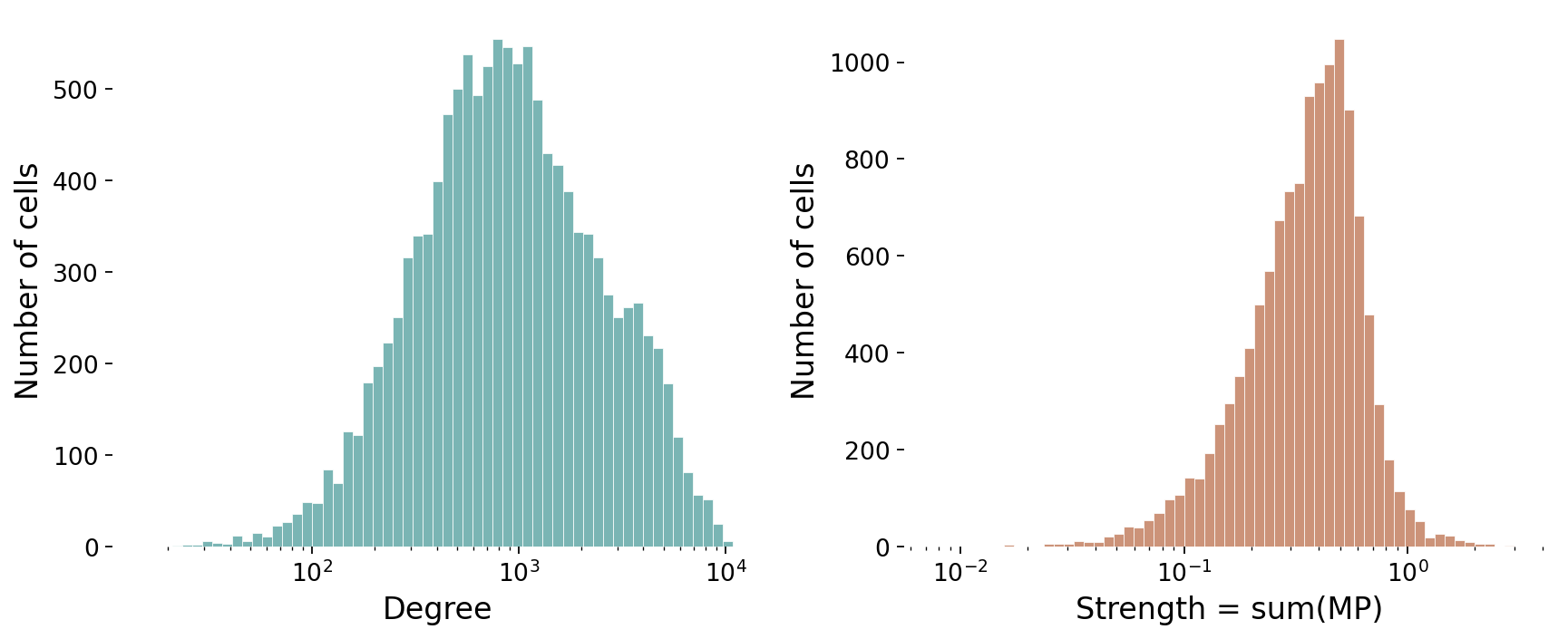}
    \caption{\textbf{Node degree and strength in the mobility network.} (a) Distribution of node degree, defined as the number of other firm cells connected to each firm cell through a non-zero meeting probability ($\mathrm{MP}>0$). (b) Distribution of node strength, defined as the sum of meeting probabilities across all connections of each firm cell.}
    \label{fig:degree}
\end{figure}

We examine the network structure induced by positive meeting probabilities, where nodes represent firm cells and edges connect pairs of firm cells with non-zero meeting probabilities. The complementary cumulative degree distribution (Supplementary Figure \ref{fig:degree}) exhibits a heavy-tailed pattern. The median firm cell is connected to 867 other cells, whereas the most connected cells have links to more than 10,000 cells, revealing substantial heterogeneity in the breadth of encounter connections across the city. This unequal connectivity is accompanied by substantial variation in node strength, defined as the sum of meeting probabilities across all links incident to a firm cell. Together, the degree and strength distributions show that firm cells differ markedly not only in how many other cells they encounter through mobility, but also in the aggregate intensity of these encounters.

\subsection{Alternative MP measurements}

To examine whether the observed meeting probability (MP) is disproportionately driven by incidental co-presence in crowded or highly frequented locations, we construct four alternative MP measures: (1) crowding-weighted MP, (2) popularity-weighted MP, (3) repeated MP, and (4) firm-cell-excluded MP. These alternatives test whether the observed MP remains similar after accounting for four possible sources of measurement bias: (1) encounters occurring when a location is unusually crowded at a particular time; (2) encounters occurring in locations that are consistently highly frequented throughout the observation period; (3) encounters observed only once; and (4) encounters occurring within the firm cell of either worker.

\paragraph{Crowding-weighted MP}
Some encounters occur in places that are unusually crowded at a particular time, such as major stations during rush hour or large venues immediately before or after an event. In these settings, workers from two firm cells may be observed together simply because many people are present simultaneously. To reduce this influence, we construct a crowding-adjusted MP that assigns less weight to encounters that occur in cell-time units with large numbers of mobility records. Let \(N_{ct}=\sum_k n_{kct}\) denote the total number of workers observed in cell \(c\) during time window \(t\), where \(k\) indexes all firm cells. We calculate the adjusted number of meetings as \(M_{ij}^{{crowd}}=\sum_{c,t}\frac{n_{ict}n_{jct}}{N_{ct}-1}\), and define the corresponding meeting probability as \(MP_{ij}^{\mathrm{crowd}}=M_{ij}^{\mathrm{crowd}}/PMM_{ij}\).

\paragraph{Popularity-weighted MP}

Some locations attract a large number of different visitors throughout the observation period. Examples include major transport hubs, shopping centers, and other widely used public facilities. Encounters in these locations may partly reflect their broad visitor base. We therefore construct a popularity-weighted MP that assigns less weight to encounters occurring in cells visited by many users over the full observation period. Let \(U_c\) denote the total number of distinct users who visited cell \(c\) during the full observation period. We calculate the adjusted number of meetings as \(M_{ij}^{\mathrm{pop}}=\sum_{c,t}\frac{n_{ict}n_{jct}}{U_c}\), and define the corresponding meeting probability as \(MP_{ij}^{\mathrm{pop}}=M_{ij}^{\mathrm{pop}}/PMM_{ij}\). 

\paragraph{Repeated MP}

Co-presence observed only once in a particular location may reflect incidental overlap or measurement noise. We therefore construct repeated MP using locations where workers from firm cells \(i\) and \(j\) are jointly observed in at least two time bins. Let \(R_{ijc}\) denote the number of time bins in which workers from firm cells \(i\) and \(j\) are jointly observed in cell \(c\). We calculate the adjusted number of meetings by retaining all co-presence contributions from cells satisfying \(R_{ijc}\geq 2\), and denote the resulting value by \(M_{ij}^{\mathrm{repeat}}\). We then define the corresponding meeting probability as \(MP_{ij}^{\mathrm{repeat}}=M_{ij}^{\mathrm{repeat}}/PMM_{ij}\).

\paragraph{Firm-cell-excluded MP}

Some encounters may arise mechanically because workers spend substantial time in the cell containing their own workplace. We therefore construct a firm-cell-excluded MP. For each user, we remove all stay records observed in the user’s assigned workplace, while retaining stays at all other locations, including the workplace cells of other firms. Using the filtered stay records, we recalculate both the number of meetings, \(M_{ij}^{\mathrm{ex}}\), and the corresponding opportunity denominator, \(PMM_{ij}^{\mathrm{ex}}\). We then define the adjusted meeting probability as \(MP_{ij}^{\mathrm{ex}}=M_{ij}^{\mathrm{ex}}/PMM_{ij}^{\mathrm{ex}}\). This measure examines whether the observed MP measurement is primarily driven by workers remaining in their own workplace cells.

\begin{figure}
    \centering
    \includegraphics[width=1\linewidth]{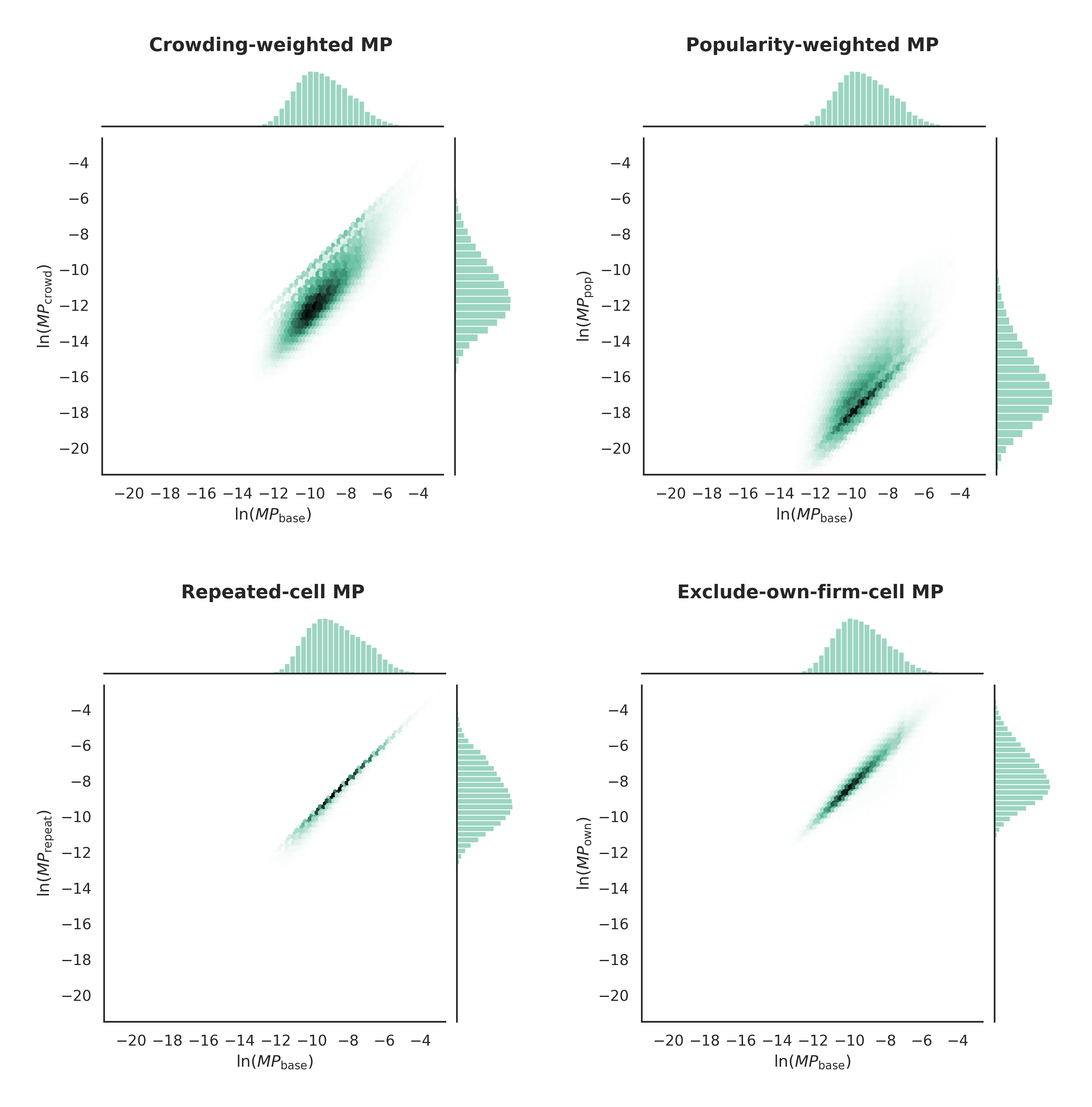}
    \caption{\textbf{Robustness of meeting probability to alternative measures.} Each panel compares the observed meeting probability (MP) with one alternative MP measure: crowding-weighted MP, popularity-weighted MP, repeated MP, or firm-cell-excluded MP. The central hexagonal layer displays the local density of cell pairs, while the marginal histograms show the distributions of baseline and alternative MP values.}
    \label{fig:alternative_mp}
\end{figure}

Supplementary Figure~\ref{fig:alternative_mp} and Supplementary Table~\ref{tab:alternative_mp_correlations} compare the alternative MP measures with the observed MP. Overall, the alternative measures remain closely associated with the observed MP. The crowding-weighted MP is strongly correlated with the observed MP with Pearson and Spearman correlations of 0.819 and 0.835, respectively. The popularity-weighted MP has a lower Pearson correlation of 0.446 but retains a relatively high Spearman correlation of 0.796, indicating that down-weighting retains the relative ordering of pairs. The repeated MP shows the opposite pattern, with a Pearson correlation of 0.997 and a Spearman correlation of 0.463. This suggests that removing one-off encounters primarily affects the relative ranking of weaker ties, while leaving the overall intensity structure largely unchanged. Excluding encounters within either worker's firm cell also produces moderate correlations with the observed MP (Pearson = 0.601; Spearman = 0.669). Taken together, these comparisons show that the alternative adjustments change MP values and rankings to different degrees. No single adjustment eliminates the association with observed MP.

\begin{table}[htbp]
\centering
\caption{Correlations between the observed MP and four alternative MP measures}
\label{tab:alternative_mp_correlations}
\begin{tabular}{lcc}
\toprule
Alternative measure & Pearson correlation & Spearman correlation \\
\midrule
Crowding-weighted MP   & 0.819 & 0.835 \\
Popularity-weighted MP & 0.446 & 0.796 \\
Repeated MP            & 0.997 & 0.463 \\
Firm-cell-excluded MP  & 0.601 & 0.669 \\
\bottomrule
\end{tabular}
\begin{minipage}{0.92\textwidth}
\footnotesize
\textit{Notes:} Each row compares an alternative meeting probability (MP) measure with baseline MP across cell pairs. Pearson correlations use the MP values; Spearman correlations use their ranks.
\end{minipage}
\end{table}

\subsection{Time-shift null MP}
The previous sections show that the observed MP is robust to alternative measures that reduce the influence of crowded or widely visited locations and other specific sources of incidental co-location. We next examine whether the observed MP reflects temporal coordination, rather than simply visits to the same places. We construct a time-shift null model to break the temporal alignment between workers while preserving their real mobility patterns. We randomly shift each worker's everyday mobility trajectory forward or backward during the observation period. To preserve systematic differences between weekday and weekend mobility, we randomly reassign each worker’s mobility records to corresponding days in other weeks. Weekday records are shuffled only across weekdays, while weekend records are shuffled only across weekends. This procedure retains mobility routines but disrupts the original calendar-date alignment across workers. In this case, workers continue to visit the same actual locations, but encounters that depend on time alignment are no longer systematically preserved. For example, two workers may arrange to meet at a café at noon on a particular day. Randomly shifting their daily trajectories to different dates would disrupt this planned encounter.

We repeated this procedure independently for $B=100$ null iterations. In each iteration, a new set of worker-specific time shifts was generated, and the encounter network was reconstructed from the shifted trajectories. For each cell pair $(i,j)$, we recomputed the realized encounters $M_{ij}^{(b)}$, potential encounters $PMM_{ij}^{(b)}$, and the corresponding meeting probability,

\[
MP_{ij}^{(b)}
=
\frac{M_{ij}^{(b)}}{PMM_{ij}^{(b)}}.
\]

The resulting distribution of $\{MP_{ij}^{(b)}\}_{b=1}^{100}$ serves as the null distribution against which the observed meeting probability is compared.

The observed firm-pair encounter probability is compared with the empirical
null distribution
\[
\{MP_{ij}^{(1)}, MP_{ij}^{(2)}, \ldots, MP_{ij}^{(100)}\}.
\]
An upper-tail empirical \(p\)-value is computed as
\[
p_{ij}^{\mathrm{upper}}
=
\frac{
1 + \sum_{b=1}^{100}
\mathbf{1}\left(MP_{ij}^{(b)} \geq MP_{ij}^{\mathrm{obs}}\right)
}{
100+1
}.
\]

If the observed \(MP_{ij}^{\mathrm{obs}}\) is systematically larger than the time-shift null distribution, this indicates that observed co-location is not only a consequence of workers visiting similar places, but also depends on their synchronized timing.

As shown in Supplementary Figure \ref{fig:nullMP}a, the distribution of \(MP_{ij}^{\mathrm{obs}}\) is shifted toward higher values than the time-shift null distribution. Across 100 time-shift realizations, none produced an aggregate MP as high as the \(MP_{ij}^{\mathrm{obs}}\) ($p<0.01$). This indicates that common spatial routines explain only part of the observed encounters. The difference is consistent with additional temporal alignment among workers, without identifying whether any co-presence involved a planned meeting. We further compute the observed-to-null MP ratio for each cell pair in Supplementary Figure \ref{fig:nullMP}b. The median observed-to-null ratio is 1.61, indicating that the \(MP_{ij}^{\mathrm{obs}}\) is typically 61\% higher than null MP. The 90th percentile of the ratio reaches 7.79, showing that for the upper 10\% of cell pairs, observed MP values are nearly 8 times their null expectations. Together, these results show that temporal synchronization contributes substantially to \(MP_{ij}^{\mathrm{obs}}\).

\begin{figure}
    \centering
    \includegraphics[width=1\linewidth]{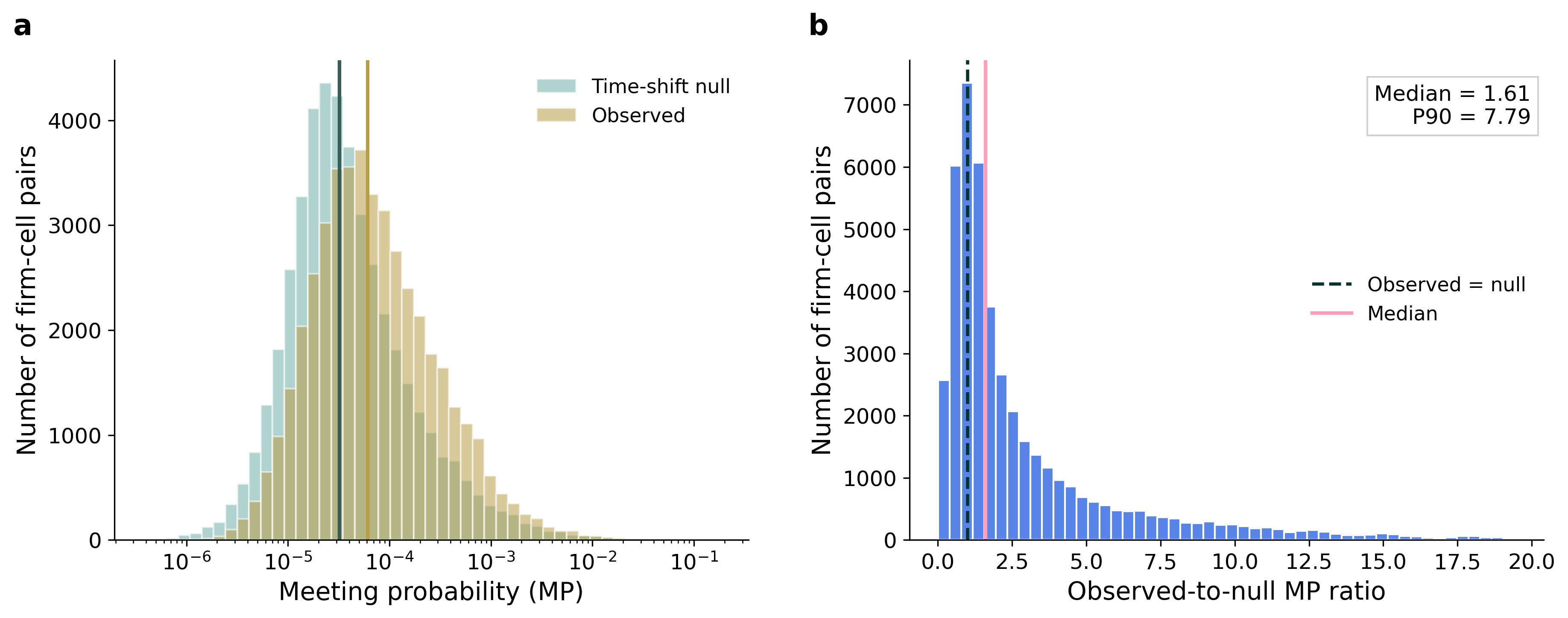}
    \caption{
\textbf{Observed and time-shift null meeting probabilities.}
(a) Distribution of observed meeting probability (MP) and mean MP from 100 realizations of the time-shift null model. (b) Distribution of the observed-to-null MP ratio for each cell pair, calculated by dividing observed MP by its mean across the 100 null realizations. The dashed line indicates equal observed and null MP ($\mathrm{ratio}=1$), and the pink line indicates the median ratio.
}
    \label{fig:nullMP}
\end{figure}

\subsection{Gravity-based null MP}

Geographical proximity may explain part of the observed encounter opportunities between firms. Workers from nearby workplaces are more likely to visit overlapping locations, even without any specific relationship between their firms. Following the gravity law, we model the expected number of meetings between workplaces \(i\) and \(j\) as

\begin{equation}
\mu_{ij}^{gravity}
=
\alpha
\frac{W_iW_j}
{(d_0+d_{ij})^\gamma},
\end{equation}

where \(W_i\) and \(W_j\) are the numbers of workers in cells \(i\) and \(j\), and \(d_{ij}\) is the geographical distance between the two cells in kilometres. The distance offset is fixed at \(d_0=0.1\) km, and \(\gamma\) controls the distance-decay effect. We estimate \(\alpha\) and \(\gamma\) using Poisson pseudo-maximum likelihood (PPML), with observed meeting counts \(M_{ij}\) as the dependent variable. The estimated parameters are \(\widehat{\alpha}=0.06365\) and \(\widehat{\gamma}=1.35413\).

We then normalize the predicted meeting counts by the same potential meeting count \(PMM_{ij}\) used for observed MP:

\begin{equation}
MP_{ij}^{gravity}
=
\frac{\widehat{\mu}_{ij}^{gravity}}{PMM_{ij}}
=
0.06365
\frac{W_iW_j}
{PMM_{ij}(0.1+d_{ij})^{1.35413}}.
\end{equation}

\section{Regression analysis and robustness checks}
\subsection{Baseline association between MP and transaction outcomes}

To examine the association between workers’ encounter opportunities and subsequent inter-firm transactions, we estimate logistic regression models using meeting probability measured in 2019 and transaction records from 2016 to 2022. Let $t$ denote the number of years after 2019, so $t=1$ denotes 2020, and let $Y_{ij,t}$ denote the binary transaction outcome for cell pair $(i,j)$ at year $t$.

We use three outcomes: (1) link existence, (2) link formation, and (3) link dissolution. (1) Link existence equals one when a cell pair has at least one buyer--supplier transaction in year $t$. (2) link formation equals one when a transaction link first appears after no transaction in any preceding year observed in the 2016--2022 data; for example, a birth at $t=1$ has no observed transaction in 2016--2019 and a transaction in 2020. (3) Link dissolution equals one at the first disappearance of a link that remained present through all preceding years after 2019. For example, dissolution at $t=1$ means that a link observed in 2019 is absent in 2020. These outcomes are modelled using fixed-effects logit models with a binomial distribution and logit link, estimated in R with the \texttt{feglm} function in the \texttt{fixest} package.

The main explanatory variable is the logarithm of meeting probability, $\ln(MP_{ij})$, which captures encounter opportunities between workers from two different firm cells during the mobility observation period.

We estimate the following specification:
\begin{equation}
\mathrm{logit}\!\left[\Pr\!\left(Y_{ij,t}=1\right)\right]
=
\beta \ln(\mathrm{MP}_{ij})
+
\mathbf{X}_{ij}'\boldsymbol{\gamma}
+
\delta_i
+
\delta_j .
\end{equation}

The model includes cell fixed effects $\delta_i$ and $\delta_j$ and a set of pair-level controls. We first control for geographic distance between cell centroids (\texttt{distance}). Firms are not randomly distributed across urban space: firms engaging in similar activities, requiring frequent coordination, or sharing latent collaboration potential may endogenously locate near one another. Under this scenario, high meeting probability could partly reflect spatial sorting into proximate locations, which would confound its association with transaction outcomes.

For firm cells $i$ and $j$, we measure the industry similarity index (\texttt{industryoverlap}) as the inner product of their industry-share vectors, $\operatorname{IndustrySimilarity}_{ij}=\sum_{k=1}^{K}s_{ik}s_{jk}$. The index ranges from 0 to 1 and can be interpreted as the probability that two firms drawn independently from cells $i$ and $j$ belong to the same industry. It equals 0 when the two cells share no industries and reaches 1 when both cells consist entirely of firms from the same industry. Higher values therefore indicate greater overlap in their industry composition.

Finally, we control for differences in workforce composition and local economic structure using the 2021 Economic Census for Business Activity. Specifically, we include the differences between the two firm cells in total employment (\texttt{d\_emp}), establishment density (\texttt{d\_est}), and workforce gender composition (\texttt{d\_maleshare}). These controls account for the possibility that cell pairs with similar economic and demographic characteristics may have both overlapping worker mobility and a higher probability of a transaction link. Together with the cell fixed effects, the pair-level controls adjust for measured spatial, industrial, and socioeconomic differences. Descriptive statistics for the transaction outcomes and pair-level covariates are shown in Supplementary Table~\ref{tab:descriptive_stats}. All variance inflation factors (VIFs) were below 2, indicating no substantial multicollinearity among the explanatory variables. 

The regression results are shown in Supplementary Table~\ref{tab:main}.

\begin{table}[htbp]
\centering
\caption{Descriptive statistics}
\label{tab:descriptive_stats}
\begin{tabular}{lrrrrr}
\hline
Variable & N & Mean & SD & Min & Max \\
\hline
$\ln(MP)$                 & 76,106,953 & -13.318 & 1.457 & -13.816 & -1.187 \\
$\ln(\mathrm{Distance})$              & 76,106,953 & 9.174  & 0.778 & 4.443   & 10.917 \\
Industry similarity (firm weighted) & 76,106,953 & 0.101 & 0.133 & 0.000 & 1.000 \\
Industry similarity (employee weighted) & 76,106,953 & 0.095 & 0.185 & 0.000 & 1.000 \\
d\_emp             & 66,349,440 & 1.430  & 1.093 & 0.000   & 11.543 \\
d\_est             & 66,349,440 & 1.041  & 0.787 & 0.000   & 7.591 \\
d\_maleshare          & 66,337,921 & 0.116  & 0.088 & 0.000   & 0.918 \\
\hline
\end{tabular}
\end{table}

\begin{table*}[!htbp]
\centering
\caption{Baseline associations between meeting probability and transaction outcomes, 2017--2022}
\label{tab:main}
\resizebox{\textwidth}{!}{%
\begin{tabular}{l*{6}{c}}
\toprule
Year & 2017 & 2018 & 2019 & 2020 & 2021 & 2022 \\
\midrule
\multicolumn{7}{l}{\textit{Panel A: Link existence}} \\
$\ln(\mathrm{MP})$ & 0.0202*** & 0.0206*** & 0.0190*** & 0.0200*** & 0.0199*** & 0.0203*** \\
 & (0.0023) & (0.0023) & (0.0022) & (0.0022) & (0.0022) & (0.0021) \\
Industry similarity & 2.6381*** & 2.7350*** & 2.7590*** & 2.7482*** & 2.7815*** & 2.8314*** \\
 & (0.1828) & (0.1734) & (0.1703) & (0.1642) & (0.1640) & (0.1653) \\
$\ln(\mathrm{distance})$ & -0.3481*** & -0.3468*** & -0.3548*** & -0.3521*** & -0.3508*** & -0.3433*** \\
 & (0.0114) & (0.0112) & (0.0109) & (0.0108) & (0.0108) & (0.0109) \\
Controls & Yes & Yes & Yes & Yes & Yes & Yes \\
Cell fixed effects ($i$ and $j$) & Yes & Yes & Yes & Yes & Yes & Yes \\
Observations & 25,105,039 & 25,776,838 & 26,130,069 & 26,409,622 & 26,054,146 & 25,737,824 \\
Pseudo $R^2$ & 0.3137 & 0.3165 & 0.3163 & 0.3183 & 0.3183 & 0.3191 \\
\midrule
\multicolumn{7}{l}{\textit{Panel B: Link formation}} \\
$\ln(\mathrm{MP})$ & 0.0116** & 0.0183*** & 0.0140*** & 0.0272*** & 0.0113** & 0.0151*** \\
 & (0.0050) & (0.0048) & (0.0051) & (0.0055) & (0.0052) & (0.0052) \\
Industry similarity & 3.1580*** & 3.1238*** & 3.0327*** & 2.7347*** & 3.2137*** & 3.2694*** \\
 & (0.2172) & (0.1876) & (0.2147) & (0.1973) & (0.2289) & (0.2317) \\
$\ln(\mathrm{distance})$ & -0.3863*** & -0.3235*** & -0.3602*** & -0.3284*** & -0.3423*** & -0.3165*** \\
 & (0.0172) & (0.0175) & (0.0177) & (0.0185) & (0.0192) & (0.0195) \\
Controls & Yes & Yes & Yes & Yes & Yes & Yes \\
Cell fixed effects ($i$ and $j$) & Yes & Yes & Yes & Yes & Yes & Yes \\
Observations & 9,142,301 & 8,304,440 & 8,285,680 & 7,892,753 & 6,956,230 & 7,658,881 \\
Pseudo $R^2$ & 0.1656 & 0.1637 & 0.1584 & 0.1606 & 0.1541 & 0.1642 \\
\midrule
\multicolumn{7}{l}{\textit{Panel C: Link dissolution}} \\
$\ln(\mathrm{MP})$ & -0.0002 & -0.0046 & 0.0038 & -0.0017 & 0.0014 & -0.0073 \\
 & (0.0059) & (0.0055) & (0.0053) & (0.0050) & (0.0049) & (0.0047) \\
Industry similarity & -0.0643 & 0.0202 & 0.3749 & -0.0452 & 0.1731 & 0.4844** \\
 & (0.2965) & (0.2783) & (0.2868) & (0.2611) & (0.2628) & (0.2457) \\
$\ln(\mathrm{distance})$ & 0.0029 & 0.0032 & -0.0024 & 0.0009 & 0.0221 & -0.0137 \\
 & (0.0175) & (0.0176) & (0.0163) & (0.0150) & (0.0151) & (0.0153) \\
Controls & Yes & Yes & Yes & Yes & Yes & Yes \\
Cell fixed effects ($i$ and $j$) & Yes & Yes & Yes & Yes & Yes & Yes \\
Observations & 161,808 & 174,936 & 184,636 & 196,207 & 200,815 & 203,487 \\
Pseudo $R^2$ & 0.1806 & 0.1790 & 0.1606 & 0.1562 & 0.1699 & 0.1671 \\
\bottomrule
\end{tabular}%
}
\begin{minipage}{\textwidth}
\footnotesize \textit{Notes:} All panels report fixed-effects logit models at the cell-pair level for 2017--2022. Panel A indicates whether a transaction link exists in the displayed year. Panel B indicates whether a link exists in the displayed year among cell pairs with no transaction records before. Panel C indicates whether at least one firm-pair relationship active in the preceding year dissolves within the cell pair. Standard errors are two-way clustered by endpoint cells $i$ and $j$ and are reported in parentheses. $^{*}p<0.10$, $^{**}p<0.05$, $^{***}p<0.01$.
\end{minipage}
\end{table*}

\subsection{Robustness to firm-count controls}
Cell pairs containing more firms have more potential firm combinations and may therefore be more likely to exhibit both worker encounters and transaction ties. To account for this, we include decile indicators for the product of firm counts in the two cells. Supplementary Table \ref{tab:firm-count-product-deciles} shows that the coefficient on meeting probability remains highly stable across all three years, ranging from 0.0199 to 0.0202. 

\begin{table}[!htbp]
\centering
\caption{Robustness to firm-count controls}
\label{tab:firm-count-product-deciles}

\begin{tabular}{lccc}
\toprule
& \textit{t}=1 & \textit{t}=2 & \textit{t}=3 \\
\midrule

$\ln(MP_{ij})$
& 0.0200*** & 0.0199*** & 0.0202*** \\
& (0.0022) & (0.0022) & (0.0021) \\

$\ln(\mathrm{Distance}_{ij})$
& -0.3510*** & -0.3495*** & -0.3419*** \\
& (0.0108) & (0.0108) & (0.0109) \\

Industry similarity
& 2.7731*** & 2.7958*** & 2.8318*** \\
& (0.1532) & (0.1540) & (0.1551) \\

\midrule

Observations
& 26,409,622 & 26,054,146 & 25,737,824 \\

Firm-count-product decile controls
& Yes & Yes & Yes \\

Cell $i$ and $j$ fixed effects
& Yes & Yes & Yes \\
Pair-level controls
& Yes & Yes & Yes \\
Two-way clustered SEs
& Yes & Yes & Yes \\

\bottomrule
\end{tabular}

\vspace{0.3em}

\begin{minipage}{0.95\linewidth}
\footnotesize
\textit{Notes:} The binary outcome is link existence between firm cells $i$ and $j$ in the indicated year. cell pairs are divided into deciles based on the product of the numbers of firms in the two cells. Decile indicators are included in all specifications, with the lowest decile omitted as the reference category. All other baseline controls are included. All specifications are fixed-effects logit models with cell fixed effects for $i$ and $j$. Standard errors are two-way clustered by firm cells $i$ and $j$ and are reported in parentheses. $^{*}p<0.10$, $^{**}p<0.05$, $^{***}p<0.01$.
\end{minipage}

\end{table}

\subsection{Robustness to alternative MP measurements}

The alternative MP measures (Supplementary Note 2.3) down-weight encounters in (1) crowded cell–time units and (2) highly frequented locations, (3) retain co-presence only at locations are observed at least two times, and (4) exclude stays within each worker’s own workplace cell. We replace the observed $MP_{ij}$ with each alternative measure in the baseline regression model (link existence in 2020) and compare the resulting estimates.

Across all four alternative specifications, the estimated coefficient on meeting probability remains positive and statistically significant, although its magnitude varies across measures (Supplementary Table~\ref{tab:mp-robustness}). These results indicate that the positive association is robust to alternative encounter definitions, while its estimated magnitude is sensitive to how encounters are weighted and which locations are included.

\begin{table}[!htbp]
\centering
\caption{Robustness to alternative meeting probability measures}
\label{tab:mp-robustness}

\resizebox{\textwidth}{!}{%
\begin{tabular}{lccccc}
\toprule
& Observed MP
& Crowding-weighted MP
& Popularity-weighted MP
& Repeated MP
& Firm-cell-excluded MP \\
\midrule

$\ln(MP_{ij})$
& 0.0200*** & 0.0335*** & 0.1978*** & 0.0203*** & 0.0046*** \\
& (0.0022) & (0.0032) & (0.0137) & (0.0022) & (0.0017) \\

$\ln(\mathrm{Distance}_{ij})$
& -0.3521*** & -0.3459*** & -0.3310*** & -0.3511*** & -0.3676*** \\
& (0.0108) & (0.0108) & (0.0111) & (0.0108) & (0.0107) \\

Industry similarity
& 2.7482*** & 2.7524*** & 2.7654*** & 2.7487*** & 2.7508*** \\
& (0.1642) & (0.1644) & (0.1653) & (0.1637) & (0.1642) \\

\midrule

Observations
& 26,409,622 & 26,409,622 & 26,409,622 & 26,409,622 & 26,409,622 \\

McFadden pseudo $R^2$
& 0.3183 & 0.3183 & 0.3184 & 0.3183 & 0.3182 \\

Cell $i$ and $j$ fixed effects
& Yes & Yes & Yes & Yes & Yes \\
Pair-level controls
& Yes & Yes & Yes & Yes & Yes  \\
Two-way clustered SEs
& Yes & Yes & Yes & Yes & Yes \\

\bottomrule
\end{tabular}%
}

\vspace{0.3em}

\begin{minipage}{0.98\linewidth}
\footnotesize
\textit{Notes:} The binary outcome is link existence between firm cells $i$ and $j$ in 2020. Each column reports a fixed-effects logit model using observed meeting probability (MP) or one of four alternative MP measures: crowding-weighted MP, popularity-weighted MP, repeated MP, and firm-cell-excluded MP. All specifications include the same baseline controls and cell fixed effects for $i$ and $j$. Standard errors are two-way clustered by firm cells $i$ and $j$ and are reported in parentheses. $^{*}p<0.10$, $^{**}p<0.05$, $^{***}p<0.01$.
\end{minipage}

\end{table}

\subsection{Null test using the time-shift null MP}

\begin{figure}
    \centering
    \includegraphics[width=1\linewidth]{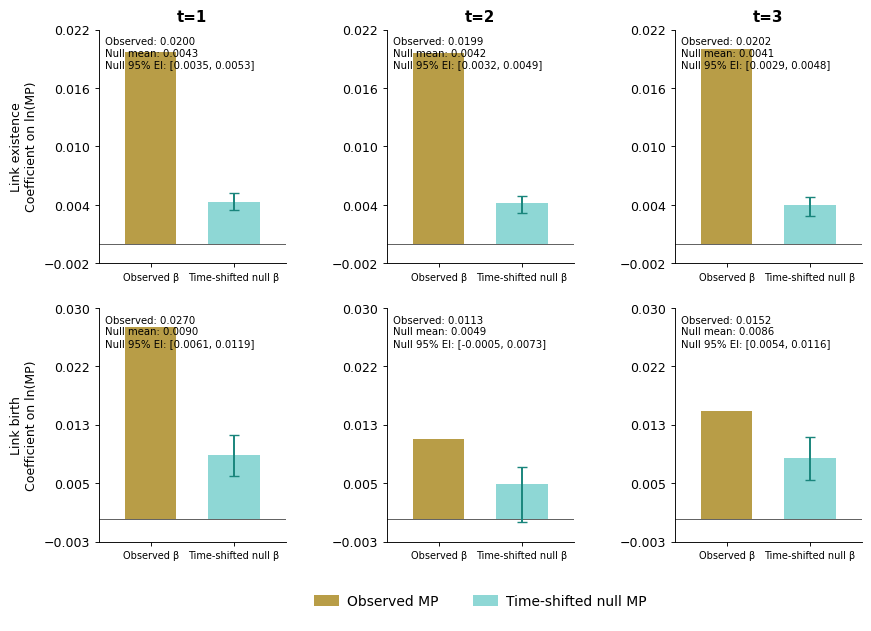}
    \caption{\textbf{Observed and time-shift null meeting-probability coefficients.} Each panel compares the estimated coefficient of observed MP with time-shift null MP. Rows correspond to the link existence and link formation models. The null bars show the mean coefficient across 100 null realizations, and the error bars indicate the empirical 95\% interval.}
    \label{fig:timeshift}
\end{figure}

Having established positive associations between observed MP and both link existence and link formation, we next examined whether shared spatial activity patterns alone could reproduce these associations. We applied the time-shift null model described in Supplementary Note 2.4, which reassigns each worker’s daily mobility trajectory to another date while preserving visited locations and within-day mobility patterns. This procedure disrupts the original calendar-date alignment across workers. For each of 100 shuffled realizations, we reconstructed the encounter network, recalculated MP, and re-estimated the same regression specifications used for observed MP. Supplementary Figure~\ref{fig:timeshift} compares the observed coefficients with the means and empirical 95\% intervals of the null estimates. For both link existence and link formation, the observed coefficients exceed the upper bounds of the corresponding null intervals across all three post-observation years. For link existence, the observed coefficients remain close to 0.020, approximately 4.7–4.9 times as large as the null means of 0.0041–0.0043. For link formation, the observed coefficients are 0.0270, 0.0113, and 0.0152, respectively, corresponding to 3.0, 2.3, and 1.8 times the null means of 0.0090, 0.0049, and 0.0086. These reductions suggest that shared visitation patterns alone do not reproduce the associations observed when workers’ original temporal alignment is retained.

\subsection{Degree-preserving transaction-network null model}

\begin{figure}
    \centering
    \includegraphics[width=1\linewidth]{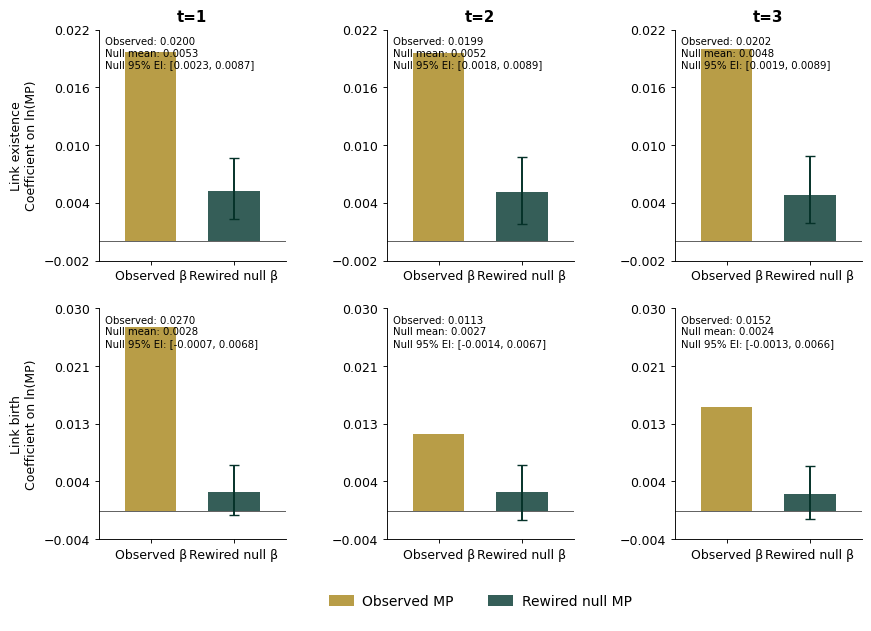}
 
    \caption{\textbf{Meeting-probability coefficients in observed and rewired transaction networks.} Each panel compares the estimated coefficient of observed MP with rewired null MP. Rows correspond to the link existence and link formation. The null bars show the mean coefficient across 100 rewired realizations, and the error bars indicate the empirical 95\% interval.}
    \label{fig:rewired}
\end{figure}

We constructed a network-based null model to assess whether the association between MP and transactions could be explained by general connectivity in the transaction network. For each year, the observed transaction relationships were represented as a network in which nodes corresponded to firm cells and an edge indicated the presence of at least one transaction relationship between two firm cells during that year. We randomized each annual transaction network using degree-preserving edge swaps. In each swap, two existing edges, $(a,b)$ and $(c,d)$, were selected and replaced by $(a,d)$ and $(c,b)$, provided that the proposed edges did not create self-loops or duplicate edges. The procedure was repeated 100 times to randomize the identities of transaction partners while preserving the degree of every node and the total number of edges in the network. We generated 100 independently rewired realizations for each network. For each rewired realization, we reconstructed the corresponding binary transaction outcome and re-estimated the same regression specification used in the main analysis. Supplementary Figure \ref{fig:rewired} shows the estimated coefficients. For the link existence model in 2020, the observed MP coefficient (0.020) is approximately 3.8 times larger than the mean MP coefficient under transaction-network rewiring (0.0054). A similar pattern is observed for the link formation models. A smaller MP estimates under transaction-network rewiring indicates that mobility-based encounters are associated with whom firms transact with, rather than merely identifying firms that transact more frequently.

\subsection{Robustness to alternative transaction-history windows}

To assess whether the estimated associations depend on how recent transaction histories are defined, we construct alternative birth and dissolution outcomes using one- to four-year transaction-history windows. These specifications vary the amount of prior transaction history required before a transition is identified, allowing us to test whether the results are sensitive to a particular window definition.

\begin{figure}
    \centering
    \includegraphics[width=1\linewidth]{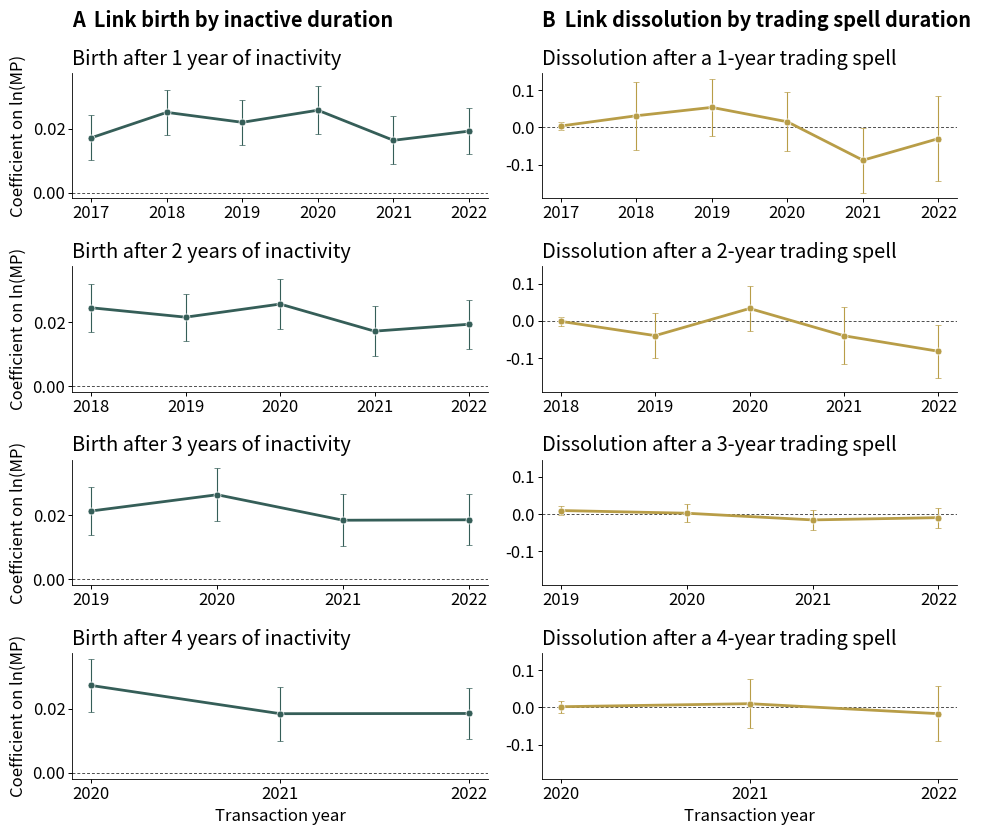}
    \caption{\textbf{Robustness to alternative transaction-history windows.}
    Estimated coefficients of $\ln(\mathrm{MP})$, for link formation and link dissolution  across one- to four-year transaction-history windows. For link formation, each window requires no transactions during the preceding one to four consecutive years. For link dissolution, the window captures relationships with an observed continuous trading spell of the corresponding length before the target year. Points show coefficient estimates and error bars indicate 95\% confidence intervals.}
    \label{fig:transaction_history_windows}
\end{figure}

For link formation, the windows require no transactions during the preceding one to four consecutive years. A birth event occurs when a previously inactive firm-cell pair starts trading in the target year. For link dissolution, the windows capture relationships with continuous trading histories of one to four years before the target year. A dissolution event occurs when a pair stops trading in the target year. Because transaction histories are observed from 2016 onward, longer windows are available only for later target years. For example, a four-year window can first be constructed for a 2020 outcome. These alternative specifications therefore provide a robustness check on whether the estimated relationship between meeting probability and transaction transitions depends on the length of prior inactivity or prior trading history (Supplementary Figure~\ref{fig:transaction_history_windows}). 

Across the alternative transaction-history windows, the association between MP and link formation remains consistently positive and statistically significant. This pattern is stable across one- to four-year inactivity windows, indicating that the main birth result is not sensitive to a particular definition of prior inactivity. In contrast, the estimates for link dissolution are substantially less stable across years and window definitions and are generally not statistically significant. This weaker and less consistent pattern suggests that the termination of existing transaction relationships may depend more strongly on other factors.

\subsection{Temporal variation in the MP--link formation association}

To assess whether the association between MP and link formation differs across years, we pool cell-pair transaction observations from 2017 to 2022. The dependent variable equals one if a transaction link first appears in that year. We estimate the following pooled logit model:

\begin{equation}
\operatorname{logit}\left[
P(Y_{ij,t}^{\mathrm{birth}}=1)
\right]
=
\sum_{s=2017}^{2022}
\beta_s
\left[
\ln(\mathrm{MP}_{ij})
\times \mathbb{1}(t=s)
\right]
+
\mathbf{X}_{ij}'\boldsymbol{\gamma}
+
\delta_i+\delta_j+\lambda_t,
\label{eq:pooled_birth_annual}
\end{equation}

where $i$ and $j$ denote the two cells forming a cell pair. $\mathrm{MP}_{ij}$ is measured using mobility data from 2019 and remains fixed across outcome years. The coefficients $\beta_s$ capture the year-specific associations between MP and link formation. The model includes baseline cell-pair controls $\mathbf{X}_{ij}$, cell fixed effects $\delta_i$ and $\delta_j$, and year fixed effects $\lambda_t$. We use a Wald test to assess the joint null hypothesis that all six year-specific coefficients are equal.

To examine whether this association differs between the periods before and after the mobility measurement, we additionally estimate a specification with separate MP coefficients for 2017--2019 and 2020--2022:

\begin{equation}
\begin{aligned}
\operatorname{logit}\left[
P(Y_{ij,t}^{\mathrm{birth}}=1)
\right]
={}&
\beta_{\mathrm{pre}}
\ln(\mathrm{MP}_{ij})
\times \mathbb{1}(t\leq 2019)
\\
&+
\beta_{\mathrm{post}}
\ln(\mathrm{MP}_{ij})
\times \mathbb{1}(t\geq 2020)
\\
&+
\mathbf{X}_{ij}'\boldsymbol{\gamma}
+
\delta_i+\delta_j+\lambda_t.
\end{aligned}
\label{eq:pooled_birth_pre_post}
\end{equation}

Here, $\beta_{\mathrm{pre}}$ captures the association for outcomes observed before or during the mobility measurement year, while $\beta_{\mathrm{post}}$ captures the association for subsequent outcomes. We test whether the two coefficients differ using a Wald test of $H_0:\beta_{\mathrm{pre}}=\beta_{\mathrm{post}}$. Standard errors in both specifications are two-way clustered by cells $i$ and $j$.

The year-specific MP coefficients are positive throughout 2017--2022, although their magnitudes and statistical significance vary across years (Supplementary Table~\ref{tab:pooled_birth_annual}). The largest coefficient is observed in 2020 ($\beta=0.0299$, $p<0.001$), the first year after mobility measurement. The joint hypothesis of equal coefficients across years is rejected ($\chi^2(5)=16.86$, $p=0.0048$). In the period-based specification, the coefficient is larger for 2020--2022 ($\beta=0.0205$, $p<0.001$) than for 2017--2019 ($\beta=0.0115$, $p<0.001$), with a statistically significant difference ($\chi^2(1)=4.63$, $p=0.0315$; Supplementary Table~\ref{tab:pooled_birth_pre_post}). Thus, the association is stronger for transaction formation observed after mobility measurement. However, the positive association with earlier outcomes suggests that MP may also capture persistent pair-level characteristics or relationships.

\begin{table}[!htbp]
\centering
\caption{Pooled associations between meeting probability and link formation: year-specific coefficients}
\label{tab:pooled_birth_annual}
\begin{tabular}{lc}
\toprule
& link formation \\
\midrule
$\ln(\mathrm{MP}) \times 2017$
    & 0.0097** \\
    & (0.0046) \\

$\ln(\mathrm{MP}) \times 2018$
    & 0.0156*** \\
    & (0.0048) \\

$\ln(\mathrm{MP}) \times 2019$
    & 0.0091* \\
    & (0.0050) \\

$\ln(\mathrm{MP}) \times 2020$
    & 0.0299*** \\
    & (0.0052) \\

$\ln(\mathrm{MP}) \times 2021$
    & 0.0066 \\
    & (0.0058) \\

$\ln(\mathrm{MP}) \times 2022$
    & 0.0227*** \\
    & (0.0059) \\

\midrule
Controls & Yes \\
Year fixed effects & Yes \\
Cell fixed effects ($i$ and $j$) & Yes \\
Observations & 154,105,210 \\
Pseudo $R^2$ & 0.2087 \\

\midrule
\multicolumn{2}{l}{\textit{Wald test}} \\
$\chi^2(5)$ & 16.8634 \\
$p$-value & 0.0048 \\
\bottomrule
\end{tabular}

\vspace{0.5em}
\begin{minipage}{0.85\textwidth}
\footnotesize
\textit{Notes:}
The dependent variable indicates the first observed formation of a transaction link in year $t$, conditional on no transaction link in any preceding observed year since 2016. The model includes interaction terms between $\ln(\mathrm{MP})$ and year indicators, year and cell fixed effects, and baseline controls whose coefficients are omitted. Standard errors are two-way clustered by cells $i$ and $j$ and reported in parentheses. Pseudo $R^2$ denotes McFadden's pseudo $R^2$. The Wald test evaluates the joint null hypothesis that all year-specific MP coefficients are equal ($\beta_{2017}=\cdots=\beta_{2022}$), using the full two-way-clustered covariance matrix. $^{*}p<0.10$, $^{**}p<0.05$, $^{***}p<0.01$.
\end{minipage}
\end{table}

\begin{table}[!htbp]
\centering
\caption{Pooled associations between meeting probability and link formation: pre- and post-period coefficients}
\label{tab:pooled_birth_pre_post}
\begin{tabular}{lc}
\toprule
& link formation \\
\midrule
$\ln(\mathrm{MP}) \times \mathrm{Pre}\ (2017\text{--}2019)$
    & 0.0115*** \\
    & (0.0029) \\

$\ln(\mathrm{MP}) \times \mathrm{Post}\ (2020\text{--}2022)$
    & 0.0205*** \\
    & (0.0035) \\

\midrule
Controls & Yes \\
Year fixed effects & Yes \\
Cell fixed effects ($i$ and $j$) & Yes \\
Observations & 154,105,210 \\
Pseudo $R^2$ & 0.2087 \\

\midrule
\multicolumn{2}{l}{\textit{Wald test}} \\
$\chi^2(1)$ & 4.6259 \\
$p$-value & 0.0315 \\
\bottomrule
\end{tabular}

\vspace{0.5em}
\begin{minipage}{0.85\textwidth}
\footnotesize
\textit{Notes:}
The dependent variable indicates the first observed formation of a transaction link in year $t$, conditional on no transaction link in any preceding observed year since 2016. The model includes interactions between $\ln(\mathrm{MP})$ and pre- and post-period indicators, year and cell fixed effects, and baseline pair-level controls whose coefficients are omitted. Standard errors are two-way clustered by cells $i$ and $j$ and reported in parentheses. Pseudo $R^2$ denotes McFadden's pseudo $R^2$. The Wald test evaluates the null hypothesis that the pre- and post-period MP coefficients are equal ($\beta_{\mathrm{pre}}=\beta_{\mathrm{post}}$), using the full two-way-clustered covariance matrix. $^{*}p<0.10$, $^{**}p<0.05$, $^{***}p<0.01$.
\end{minipage}
\end{table}

\subsection{Distance heterogeneity}

To examine distance heterogeneity in the association between meeting probability and link existence, we divide cell pairs into quartiles based on geographic distance between cell centroids. Let $Q_{ij}^{q}$ equal one when pair $(i,j)$ belongs to quartile $q$, with the nearest quartile (Q1) as the reference group. We estimate a fixed-effects logit model with interactions between log meeting probability and the other three quartiles:
\begin{equation}
\mathrm{logit}\!\left[\Pr\!\left(Y_{ij,t}=1\right)\right]
=
\beta_1\ln(\mathrm{MP}_{ij})
+
\sum_{q=2}^{4}\beta_q\left[\ln(\mathrm{MP}_{ij})\times Q_{ij}^{q}\right]
+
\sum_{q=2}^{4}
\theta_q Q_{ij}^{q}
+
\mathbf{X}_{ij}'\boldsymbol{\gamma}
+
\delta_i
+
\delta_j .
\end{equation}

The coefficient $\beta_1$ is the log-MP slope for Q1; the slope for quartile $q>1$ is $\beta_1+\beta_q$. Supplementary Table~\ref{tab:logit_dist_het} reports the interaction-model estimates for link existence. A Wald test rejects the null hypothesis that the log-MP slopes are equal across the four distance quartiles ($\chi^{2}(3)=26.99$, $p<0.001$).

\begin{table}[htbp]
\centering
\caption{Distance heterogeneity in the association between meeting probability and link existence}
\label{tab:logit_dist_het}

\begin{tabular}{lccc}
\toprule
& (1) \textit{t}=1 & (2) \textit{t}=2 & (3) \textit{t}=3 \\
\midrule
$\ln(MP_{ij})$
& 0.0539*** & 0.0536*** & 0.0537*** \\
& (0.0028) & (0.0027) & (0.0027) \\

Distance Q2
& -0.4726*** & -0.4796*** & -0.4671*** \\
& (0.0630) & (0.0630) & (0.0632) \\

Distance Q3
& -0.7769*** & -0.8078*** & -0.8345*** \\
& (0.1237) & (0.1234) & (0.1186) \\

Distance Q4
& -0.7825*** & -0.7040*** & -0.7127*** \\
& (0.1669) & (0.1559) & (0.1516) \\

$\ln(MP_{ij}) \times$ Distance Q2
& -0.0215*** & -0.0218*** & -0.0219*** \\
& (0.0049) & (0.0049) & (0.0050) \\

$\ln(MP_{ij}) \times$ Distance Q3
& -0.0294*** & -0.0305*** & -0.0354*** \\
& (0.0090) & (0.0090) & (0.0087) \\

$\ln(MP_{ij}) \times$ Distance Q4
& 0.0011 & 0.0062 & 0.0039 \\
& (0.0119) & (0.0112) & (0.0109) \\

\midrule
Observations
& 26,409,622 & 26,054,146 & 25,737,824 \\

Cell $i$ and $j$ FE
& Yes & Yes & Yes \\

Pair-level controls
& Yes & Yes & Yes \\

Adj. Pseudo $R^2$
& 0.301 & 0.301 & 0.302 \\

\bottomrule
\end{tabular}

\vspace{0.3em}
\footnotesize
\textit{Notes:} The binary outcome is link existence between firm cells $i$ and $j$ in the indicated year. Distance quartiles are based on geographic distance between; Q1 (the nearest quartile) is the reference group. The coefficient on $\ln(\mathrm{MP}_{ij})$ is the Q1 slope, and each interaction coefficient is the difference from that slope. The fixed-effects logit model includes cell fixed effects and pair-level controls. Standard errors are two-way clustered by firm cells $i$ and $j$ and are reported in parentheses. $^{*}p<0.10$, $^{**}p<0.05$, $^{***}p<0.01$.
\end{table}

We next test whether this non-monotonic pattern depends on the distance-bin definitions or on pairs at the extremes of the distance distribution.

\paragraph{Alternative distance bins.}
Supplementary Table~\ref{tab:supp_distance_bins_2020} reports the estimated log-MP slope within each alternative distance bin. For each binning, the slope is the reference-bin coefficient plus the relevant interaction coefficient. The pattern is similar under quintile and decile definitions: the association is stronger for the closest pairs, weaker for middle-distance pairs, and stronger again for the farthest pairs. This pattern is inconsistent with a simple monotonic distance-decay interpretation.

\begin{table}[!htbp]
\centering
\caption{Robustness to alternative distance-bin definitions}
\label{tab:supp_distance_bins_2020}
\begin{adjustbox}{max width=\textwidth}
\begin{tabular}{llccc}
\toprule
Distance binning & Bin & Median distance (m) & Coefficient & S.E. \\
\midrule
Quintile & 1 & 3,570 & 0.054$^{***}$ & (0.003) \\
Quintile & 2 & 6,970 & 0.033$^{***}$ & (0.004) \\
Quintile & 3 & 10,285 & 0.025$^{***}$ & (0.007) \\
Quintile & 4 & 15,045 & 0.036$^{***}$ & (0.011) \\
Quintile & 5 & 25,415 & 0.058$^{***}$ & (0.012) \\
\midrule
Decile & 1 & 2,380 & 0.061$^{***}$ & (0.003) \\
Decile & 2 & 4,505 & 0.020$^{***}$ & (0.004) \\
Decile & 3 & 6,120 & 0.033$^{***}$ & (0.005) \\
Decile & 4 & 7,735 & 0.026$^{***}$ & (0.006) \\
Decile & 5 & 9,350 & 0.030$^{***}$ & (0.008) \\
Decile & 6 & 11,220 & 0.013 & (0.011) \\
Decile & 7 & 13,515 & 0.032$^{**}$ & (0.013) \\
Decile & 8 & 16,915 & 0.030$^{*}$ & (0.017) \\
Decile & 9 & 22,185 & 0.043$^{***}$ & (0.016) \\
Decile & 10 & 29,495 & 0.088$^{***}$ & (0.019) \\
\bottomrule
\end{tabular}
\end{adjustbox}

\vspace{0.5em}
\begin{minipage}{0.95\textwidth}
\footnotesize
\textit{Notes:} Entries report the within-bin slope of log meeting probability, $\ln(\mathrm{MP}_{ij})$, for link existence. For each quintile or decile definition, a fixed-effects logit model interacts log MP with the distance-bin indicators; each reported slope is the reference-bin coefficient plus the relevant interaction coefficient. All models include the baseline pair-level controls and cell fixed effects. Standard errors are two-way clustered by cells $i$ and $j$ and are reported in parentheses. $^{*}p<0.10$, $^{**}p<0.05$, $^{***}p<0.01$.
\end{minipage}
\end{table}

\paragraph{Excluding extreme distances.}
Supplementary Table~\ref{tab:supp_trim_distance_2020} reports quartile-specific estimates after excluding observations at the tails of the distance distribution. Dropping only the longest-distance observations leaves the observed pattern largely unchanged: after excluding the top 1\% or top 5\% of distances, the first and fourth quartiles remain larger than the middle quartiles. When both tails are removed, the first-quartile estimate becomes smaller, indicating that very close pairs contribute to the large short-distance effect. The farthest remaining quartile remains larger than the middle-distance quartiles, so the non-monotonic pattern does not disappear.

\begin{table}[!htbp]
\centering
\caption{Robustness after excluding extreme distances}
\label{tab:supp_trim_distance_2020}
\begin{adjustbox}{max width=\textwidth}
\begin{tabular}{llccc}
\toprule
Sample restriction & Quartile & Median distance (m) & Coefficient & S.E. \\
\midrule
Full sample & Q1 & 3,995 & 0.054$^{***}$ & (0.003) \\
Full sample & Q2 & 8,160 & 0.032$^{***}$ & (0.004) \\
Full sample & Q3 & 12,920 & 0.024$^{***}$ & (0.009) \\
Full sample & Q4 & 23,715 & 0.055$^{***}$ & (0.012) \\
\midrule
Drop top 1\% & Q1 & 3,995 & 0.054$^{***}$ & (0.003) \\
Drop top 1\% & Q2 & 8,075 & 0.033$^{***}$ & (0.004) \\
Drop top 1\% & Q3 & 12,750 & 0.026$^{***}$ & (0.008) \\
Drop top 1\% & Q4 & 23,205 & 0.052$^{***}$ & (0.011) \\
\midrule
Drop top 5\% & Q1 & 3,910 & 0.054$^{***}$ & (0.003) \\
Drop top 5\% & Q2 & 7,820 & 0.033$^{***}$ & (0.004) \\
Drop top 5\% & Q3 & 12,155 & 0.025$^{***}$ & (0.008) \\
Drop top 5\% & Q4 & 20,995 & 0.051$^{***}$ & (0.011) \\
\midrule
Keep P1--P99 & Q1 & 4,165 & 0.021$^{***}$ & (0.003) \\
Keep P1--P99 & Q2 & 8,160 & 0.026$^{***}$ & (0.004) \\
Keep P1--P99 & Q3 & 12,835 & 0.017$^{**}$ & (0.008) \\
Keep P1--P99 & Q4 & 23,290 & 0.046$^{***}$ & (0.011) \\
\midrule
Keep P5--P95 & Q1 & 4,675 & 0.016$^{***}$ & (0.003) \\
Keep P5--P95 & Q2 & 8,330 & 0.021$^{***}$ & (0.005) \\
Keep P5--P95 & Q3 & 12,580 & 0.014 & (0.009) \\
Keep P5--P95 & Q4 & 21,335 & 0.042$^{***}$ & (0.012) \\
\bottomrule
\end{tabular}
\end{adjustbox}

\vspace{0.5em}
\begin{minipage}{0.95\textwidth}
\footnotesize
\textit{Notes:} Entries report the within-quartile slope of log meeting probability, $\ln(\mathrm{MP}_{ij})$, for link existence under each sample restriction. In each restricted sample, the fixed-effects logit model interacts log MP with distance-quartile indicators; each reported slope is the reference-quartile coefficient plus the relevant interaction coefficient. All models include the baseline pair-level controls and -cell fixed effects. Standard errors are two-way clustered by cells $i$ and $j$ and are reported in parentheses. $^{*}p<0.10$, $^{**}p<0.05$, $^{***}p<0.01$.
\end{minipage}
\end{table}

\subsection{Industry heterogeneity}

The full regression results are reported in Supplementary Tables~\ref{tab:industry-matrix-main}. 

\begin{industryregressiontable}
\caption{Industry-pair regression results displayed in the industry matrix (firm-count specification).}\label{tab:industry-matrix-main} \\
\toprule
Pair & $\hat\beta$ & SE & 95\% CI & $p$ & BH $q$ & $N_{\mathrm{pair}}$ \\
\midrule
\endfirsthead
\multicolumn{7}{l}{\textit{Continued from previous page}} \\
\toprule
Pair & $\hat\beta$ & SE & 95\% CI & $p$ & BH $q$ & $N_{\mathrm{pair}}$ \\
\midrule
\endhead
\midrule
\multicolumn{7}{r}{\textit{Continued on next page}} \\
\endfoot
\bottomrule
\endlastfoot
\shortstack[l]{Agric. $\times$\\Constr.} & $\mathbf{-0.0011}$ & $0.0002$ & \shortstack[r]{$[-0.0015,$\\$-0.0007]$} & $<0.001$ & $<0.001$ & 60,648 \\
\shortstack[l]{Agric. $\times$\\Manuf.} & $\mathbf{-0.0006}$ & $0.0002$ & \shortstack[r]{$[-0.0010,$\\$-0.0002]$} & $0.006$ & $0.010$ & 31,752 \\
\shortstack[l]{Agric. $\times$\\Whsl.} & $0.0003$ & $0.0002$ & \shortstack[r]{$[-0.0001,$\\$0.0007]$} & $0.107$ & $0.124$ & 52,479 \\
\shortstack[l]{Agric. $\times$\\Retail} & $\mathbf{-0.0011}$ & $0.0002$ & \shortstack[r]{$[-0.0015,$\\$-0.0007]$} & $<0.001$ & $<0.001$ & 14,028 \\
\shortstack[l]{Agric. $\times$\\Transp.} & $\mathbf{-0.0006}$ & $0.0003$ & \shortstack[r]{$[-0.0012,$\\$-0.0001]$} & $0.014$ & $0.021$ & 4,515 \\
\shortstack[l]{Agric. $\times$\\Info.} & $\mathbf{0.0008}$ & $0.0002$ & \shortstack[r]{$[0.0003,$\\$0.0012]$} & $<0.001$ & $0.001$ & 11,676 \\
\shortstack[l]{Agric. $\times$\\Finance} & $\mathbf{0.0024}$ & $0.0006$ & \shortstack[r]{$[0.0013,$\\$0.0035]$} & $<0.001$ & $<0.001$ & 1,575 \\
\shortstack[l]{Agric. $\times$\\Real est.} & $\mathbf{-0.0006}$ & $0.0002$ & \shortstack[r]{$[-0.0010,$\\$-0.0002]$} & $0.002$ & $0.004$ & 23,205 \\
\shortstack[l]{Agric. $\times$\\Prof./tech.} & $\mathbf{0.0005}$ & $0.0002$ & \shortstack[r]{$[0.0001,$\\$0.0010]$} & $0.019$ & $0.027$ & 16,653 \\
\shortstack[l]{Agric. $\times$\\Admin. supp.} & $0.0001$ & $0.0002$ & \shortstack[r]{$[-0.0004,$\\$0.0006]$} & $0.607$ & $0.627$ & 5,985 \\
\shortstack[l]{Agric. $\times$\\Educ.} & $\mathbf{-0.0014}$ & $0.0002$ & \shortstack[r]{$[-0.0018,$\\$-0.0010]$} & $<0.001$ & $<0.001$ & 1,281 \\
\shortstack[l]{Agric. $\times$\\Health} & $\mathbf{-0.0010}$ & $0.0002$ & \shortstack[r]{$[-0.0015,$\\$-0.0006]$} & $<0.001$ & $<0.001$ & 8,379 \\
\shortstack[l]{Agric. $\times$\\Accom./food} & $\mathbf{-0.0011}$ & $0.0003$ & \shortstack[r]{$[-0.0016,$\\$-0.0006]$} & $<0.001$ & $<0.001$ & 1,638 \\
\shortstack[l]{Agric. $\times$\\Other serv.} & $\mathbf{-0.0006}$ & $0.0002$ & \shortstack[r]{$[-0.0010,$\\$-0.0001]$} & $0.013$ & $0.019$ & 6,342 \\
\shortstack[l]{Util. $\times$\\Constr.} & $0.0016$ & $0.0010$ & \shortstack[r]{$[-0.0003,$\\$0.0035]$} & $0.097$ & $0.116$ & 54,872 \\
\shortstack[l]{Util. $\times$\\Manuf.} & $\mathbf{0.0020}$ & $0.0009$ & \shortstack[r]{$[0.0002,$\\$0.0038]$} & $0.033$ & $0.043$ & 28,728 \\
\shortstack[l]{Util. $\times$\\Whsl.} & $\mathbf{0.0026}$ & $0.0008$ & \shortstack[r]{$[0.0009,$\\$0.0042]$} & $0.002$ & $0.003$ & 47,481 \\
\shortstack[l]{Util. $\times$\\Retail} & $0.0016$ & $0.0010$ & \shortstack[r]{$[-0.0003,$\\$0.0035]$} & $0.104$ & $0.123$ & 12,692 \\
\shortstack[l]{Util. $\times$\\Transp.} & $0.0020$ & $0.0010$ & \shortstack[r]{$[0.0000,$\\$0.0039]$} & $0.046$ & $0.058$ & 4,085 \\
\shortstack[l]{Util. $\times$\\Info.} & $\mathbf{0.0031}$ & $0.0008$ & \shortstack[r]{$[0.0015,$\\$0.0047]$} & $<0.001$ & $<0.001$ & 10,564 \\
\shortstack[l]{Util. $\times$\\Finance} & $\mathbf{0.0043}$ & $0.0007$ & \shortstack[r]{$[0.0029,$\\$0.0056]$} & $<0.001$ & $<0.001$ & 1,425 \\
\shortstack[l]{Util. $\times$\\Real est.} & $\mathbf{0.0020}$ & $0.0009$ & \shortstack[r]{$[0.0002,$\\$0.0038]$} & $0.033$ & $0.043$ & 20,995 \\
\shortstack[l]{Util. $\times$\\Prof./tech.} & $\mathbf{0.0028}$ & $0.0008$ & \shortstack[r]{$[0.0012,$\\$0.0044]$} & $<0.001$ & $0.001$ & 15,067 \\
\shortstack[l]{Util. $\times$\\Admin. supp.} & $\mathbf{0.0026}$ & $0.0009$ & \shortstack[r]{$[0.0009,$\\$0.0044]$} & $0.003$ & $0.005$ & 5,415 \\
\shortstack[l]{Util. $\times$\\Educ.} & $0.0014$ & $0.0010$ & \shortstack[r]{$[-0.0006,$\\$0.0033]$} & $0.161$ & $0.179$ & 1,159 \\
\shortstack[l]{Util. $\times$\\Health} & $0.0017$ & $0.0010$ & \shortstack[r]{$[-0.0002,$\\$0.0036]$} & $0.088$ & $0.108$ & 7,581 \\
\shortstack[l]{Util. $\times$\\Accom./food} & $0.0018$ & $0.0010$ & \shortstack[r]{$[-0.0002,$\\$0.0039]$} & $0.081$ & $0.100$ & 1,482 \\
\shortstack[l]{Util. $\times$\\Other serv.} & $\mathbf{0.0021}$ & $0.0009$ & \shortstack[r]{$[0.0002,$\\$0.0039]$} & $0.027$ & $0.036$ & 5,738 \\
\shortstack[l]{Constr. $\times$\\Constr.} & $\mathbf{-0.0012}$ & $0.0001$ & \shortstack[r]{$[-0.0013,$\\$-0.0010]$} & $<0.001$ & $<0.001$ & 4,168,828 \\
\shortstack[l]{Constr. $\times$\\Manuf.} & $\mathbf{-0.0007}$ & $0.0001$ & \shortstack[r]{$[-0.0008,$\\$-0.0005]$} & $<0.001$ & $<0.001$ & 4,366,656 \\
\shortstack[l]{Constr. $\times$\\Whsl.} & $\mathbf{0.0002}$ & $0.0001$ & \shortstack[r]{$[0.0001,$\\$0.0003]$} & $<0.001$ & $<0.001$ & 7,217,112 \\
\shortstack[l]{Constr. $\times$\\Retail} & $\mathbf{-0.0012}$ & $0.0001$ & \shortstack[r]{$[-0.0014,$\\$-0.0011]$} & $<0.001$ & $<0.001$ & 1,929,184 \\
\shortstack[l]{Constr. $\times$\\Transp.} & $\mathbf{-0.0007}$ & $0.0002$ & \shortstack[r]{$[-0.0011,$\\$-0.0004]$} & $<0.001$ & $<0.001$ & 620,920 \\
\shortstack[l]{Constr. $\times$\\Info.} & $\mathbf{0.0007}$ & $0.0001$ & \shortstack[r]{$[0.0004,$\\$0.0009]$} & $<0.001$ & $<0.001$ & 1,605,728 \\
\shortstack[l]{Constr. $\times$\\Finance} & $\mathbf{0.0024}$ & $0.0006$ & \shortstack[r]{$[0.0013,$\\$0.0034]$} & $<0.001$ & $<0.001$ & 216,600 \\
\shortstack[l]{Constr. $\times$\\Real est.} & $\mathbf{-0.0007}$ & $0.0001$ & \shortstack[r]{$[-0.0009,$\\$-0.0006]$} & $<0.001$ & $<0.001$ & 3,191,240 \\
\shortstack[l]{Constr. $\times$\\Prof./tech.} & $\mathbf{0.0004}$ & $0.0001$ & \shortstack[r]{$[0.0002,$\\$0.0006]$} & $<0.001$ & $<0.001$ & 2,290,184 \\
\shortstack[l]{Constr. $\times$\\Admin. supp.} & $0.0000$ & $0.0002$ & \shortstack[r]{$[-0.0003,$\\$0.0003]$} & $0.863$ & $0.868$ & 823,080 \\
\shortstack[l]{Constr. $\times$\\Educ.} & $\mathbf{-0.0015}$ & $0.0001$ & \shortstack[r]{$[-0.0017,$\\$-0.0014]$} & $<0.001$ & $<0.001$ & 176,168 \\
\shortstack[l]{Constr. $\times$\\Health} & $\mathbf{-0.0012}$ & $0.0001$ & \shortstack[r]{$[-0.0014,$\\$-0.0009]$} & $<0.001$ & $<0.001$ & 1,152,312 \\
\shortstack[l]{Constr. $\times$\\Arts/rec.} & $\mathbf{-0.0011}$ & $0.0002$ & \shortstack[r]{$[-0.0016,$\\$-0.0006]$} & $<0.001$ & $<0.001$ & 66,424 \\
\shortstack[l]{Constr. $\times$\\Accom./food} & $\mathbf{-0.0011}$ & $0.0002$ & \shortstack[r]{$[-0.0015,$\\$-0.0007]$} & $<0.001$ & $<0.001$ & 225,264 \\
\shortstack[l]{Constr. $\times$\\Other serv.} & $\mathbf{-0.0007}$ & $0.0001$ & \shortstack[r]{$[-0.0009,$\\$-0.0004]$} & $<0.001$ & $<0.001$ & 872,176 \\
\shortstack[l]{Manuf. $\times$\\Manuf.} & $-0.0002$ & $0.0001$ & \shortstack[r]{$[-0.0004,$\\$0.0000]$} & $0.074$ & $0.092$ & 1,142,316 \\
\shortstack[l]{Manuf. $\times$\\Whsl.} & $\mathbf{0.0006}$ & $0.0001$ & \shortstack[r]{$[0.0005,$\\$0.0008]$} & $<0.001$ & $<0.001$ & 3,778,488 \\
\shortstack[l]{Manuf. $\times$\\Retail} & $\mathbf{-0.0007}$ & $0.0001$ & \shortstack[r]{$[-0.0009,$\\$-0.0006]$} & $<0.001$ & $<0.001$ & 1,010,016 \\
\shortstack[l]{Manuf. $\times$\\Transp.} & $-0.0002$ & $0.0002$ & \shortstack[r]{$[-0.0006,$\\$0.0001]$} & $0.174$ & $0.192$ & 325,080 \\
\shortstack[l]{Manuf. $\times$\\Info.} & $\mathbf{0.0011}$ & $0.0001$ & \shortstack[r]{$[0.0009,$\\$0.0014]$} & $<0.001$ & $<0.001$ & 840,672 \\
\shortstack[l]{Manuf. $\times$\\Finance} & $\mathbf{0.0028}$ & $0.0005$ & \shortstack[r]{$[0.0017,$\\$0.0038]$} & $<0.001$ & $<0.001$ & 113,400 \\
\shortstack[l]{Manuf. $\times$\\Real est.} & $\mathbf{-0.0002}$ & $0.0001$ & \shortstack[r]{$[-0.0004,$\\$-0.0001]$} & $0.002$ & $0.004$ & 1,670,760 \\
\shortstack[l]{Manuf. $\times$\\Prof./tech.} & $\mathbf{0.0009}$ & $0.0001$ & \shortstack[r]{$[0.0007,$\\$0.0011]$} & $<0.001$ & $<0.001$ & 1,199,016 \\
\shortstack[l]{Manuf. $\times$\\Admin. supp.} & $\mathbf{0.0005}$ & $0.0002$ & \shortstack[r]{$[0.0002,$\\$0.0008]$} & $<0.001$ & $0.002$ & 430,920 \\
\shortstack[l]{Manuf. $\times$\\Educ.} & $\mathbf{-0.0010}$ & $0.0001$ & \shortstack[r]{$[-0.0012,$\\$-0.0008]$} & $<0.001$ & $<0.001$ & 92,232 \\
\shortstack[l]{Manuf. $\times$\\Health} & $\mathbf{-0.0007}$ & $0.0001$ & \shortstack[r]{$[-0.0009,$\\$-0.0004]$} & $<0.001$ & $<0.001$ & 603,288 \\
\shortstack[l]{Manuf. $\times$\\Arts/rec.} & $\mathbf{-0.0006}$ & $0.0002$ & \shortstack[r]{$[-0.0010,$\\$-0.0001]$} & $0.021$ & $0.029$ & 34,776 \\
\shortstack[l]{Manuf. $\times$\\Accom./food} & $\mathbf{-0.0006}$ & $0.0002$ & \shortstack[r]{$[-0.0010,$\\$-0.0002]$} & $0.004$ & $0.006$ & 117,936 \\
\shortstack[l]{Manuf. $\times$\\Other serv.} & $-0.0002$ & $0.0001$ & \shortstack[r]{$[-0.0004,$\\$0.0001]$} & $0.232$ & $0.249$ & 456,624 \\
\shortstack[l]{Whsl. $\times$\\Whsl.} & $\mathbf{0.0014}$ & $0.0001$ & \shortstack[r]{$[0.0012,$\\$0.0016]$} & $<0.001$ & $<0.001$ & 3,121,251 \\
\shortstack[l]{Whsl. $\times$\\Retail} & $\mathbf{0.0002}$ & $0.0001$ & \shortstack[r]{$[0.0000,$\\$0.0003]$} & $0.009$ & $0.014$ & 1,669,332 \\
\shortstack[l]{Whsl. $\times$\\Transp.} & $\mathbf{0.0006}$ & $0.0002$ & \shortstack[r]{$[0.0003,$\\$0.0009]$} & $<0.001$ & $<0.001$ & 537,285 \\
\shortstack[l]{Whsl. $\times$\\Info.} & $\mathbf{0.0019}$ & $0.0001$ & \shortstack[r]{$[0.0016,$\\$0.0021]$} & $<0.001$ & $<0.001$ & 1,389,444 \\
\shortstack[l]{Whsl. $\times$\\Finance} & $\mathbf{0.0032}$ & $0.0005$ & \shortstack[r]{$[0.0023,$\\$0.0041]$} & $<0.001$ & $<0.001$ & 187,425 \\
\shortstack[l]{Whsl. $\times$\\Real est.} & $\mathbf{0.0006}$ & $0.0001$ & \shortstack[r]{$[0.0005,$\\$0.0008]$} & $<0.001$ & $<0.001$ & 2,761,395 \\
\shortstack[l]{Whsl. $\times$\\Prof./tech.} & $\mathbf{0.0017}$ & $0.0001$ & \shortstack[r]{$[0.0015,$\\$0.0019]$} & $<0.001$ & $<0.001$ & 1,981,707 \\
\shortstack[l]{Whsl. $\times$\\Admin. supp.} & $\mathbf{0.0013}$ & $0.0001$ & \shortstack[r]{$[0.0011,$\\$0.0016]$} & $<0.001$ & $<0.001$ & 712,215 \\
\shortstack[l]{Whsl. $\times$\\Educ.} & $-0.0001$ & $0.0001$ & \shortstack[r]{$[-0.0003,$\\$0.0000]$} & $0.043$ & $0.056$ & 152,439 \\
\shortstack[l]{Whsl. $\times$\\Health} & $0.0002$ & $0.0001$ & \shortstack[r]{$[0.0000,$\\$0.0005]$} & $0.068$ & $0.086$ & 997,101 \\
\shortstack[l]{Whsl. $\times$\\Arts/rec.} & $0.0003$ & $0.0002$ & \shortstack[r]{$[-0.0001,$\\$0.0008]$} & $0.130$ & $0.147$ & 57,477 \\
\shortstack[l]{Whsl. $\times$\\Accom./food} & $0.0003$ & $0.0002$ & \shortstack[r]{$[0.0000,$\\$0.0006]$} & $0.095$ & $0.115$ & 194,922 \\
\shortstack[l]{Whsl. $\times$\\Other serv.} & $\mathbf{0.0007}$ & $0.0001$ & \shortstack[r]{$[0.0005,$\\$0.0009]$} & $<0.001$ & $<0.001$ & 754,698 \\
\shortstack[l]{Retail $\times$\\Retail} & $\mathbf{-0.0013}$ & $0.0001$ & \shortstack[r]{$[-0.0015,$\\$-0.0011]$} & $<0.001$ & $<0.001$ & 222,778 \\
\shortstack[l]{Retail $\times$\\Transp.} & $\mathbf{-0.0008}$ & $0.0002$ & \shortstack[r]{$[-0.0012,$\\$-0.0004]$} & $<0.001$ & $<0.001$ & 143,620 \\
\shortstack[l]{Retail $\times$\\Info.} & $\mathbf{0.0006}$ & $0.0001$ & \shortstack[r]{$[0.0004,$\\$0.0009]$} & $<0.001$ & $<0.001$ & 371,408 \\
\shortstack[l]{Retail $\times$\\Finance} & $\mathbf{0.0024}$ & $0.0006$ & \shortstack[r]{$[0.0013,$\\$0.0035]$} & $<0.001$ & $<0.001$ & 50,100 \\
\shortstack[l]{Retail $\times$\\Real est.} & $\mathbf{-0.0008}$ & $0.0001$ & \shortstack[r]{$[-0.0009,$\\$-0.0006]$} & $<0.001$ & $<0.001$ & 738,140 \\
\shortstack[l]{Retail $\times$\\Prof./tech.} & $\mathbf{0.0004}$ & $0.0001$ & \shortstack[r]{$[0.0002,$\\$0.0006]$} & $<0.001$ & $<0.001$ & 529,724 \\
\shortstack[l]{Retail $\times$\\Admin. supp.} & $0.0000$ & $0.0002$ & \shortstack[r]{$[-0.0003,$\\$0.0003]$} & $0.940$ & $0.940$ & 190,380 \\
\shortstack[l]{Retail $\times$\\Educ.} & $\mathbf{-0.0016}$ & $0.0001$ & \shortstack[r]{$[-0.0018,$\\$-0.0014]$} & $<0.001$ & $<0.001$ & 40,748 \\
\shortstack[l]{Retail $\times$\\Health} & $\mathbf{-0.0012}$ & $0.0001$ & \shortstack[r]{$[-0.0015,$\\$-0.0009]$} & $<0.001$ & $<0.001$ & 266,532 \\
\shortstack[l]{Retail $\times$\\Arts/rec.} & $\mathbf{-0.0011}$ & $0.0002$ & \shortstack[r]{$[-0.0016,$\\$-0.0006]$} & $<0.001$ & $<0.001$ & 15,364 \\
\shortstack[l]{Retail $\times$\\Accom./food} & $\mathbf{-0.0012}$ & $0.0002$ & \shortstack[r]{$[-0.0016,$\\$-0.0007]$} & $<0.001$ & $<0.001$ & 52,104 \\
\shortstack[l]{Retail $\times$\\Other serv.} & $\mathbf{-0.0007}$ & $0.0001$ & \shortstack[r]{$[-0.0010,$\\$-0.0004]$} & $<0.001$ & $<0.001$ & 201,736 \\
\shortstack[l]{Transp. $\times$\\Transp.} & $-0.0005$ & $0.0003$ & \shortstack[r]{$[-0.0011,$\\$0.0002]$} & $0.159$ & $0.178$ & 23,005 \\
\shortstack[l]{Transp. $\times$\\Info.} & $\mathbf{0.0011}$ & $0.0002$ & \shortstack[r]{$[0.0007,$\\$0.0015]$} & $<0.001$ & $<0.001$ & 119,540 \\
\shortstack[l]{Transp. $\times$\\Finance} & $\mathbf{0.0027}$ & $0.0006$ & \shortstack[r]{$[0.0016,$\\$0.0038]$} & $<0.001$ & $<0.001$ & 16,125 \\
\shortstack[l]{Transp. $\times$\\Real est.} & $-0.0003$ & $0.0002$ & \shortstack[r]{$[-0.0006,$\\$0.0001]$} & $0.113$ & $0.130$ & 237,575 \\
\shortstack[l]{Transp. $\times$\\Prof./tech.} & $\mathbf{0.0008}$ & $0.0002$ & \shortstack[r]{$[0.0005,$\\$0.0012]$} & $<0.001$ & $<0.001$ & 170,495 \\
\shortstack[l]{Transp. $\times$\\Admin. supp.} & $\mathbf{0.0005}$ & $0.0002$ & \shortstack[r]{$[0.0000,$\\$0.0009]$} & $0.038$ & $0.049$ & 61,275 \\
\shortstack[l]{Transp. $\times$\\Educ.} & $\mathbf{-0.0011}$ & $0.0002$ & \shortstack[r]{$[-0.0015,$\\$-0.0007]$} & $<0.001$ & $<0.001$ & 13,115 \\
\shortstack[l]{Transp. $\times$\\Health} & $\mathbf{-0.0007}$ & $0.0002$ & \shortstack[r]{$[-0.0011,$\\$-0.0003]$} & $<0.001$ & $0.001$ & 85,785 \\
\shortstack[l]{Transp. $\times$\\Arts/rec.} & $\mathbf{-0.0006}$ & $0.0003$ & \shortstack[r]{$[-0.0012,$\\$-0.0001]$} & $0.026$ & $0.036$ & 4,945 \\
\shortstack[l]{Transp. $\times$\\Accom./food} & $\mathbf{-0.0007}$ & $0.0003$ & \shortstack[r]{$[-0.0012,$\\$-0.0001]$} & $0.014$ & $0.021$ & 16,770 \\
\shortstack[l]{Transp. $\times$\\Other serv.} & $-0.0002$ & $0.0002$ & \shortstack[r]{$[-0.0007,$\\$0.0002]$} & $0.289$ & $0.305$ & 64,930 \\
\shortstack[l]{Info. $\times$\\Info.} & $\mathbf{0.0019}$ & $0.0002$ & \shortstack[r]{$[0.0016,$\\$0.0023]$} & $<0.001$ & $<0.001$ & 154,290 \\
\shortstack[l]{Info. $\times$\\Finance} & $\mathbf{0.0033}$ & $0.0004$ & \shortstack[r]{$[0.0025,$\\$0.0041]$} & $<0.001$ & $<0.001$ & 41,700 \\
\shortstack[l]{Info. $\times$\\Real est.} & $\mathbf{0.0011}$ & $0.0001$ & \shortstack[r]{$[0.0009,$\\$0.0014]$} & $<0.001$ & $<0.001$ & 614,380 \\
\shortstack[l]{Info. $\times$\\Prof./tech.} & $\mathbf{0.0020}$ & $0.0001$ & \shortstack[r]{$[0.0017,$\\$0.0022]$} & $<0.001$ & $<0.001$ & 440,908 \\
\shortstack[l]{Info. $\times$\\Admin. supp.} & $\mathbf{0.0016}$ & $0.0002$ & \shortstack[r]{$[0.0013,$\\$0.0019]$} & $<0.001$ & $<0.001$ & 158,460 \\
\shortstack[l]{Info. $\times$\\Educ.} & $\mathbf{0.0004}$ & $0.0001$ & \shortstack[r]{$[0.0001,$\\$0.0006]$} & $0.003$ & $0.005$ & 33,916 \\
\shortstack[l]{Info. $\times$\\Health} & $\mathbf{0.0007}$ & $0.0002$ & \shortstack[r]{$[0.0004,$\\$0.0010]$} & $<0.001$ & $<0.001$ & 221,844 \\
\shortstack[l]{Info. $\times$\\Arts/rec.} & $\mathbf{0.0007}$ & $0.0002$ & \shortstack[r]{$[0.0003,$\\$0.0011]$} & $<0.001$ & $<0.001$ & 12,788 \\
\shortstack[l]{Info. $\times$\\Accom./food} & $\mathbf{0.0008}$ & $0.0002$ & \shortstack[r]{$[0.0004,$\\$0.0012]$} & $<0.001$ & $<0.001$ & 43,368 \\
\shortstack[l]{Info. $\times$\\Other serv.} & $\mathbf{0.0012}$ & $0.0002$ & \shortstack[r]{$[0.0009,$\\$0.0015]$} & $<0.001$ & $<0.001$ & 167,912 \\
\shortstack[l]{Finance $\times$\\Finance} & $\mathbf{0.0039}$ & $0.0008$ & \shortstack[r]{$[0.0023,$\\$0.0056]$} & $<0.001$ & $<0.001$ & 2,775 \\
\shortstack[l]{Finance $\times$\\Real est.} & $\mathbf{0.0027}$ & $0.0005$ & \shortstack[r]{$[0.0017,$\\$0.0038]$} & $<0.001$ & $<0.001$ & 82,875 \\
\shortstack[l]{Finance $\times$\\Prof./tech.} & $\mathbf{0.0033}$ & $0.0005$ & \shortstack[r]{$[0.0024,$\\$0.0042]$} & $<0.001$ & $<0.001$ & 59,475 \\
\shortstack[l]{Finance $\times$\\Admin. supp.} & $\mathbf{0.0032}$ & $0.0005$ & \shortstack[r]{$[0.0022,$\\$0.0042]$} & $<0.001$ & $<0.001$ & 21,375 \\
\shortstack[l]{Finance $\times$\\Educ.} & $\mathbf{0.0022}$ & $0.0006$ & \shortstack[r]{$[0.0011,$\\$0.0033]$} & $<0.001$ & $<0.001$ & 4,575 \\
\shortstack[l]{Finance $\times$\\Health} & $\mathbf{0.0024}$ & $0.0006$ & \shortstack[r]{$[0.0013,$\\$0.0035]$} & $<0.001$ & $<0.001$ & 29,925 \\
\shortstack[l]{Finance $\times$\\Arts/rec.} & $\mathbf{0.0025}$ & $0.0006$ & \shortstack[r]{$[0.0014,$\\$0.0037]$} & $<0.001$ & $<0.001$ & 1,725 \\
\shortstack[l]{Finance $\times$\\Accom./food} & $\mathbf{0.0025}$ & $0.0006$ & \shortstack[r]{$[0.0013,$\\$0.0037]$} & $<0.001$ & $<0.001$ & 5,850 \\
\shortstack[l]{Finance $\times$\\Other serv.} & $\mathbf{0.0029}$ & $0.0006$ & \shortstack[r]{$[0.0018,$\\$0.0040]$} & $<0.001$ & $<0.001$ & 22,650 \\
\shortstack[l]{Real est. $\times$\\Real est.} & $\mathbf{-0.0003}$ & $0.0001$ & \shortstack[r]{$[-0.0005,$\\$0.0000]$} & $0.017$ & $0.024$ & 609,960 \\
\shortstack[l]{Real est. $\times$\\Prof./tech.} & $\mathbf{0.0009}$ & $0.0001$ & \shortstack[r]{$[0.0006,$\\$0.0011]$} & $<0.001$ & $<0.001$ & 876,265 \\
\shortstack[l]{Real est. $\times$\\Admin. supp.} & $\mathbf{0.0005}$ & $0.0002$ & \shortstack[r]{$[0.0002,$\\$0.0008]$} & $0.002$ & $0.003$ & 314,925 \\
\shortstack[l]{Real est. $\times$\\Educ.} & $\mathbf{-0.0010}$ & $0.0001$ & \shortstack[r]{$[-0.0012,$\\$-0.0009]$} & $<0.001$ & $<0.001$ & 67,405 \\
\shortstack[l]{Real est. $\times$\\Health} & $\mathbf{-0.0007}$ & $0.0001$ & \shortstack[r]{$[-0.0009,$\\$-0.0004]$} & $<0.001$ & $<0.001$ & 440,895 \\
\shortstack[l]{Real est. $\times$\\Arts/rec.} & $\mathbf{-0.0006}$ & $0.0002$ & \shortstack[r]{$[-0.0010,$\\$-0.0001]$} & $0.014$ & $0.021$ & 25,415 \\
\shortstack[l]{Real est. $\times$\\Accom./food} & $\mathbf{-0.0006}$ & $0.0002$ & \shortstack[r]{$[-0.0010,$\\$-0.0002]$} & $0.005$ & $0.007$ & 86,190 \\
\shortstack[l]{Real est. $\times$\\Other serv.} & $-0.0002$ & $0.0001$ & \shortstack[r]{$[-0.0004,$\\$0.0001]$} & $0.200$ & $0.218$ & 333,710 \\
\shortstack[l]{Prof./tech. $\times$\\Prof./tech.} & $\mathbf{0.0018}$ & $0.0002$ & \shortstack[r]{$[0.0014,$\\$0.0021]$} & $<0.001$ & $<0.001$ & 314,028 \\
\shortstack[l]{Prof./tech. $\times$\\Admin. supp.} & $\mathbf{0.0015}$ & $0.0002$ & \shortstack[r]{$[0.0012,$\\$0.0018]$} & $<0.001$ & $<0.001$ & 226,005 \\
\shortstack[l]{Prof./tech. $\times$\\Educ.} & $0.0001$ & $0.0001$ & \shortstack[r]{$[-0.0001,$\\$0.0003]$} & $0.352$ & $0.368$ & 48,373 \\
\shortstack[l]{Prof./tech. $\times$\\Health} & $\mathbf{0.0004}$ & $0.0001$ & \shortstack[r]{$[0.0001,$\\$0.0007]$} & $0.003$ & $0.005$ & 316,407 \\
\shortstack[l]{Prof./tech. $\times$\\Arts/rec.} & $\mathbf{0.0005}$ & $0.0002$ & \shortstack[r]{$[0.0001,$\\$0.0009]$} & $0.023$ & $0.031$ & 18,239 \\
\shortstack[l]{Prof./tech. $\times$\\Accom./food} & $\mathbf{0.0005}$ & $0.0002$ & \shortstack[r]{$[0.0001,$\\$0.0009]$} & $0.007$ & $0.011$ & 61,854 \\
\shortstack[l]{Prof./tech. $\times$\\Other serv.} & $\mathbf{0.0009}$ & $0.0002$ & \shortstack[r]{$[0.0006,$\\$0.0012]$} & $<0.001$ & $<0.001$ & 239,486 \\
\shortstack[l]{Admin. supp. $\times$\\Admin. supp.} & $\mathbf{0.0011}$ & $0.0003$ & \shortstack[r]{$[0.0006,$\\$0.0016]$} & $<0.001$ & $<0.001$ & 40,470 \\
\shortstack[l]{Admin. supp. $\times$\\Educ.} & $-0.0003$ & $0.0002$ & \shortstack[r]{$[-0.0006,$\\$0.0000]$} & $0.073$ & $0.091$ & 17,385 \\
\shortstack[l]{Admin. supp. $\times$\\Health} & $0.0001$ & $0.0002$ & \shortstack[r]{$[-0.0003,$\\$0.0004]$} & $0.739$ & $0.748$ & 113,715 \\
\shortstack[l]{Admin. supp. $\times$\\Arts/rec.} & $0.0001$ & $0.0002$ & \shortstack[r]{$[-0.0004,$\\$0.0006]$} & $0.689$ & $0.702$ & 6,555 \\
\shortstack[l]{Admin. supp. $\times$\\Accom./food} & $0.0001$ & $0.0002$ & \shortstack[r]{$[-0.0003,$\\$0.0006]$} & $0.569$ & $0.591$ & 22,230 \\
\shortstack[l]{Admin. supp. $\times$\\Other serv.} & $\mathbf{0.0006}$ & $0.0002$ & \shortstack[r]{$[0.0002,$\\$0.0009]$} & $0.003$ & $0.005$ & 86,070 \\
\shortstack[l]{Educ. $\times$\\Educ.} & $\mathbf{-0.0019}$ & $0.0001$ & \shortstack[r]{$[-0.0021,$\\$-0.0016]$} & $<0.001$ & $<0.001$ & 1,830 \\
\shortstack[l]{Educ. $\times$\\Health} & $\mathbf{-0.0015}$ & $0.0001$ & \shortstack[r]{$[-0.0018,$\\$-0.0012]$} & $<0.001$ & $<0.001$ & 24,339 \\
\shortstack[l]{Educ. $\times$\\Arts/rec.} & $\mathbf{-0.0014}$ & $0.0003$ & \shortstack[r]{$[-0.0019,$\\$-0.0009]$} & $<0.001$ & $<0.001$ & 1,403 \\
\shortstack[l]{Educ. $\times$\\Accom./food} & $\mathbf{-0.0014}$ & $0.0002$ & \shortstack[r]{$[-0.0019,$\\$-0.0009]$} & $<0.001$ & $<0.001$ & 4,758 \\
\shortstack[l]{Educ. $\times$\\Other serv.} & $\mathbf{-0.0010}$ & $0.0002$ & \shortstack[r]{$[-0.0013,$\\$-0.0007]$} & $<0.001$ & $<0.001$ & 18,422 \\
\shortstack[l]{Health $\times$\\Health} & $\mathbf{-0.0011}$ & $0.0003$ & \shortstack[r]{$[-0.0016,$\\$-0.0006]$} & $<0.001$ & $<0.001$ & 79,401 \\
\shortstack[l]{Health $\times$\\Arts/rec.} & $\mathbf{-0.0010}$ & $0.0003$ & \shortstack[r]{$[-0.0016,$\\$-0.0005]$} & $<0.001$ & $<0.001$ & 9,177 \\
\shortstack[l]{Health $\times$\\Accom./food} & $\mathbf{-0.0011}$ & $0.0002$ & \shortstack[r]{$[-0.0016,$\\$-0.0006]$} & $<0.001$ & $<0.001$ & 31,122 \\
\shortstack[l]{Health $\times$\\Other serv.} & $\mathbf{-0.0006}$ & $0.0002$ & \shortstack[r]{$[-0.0010,$\\$-0.0003]$} & $<0.001$ & $<0.001$ & 120,498 \\
\shortstack[l]{Arts/rec. $\times$\\Accom./food} & $\mathbf{-0.0009}$ & $0.0003$ & \shortstack[r]{$[-0.0015,$\\$-0.0002]$} & $0.009$ & $0.013$ & 1,794 \\
\shortstack[l]{Arts/rec. $\times$\\Other serv.} & $-0.0005$ & $0.0003$ & \shortstack[r]{$[-0.0010,$\\$0.0001]$} & $0.103$ & $0.123$ & 6,946 \\
\shortstack[l]{Accom./food $\times$\\Accom./food} & $\mathbf{-0.0009}$ & $0.0004$ & \shortstack[r]{$[-0.0017,$\\$-0.0001]$} & $0.032$ & $0.043$ & 3,003 \\
\shortstack[l]{Accom./food $\times$\\Other serv.} & $\mathbf{-0.0005}$ & $0.0002$ & \shortstack[r]{$[-0.0010,$\\$-0.0001]$} & $0.028$ & $0.037$ & 23,556 \\
\shortstack[l]{Other serv. $\times$\\Other serv.} & $-0.0001$ & $0.0003$ & \shortstack[r]{$[-0.0006,$\\$0.0004]$} & $0.686$ & $0.702$ & 45,451 \\
\end{industryregressiontable}

To assess whether the industry heterogeneity results depend on how the dominant industry of each cell is defined, we construct two alternative cell-level industry assignments. In the firm-count assignment, we count the number of firms in each two-digit NAICS sector within a cell and define the sector with the largest number of firms as the dominant industry. In the employee-weighted assignment, we instead sum the number of employees across firms within each two-digit NAICS sector, and define the sector with the largest total employment as the dominant industry. 

\begin{figure}
    \centering
    \includegraphics[width=1\linewidth]{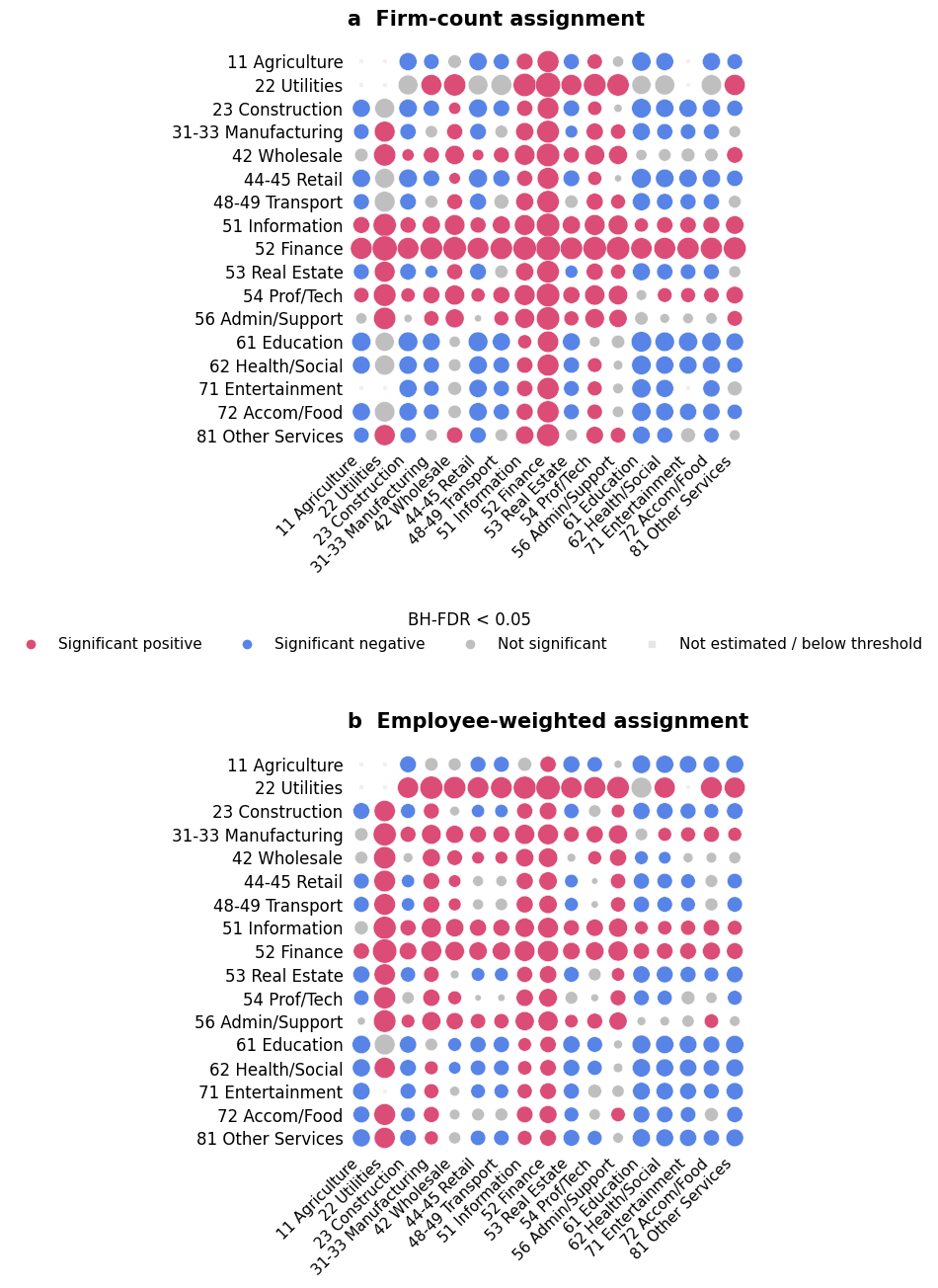}
\caption{
\textbf{Industry-pair heterogeneity under alternative industry assignments.}
Panel \textbf{a} uses firm counts, while panel \textbf{b} uses total employment to assign each firm cell's dominant industry.
Circle size indicates the absolute coefficient magnitude.
Pink and blue circles denote significant positive and negative coefficients after BH-FDR correction at $q<0.05$, while grey circles are not significant.
}
    \label{fig:industry_assignment_robustness}
\end{figure}

Supplementary Figure~\ref{fig:industry_assignment_robustness} shows that the overall pattern of industry heterogeneity is highly similar across the two definitions. Among industry-pair coefficients estimated under both specifications, the Pearson correlation is $r=0.824$ and the Spearman rank correlation is $\rho=0.836$. In addition, 83.8\% of the coefficients retain the same sign across the two assignments. These results indicate that both the magnitude and relative ordering of the industry-pair associations are broadly preserved when the dominant industry is defined using employment rather than the number of firms.

\subsection{Industry heterogeneity across functional sector groups}

The industry heterogeneity analysis suggests that the association between meeting probability and link formation differs across industries. We next ask whether this variation follows broader economic functions. The conventional primary--secondary--tertiary division groups many distinct service activities together, so we aggregate the NAICS sectors into six functional groups adapted from the framework of Browning and Singelmann \cite{browning1975emergence,browning1978transformation}. The sector assignments are listed below.

\begin{itemize}
\item \textit{1. Extractive}: Agriculture, Forestry, Fishing and Hunting (NAICS 11); and Mining, Quarrying, and Oil and Gas Extraction (21).

\item \textit{2. Transformative}: Utilities (22); Construction (23); and Manufacturing (31--33).

\item \textit{3. Distributive Services}: Wholesale Trade (42); Retail Trade (44--45); Transportation and Warehousing (48--49); and Information (51). Information is included because this sector plays a central role in producing, transmitting, and circulating information, although some activities may also provide direct inputs to production.

\item \textit{4. Producer Services}: Finance and Insurance (52); Real Estate and Rental and Leasing (53); Professional, Scientific, and Technical Services (54); Management of Companies and Enterprises (55); and Administrative and Support Services (56).

\item \textit{5. Social Services}: Educational Services (61); Health Care and Social Assistance (62); and Public Administration (92).

\item \textit{6. Personal Services}: Arts, Entertainment, and Recreation (71); Accommodation and Food Services (72); and Other Services (81).

\end{itemize}

We interact $\ln(\mathrm{MP}_{ij})$ with indicators for each functional-group pair while retaining the same controls and fixed effects as in the baseline specification. This model estimates a separate relationship between meeting probability and transaction formation for each combination of functional groups. The results are shown in Supplementary Figure~\ref{fig:functional}. The size of each dot represents the absolute magnitude of the estimated coefficient, while red and blue indicate statistically significant positive and negative associations, respectively; grey circles are not statistically significant. Positive associations are concentrated among pairs involving distributive services. Distributive--distributive and distributive--producer service pairs show the largest positive coefficients, while distributive services also exhibit significant positive associations with several other service groups. The stronger associations for pairs involving distributive services are consistent with a role for mobility-based contact in sectors that circulate goods, people, and information, although the analysis does not establish a causal mechanism. By contrast, most pairs involving extractive industries are not statistically significant.

\begin{figure}
    \centering
    \includegraphics[width=1\linewidth]{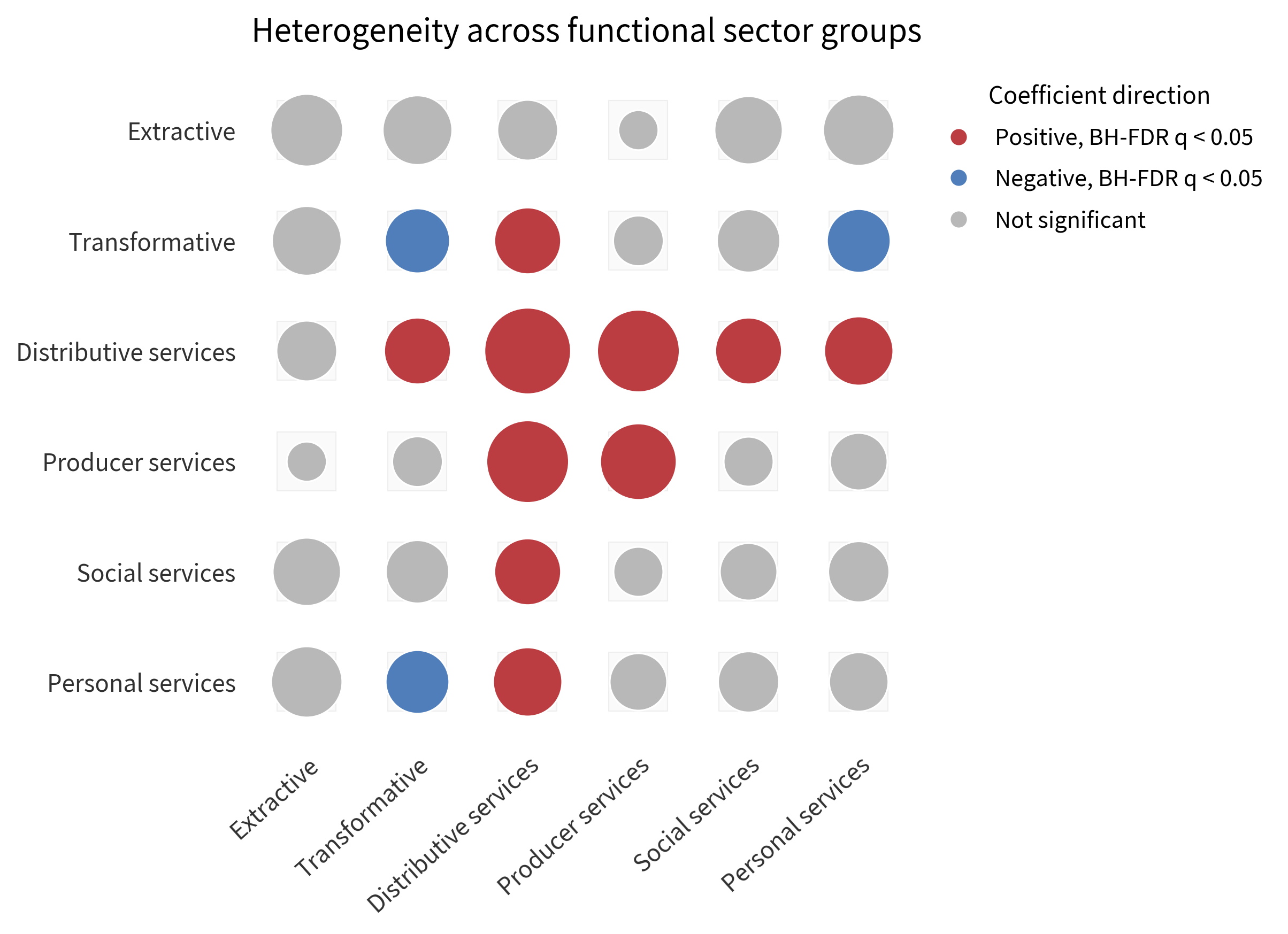}
    \caption{\textbf{Heterogeneity across functional sector groups.} Each dot represents the estimated coefficient on $\ln(\mathrm{MP}_{ij})$ for one functional sector pair. Dot area is proportional to the absolute coefficient magnitude. Coral and teal dots indicate positive and negative coefficients, respectively, whose 95\% confidence intervals exclude zero, whereas grey dots indicate statistically insignificant estimates.}
    \label{fig:functional}
\end{figure}

\subsection{Two-hop mobility exposure and link formation}

To examine whether indirect mobility-network connectivity predicts subsequent economic link formation, we consider pairs of firm cells $a$ and $c$ that are connected through at least one shared mobility neighbour $b_k$, such that both $MP_{ab_k}>0$ and $MP_{b_kc}>0$. We define the strength of the two-hop connection as
\begin{equation}
    \mathrm{TwoHop}_{ac}
    = \sum_{k \in \mathcal{N}_{ac}}
      MP_{ab_k} \times MP_{b_kc},
\end{equation}
where $\mathcal{N}_{ac}$ is the set of shared mobility neighbours of $a$ and $c$. This measure increases with both the number of shared neighbours and the strength of the two meeting-probability links forming each two-hop path. We estimate the fixed-effects logit model
\[
\begin{aligned}
\Pr(Y_{ac,t}=1)
=\Lambda\Big(
&\beta\ln(MP_{ac})
+\theta_1 Twohop_{ac}
+\theta_2 I^{\mathrm{same}}_{ac} \\
&+\theta_3 Twohop_{ac}I^{\mathrm{same}}_{ac}
+\mathbf{X}_{ac}^{\top}\boldsymbol{\gamma}
+\delta_a+\delta_c
\Big).
\end{aligned}
\]

The outcome \(Y_{ac,t}\) indicates whether a new transaction link between
\(a\) and \(c\) is first observed \(t\) years after the mobility measurement,
where \(t=1,2,3\) corresponds to 2020, 2021, and 2022, respectively.
The indicator \(I^{\mathrm{same}}_{ac}\) equals one when the cells belong to the same industry and zero otherwise.
The coefficient \(\theta_1\) captures the association between two-hop exposure
and subsequent transaction formation for cross-industry pairs, whereas
\(\theta_1+\theta_3\) gives the corresponding association for same-industry
pairs. Both estimates are conditional on direct meeting probability, the
control variables in \(\mathbf{X}_{ac}\), and fixed effects.

We assess robustness to network position by adding two pair-level controls. Let $D_i$ denote the degree of endpoint $i$ and let $W_i=\sum_j MP_{ij}$ denote its total meeting probability. We control for $ln(D_a+D_c)$ and $ln(W_a+W_c)$. These controls account for the possibility that two-hop exposure is mechanically higher for pairs involving more central or more active cells. Degree captures the number of mobility connections associated with each cell, whereas total meeting probability captures the overall strength of those connections. These controls help distinguish the association of specific two-hop connectivity from that of general network centrality and activity.

Columns~(1) and~(2) of Table~\ref{tab:twohop} report standard errors two-way clustered by $a$ and $c$. Column~(2) extends the baseline specification by adding network controls. Column~(3) retains the estimation from Column~(2) but reports dyadic cluster-robust standard errors.

\begin{table}[!htbp]
\centering
\caption{Two-hop exposure and link formation: robustness to network controls and dyadic dependence}
\label{tab:twohop}
\begin{threeparttable}
\begin{tabular}{lccc}
\toprule
 & (1) Baseline & (2) Network controls & (3) Dyadic-robust SE  \\
\midrule
\multicolumn{4}{l}{\textit{Panel A: Ever ($t \geq 1$)}} \\
Cross-industry  & 0.0103*** & 0.0100*** & 0.0100*** \\
                & (0.0022)  & (0.0021)  & (0.0022)  \\
Within-industry & 0.0003    & -0.0006   & -0.0006   \\
                & (0.0043)  & (0.0044)  & (0.0045)  \\
Observations    & 18,402,728 & 18,402,728 & 18,402,728 \\
Pseudo $R^2$    & 0.1889 & 0.1890 & 0.1890 \\
\addlinespace
\multicolumn{4}{l}{\textit{Panel B: 2020 ($t=1$)}} \\
Cross-industry  & 0.0171*** & 0.0168*** & 0.0168*** \\
                & (0.0034)  & (0.0034)  & (0.0035)  \\
Within-industry & 0.0008    & 0.0001    & 0.0001    \\
                & (0.0054)  & (0.0055)  & (0.0057)  \\
Observations    & 11,991,291 & 11,991,291 & 11,991,291 \\
Pseudo $R^2$    & 0.1355 & 0.1355 & 0.1355 \\
\addlinespace
\multicolumn{4}{l}{\textit{Panel C: 2021 ($t=2$)}} \\
Cross-industry  & 0.0095**  & 0.0090**  & 0.0090**  \\
                & (0.0041)  & (0.0044)  & (0.0043)  \\
Within-industry & -0.0039   & -0.0045   & -0.0045   \\
                & (0.0094)  & (0.0096)  & (0.0096)  \\
Observations    & 5,443,331 & 5,443,331 & 5,443,331 \\
Pseudo $R^2$    & 0.0544 & 0.0544 & 0.0544 \\
\addlinespace
\multicolumn{4}{l}{\textit{Panel D: 2022 ($t=3$)}} \\
Cross-industry  & 0.0057    & 0.0035    & 0.0035    \\
                & (0.0055)  & (0.0068)  & (0.0069)  \\
Within-industry & -0.0022   & -0.0039   & -0.0039   \\
                & (0.0077)  & (0.0080)  & (0.0082)  \\
Observations    & 5,822,859 & 5,822,859 & 5,822,859 \\
Pseudo $R^2$    & 0.0676 & 0.0678 & 0.0678 \\
\midrule
Pair-level controls        & Yes & Yes & Yes \\
Network controls   & No  & Yes & Yes \\
Cell $a$ and $c$ FE & Yes & Yes & Yes \\
Standard errors & Two-way $a/c$ & Two-way $a/c$ &  Dyadic robust  \\
\bottomrule
\end{tabular}
\begin{tablenotes}[flushleft]
\footnotesize
\item \textit{Notes:} The binary outcome is link formation between firm cells $a$ and $c$. The cross-industry slope is the main effect of two-hop exposure; the same-industry slope is the sum of the main effect and its interaction with the same-industry indicator. Column (2) adds network controls to the baseline specification. Column (3) uses the same specification and sample as column (2) but reports dyadic cluster-robust actor-CR1 standard errors. Columns (1)--(2) report standard errors two-way clustered by firm cells $a$ and $c$. Standard errors are reported in parentheses. All specifications are fixed-effects logit models and include the pair-level controls described in the methodology. $^{*}p<0.10$, $^{**}p<0.05$, $^{***}p<0.01$.
\end{tablenotes}
\end{threeparttable}
\end{table}

\bibliographystyle{unsrt} 
\bibliography{_references}